\documentclass[aps,prl,twocolumn,superscriptaddress,groupedaddress]{revtex4-2}

\usepackage{amsmath,amssymb,amsfonts}
\usepackage{bm, latexsym}
\usepackage[normalem]{ulem}
\usepackage{bbm,times}
\usepackage{braket}
\usepackage{mathtools}
\usepackage{enumerate}

\usepackage{graphicx}
\usepackage{hyperref}
\usepackage{color}

\usepackage{comment}

\newcommand{\cred}[1]{\textcolor{black}{#1}}
\newcommand{\dd}[1]{\mathrm{d}#1}
\newcommand{\credsec}[1]{\textcolor{black}{#1}}

\graphicspath{
{./image/}
}

\begin{document}

\title{Spin-Depairing-Induced Exceptional Fermionic Superfluidity}
\author{Soma Takemori}
\email{takemori.s.041d@m.isct.ac.jp}
\author{Kazuki Yamamoto}
\author{Akihisa Koga}
\affiliation{Department of Physics, Institute of Science Tokyo, Meguro, Tokyo 152-8551, Japan}
\altaffiliation{Former Tokyo Institute of Technology}

\date{\today}

\begin{abstract}
We investigate the non-Hermitian (NH) attractive Hubbard model with spin depairing, 
which is a spin-resolved asymmetric hopping that nonreciprocally operates spins in the opposite direction. 
We find that spin depairing stabilizes a superfluid state unique to the NH system. 
This phase is characterized not only by a finite order parameter, 
but also by the emergence of exceptional points (EPs) in the momentum space
%
-- a feature that starkly contrasts with previously discussed NH fermionic superfluidity, where 
EPs are absent within the superfluid state and
emerge only at the onset of the superfluid breakdown.
We uncover the rich mechanism underlying this ``\textit{exceptional fermionic superfluidity}''
by analyzing the interplay between EPs and the effective density of states of the complex energy dispersion. 
Furthermore, we reveal that 
the exceptional superfluid state breaks down induced by strong spin depairing on the cubic lattice, while 
it remains robust on the square lattice.
\end{abstract}

\maketitle

\textit{Introduction}--Open quantum systems have attracted great interest because they give rise to novel phenomena that are inaccessible in equilibrium settings~\cite{muller12, daley14, Sieberer16, fazio24}. In particular, experimental advances in ultracold atoms have realized drastic dissipative many-body phenomena~\cite{syassen08,mark12,yan13,barontini13,zhu14,patil15,labouvie16,luschen17,sponselee18,tomita19,corman19,mark20,takasu20,fahri21,benary22,ren22,huang23,huang24,tsuno24,zhao25,tao25,zhang25}, such as loss-induced superfluid-to-Mott insulator transitions \cite{tomita17}, anomalous decoherence in many-body systems \cite{bouganne20}, and dissipation-induced dynamical crossover of the magnetization \cite{honda23}. Recently, non-Hermitian (NH) systems have gained significant attention as a framework to describe many-body dynamics of open quantum systems~\cite{ashida20,ashida16,nakagawa18,resendiz20,nakagawa20,xu20,matsumoto20,zhang21eta,tajima21,yamamoto22, yamamoto23sun,han23,wang23Hal,yang24,yamamoto24}. One of the most fundamental dissipation effects is the asymmetric hopping \cite{hatano96,hatano97,hatano98,gong18, liu19}, which can induce an extreme sensitivity of eigenstates to boundary conditions, called NH skin effects~\cite{yao18, borgnia20}. The asymmetric hopping can significantly alter many-body properties \cite{fukui98B,hamazaki19,zhang20,suthar22,kawabata22,orito22dis,faugno22,li23stark,dora24,mak24,yoshida24,dupays25} as seen in NH skin effects in interacting Hatano-Nelson models~\cite{LeeE20, liu20}, breakdown of the Mott insulator \cite{fukui98}, and spin-depairing-induced phase transitions \cite{uchino12, hayata21sign, yu24}.

In NH systems, exceptional points (EPs), which are the unique nonequilibrium singularities where the eigenvalues and eigenstates of the NH Hamiltonian coalesce, can ubiquitously emerge in various physical setups \cite{kawabata19, gong18,stranky18, kawabata19X, yang21}; the laser intensity is extremely enhanced at EPs \cite{miri19}, anomalous bulk Fermi arcs can emerge due to the lifetime effect of electrons \cite{Kozii24, yoshida18, Nagai20}, and unique NH topological phenomena can occur due to the symmetry protection \cite{Budich19, Okugawa19, YoshidaT19, ZhouH19,schafer22}.
One of the salient features in NH many-body phenomena is the nonequilibrium phase transitions accompanied by EPs~\cite{wei17,yang20,fruchart21,kim24}, as seen in, e.g., parity-time symmetric systems \cite{ashida17,zhang22}, Kondo problems \cite{lourencco18}, Bose-Einstein condensates \cite{Hanai19, hanai20}, and Mott insulators \cite{nakagawa21}.
\cred{However, there are few mechanisms} 
\credsec{in which many-body phases associated with EPs emerge}
\cred{and such a realization of many-body EPs has been a major challenge even in prototypical NH many-body physics.}
In particular, NH fermionic superfluidity (NH-SF) has attracted considerable attention as it exhibits reentrant superfluid phase transitions associated with EPs protected by an emergent NH symmetry induced by the continuous quantum Zeno effect \cite{yamamoto19,gahtak18,kanazawa21,iskin21,he21,tajima23topo, li23yang, tajima24,shi24,takemori24honey,takemori24asym}. However, in NH-SF, EPs only emerge at the phase boundary, where the metastable superfluid state breaks down~\cite{yamamoto19}, the fact of which makes it difficult to experimentally access EPs on top of NH-SF.
\credsec{Since NH fermionic superfluidity can exhibit unconventional yet universal behavior at EPs in fundamental physical quantities, such as in correlation length, compressibility, and correlation functions~\cite{yamamoto19,li23yang}, the mechanism for stable NH fermionic superfluidity associated with EPs can lead to understanding essential physical responses induced by EPs in NH many-body quantum matter.}

\begin{figure}[b]
  \centering
  \includegraphics[width=\linewidth]{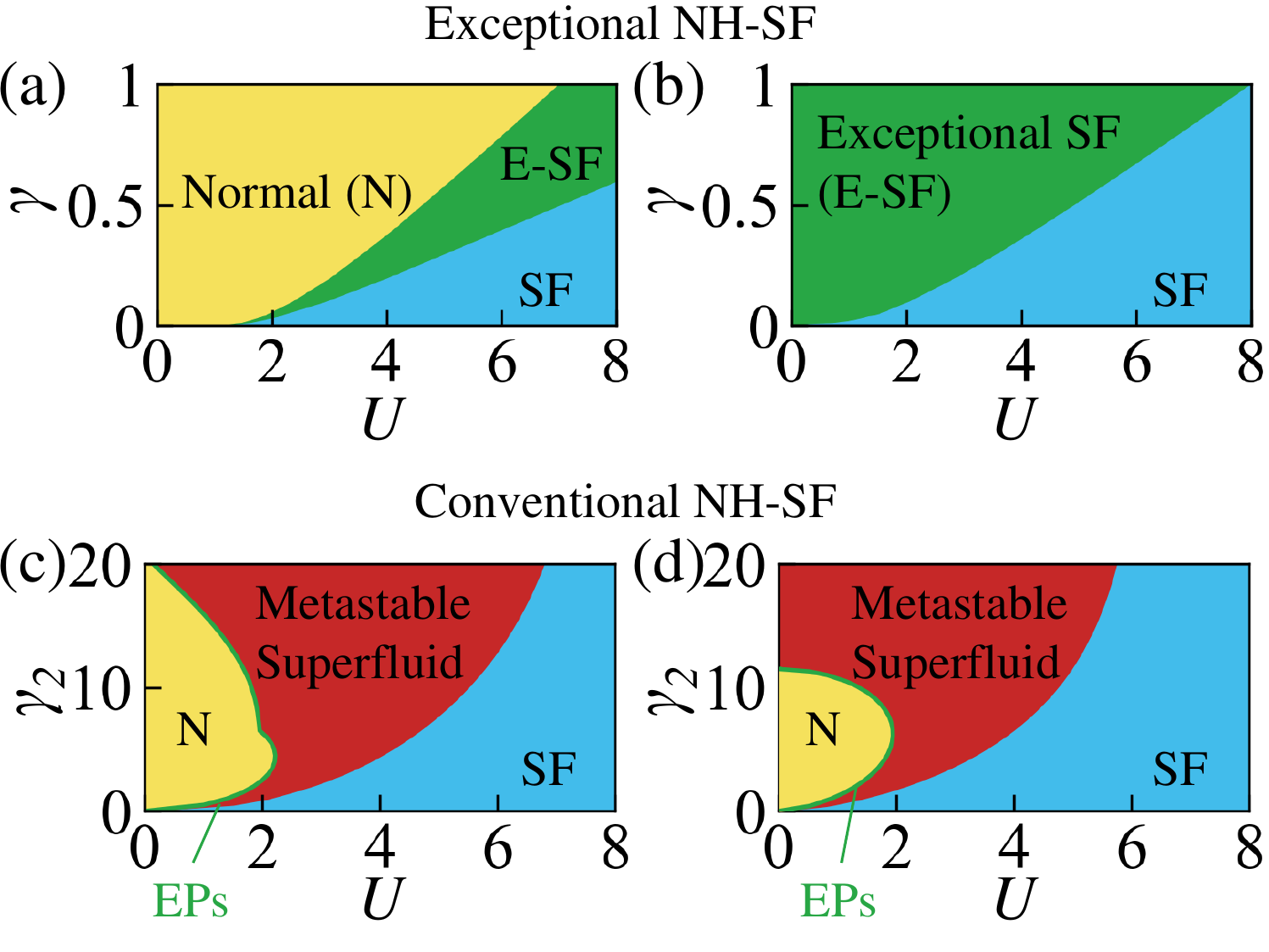}
  \caption{Phase diagrams of the attractive Hubbard model with spin depairing on (a) cubic and (b) square lattices. $\gamma$ and $U$ stand for the rate of spin depairing and the strength of attraction, respectively. 
  Spin depairing induces a stable exceptional superfluid state,
  which is not realized in the Hubbard model with two-body dissipation discussed in previous studies~\cite{yamamoto19}.
  The corresponding phase diagrams for the model with two-body dissipation $\gamma_2$ 
  on the cubic and square lattices are shown in (c) and (d), respectively \cite{footnote}.
%
  }
\label{sdBCS_PD_image}
\end{figure}

In this Letter, we study the NH attractive Hubbard model 
to find that spin depairing stabilizes the superfluid state unique to the NH system,
which we refer to as \textit{exceptional fermionic superfluidity}.
This phase is characterized not only by a finite order parameter, 
but also by the emergence of EPs in the momentum space.
This is in stark contrast to the conventional NH-SF (see Fig.~\ref{sdBCS_PD_image}). 
To clarify the origin of the exceptional superfluid state, 
we employ the NH-BCS theory, introducing the effective density of states (DOS) in the complex energy plane.
We find that the interplay between EPs and the effective DOS gives rise to the rich behavior of the exceptional SF. 
Furthermore, we demonstrate that the exceptional
superfluid state breaks down induced by strong spin depairing on the cubic lattice, while it remains robust on the square
lattice.
\cred{The exceptional fermionic superfluidity uncovers a new structure linking EPs and NH fermionic superfluidity,
  and opens a new avenue towards experimental realization of EPs in NH many-body physics.}

\begin{figure}[t]
  \centering
  \includegraphics[width=\linewidth]{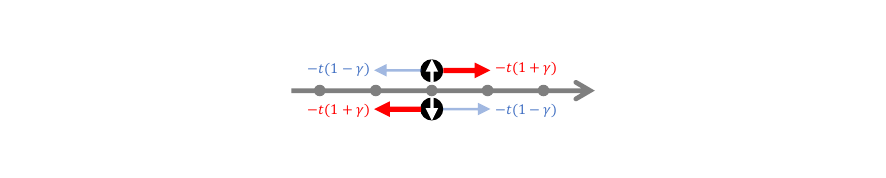}
  \caption{Spin-resolved asymmetric hopping (spin depairing) given in Eq.~\eqref{sdBCS_effNHHamiltonian_eq}.
    The hopping amplitude of the fermion with an up spin along the positive (negative) $x,y,z$ directions is described by $-t(1+\gamma)$ [$-t(1-\gamma)$].
In contrast, the hopping amplitude for down-spin fermions is the opposite of that for the up-spin fermions.
  }
  \label{sdBCS_effNHHamiltonian_image}
\end{figure}

\textit{NH-BCS theory with spin depairing}--We consider the attractive Hubbard model with spin-resolved Hatano-Nelson-type asymmetric hopping \cite{hatano96,hatano97,hatano98, uchino12, hayata21sign, yu24} as
\begin{align}
    H = &-t\sum_{\langle i,j\rangle,\sigma} \left[(1+\sigma \gamma)c_{i\sigma}^{\dagger}c_{j\sigma} + (1-\sigma \gamma)c_{j\sigma}^{\dagger}c_{i\sigma}\right] \notag \\
  &-U \sum_{i}\left(n_{i\uparrow}-\frac{1}{2}\right)\left(n_{i\downarrow}-\frac{1}{2}\right), \label{sdBCS_effNHHamiltonian_eq}
\end{align}
where $c_{i\sigma}^{\dagger}(c_{i\sigma})$ is the creation (annhilation) operator of a fermion with spin $\sigma=\uparrow,\downarrow$ at site $i$ and $n_{i\sigma}=c_{i\sigma}^{\dagger}c_{i\sigma}$ is the particle number operator. 
The summation is taken over nearest-neighbor site pair $\langle i,j\rangle$, 
$t$ is the hopping amplitude, $U\:(>0)$ is the interaction strength, and $\gamma\;(0\le \gamma\le 1)$ is the rate of spin depairing. 
The spin-resolved asymmetric hopping is schematically
shown in Fig.~\ref{sdBCS_effNHHamiltonian_image}. 
Such effective NH descriptions have been employed in the context of dissipative ultracold atoms on an optical lattice~\cite{lindblad76, daley14}, and the asymmetric hopping can be implemented by the gauge flux \cite{hatano96,hatano97,hatano98, fukui98, uchino12} or the collective one-body loss \cite{gong18, takemori24asym, yu24} (see Appendix A in End Matter for the experimental implementation of our system).

By performing the Fourier transformation, the kinetic term of Eq.~\eqref{sdBCS_effNHHamiltonian_eq} is rewritten as 
\begin{gather}
H^\mathrm{kin} = \sum_{\bm{k}}\left(\epsilon_{\bm{k}}c_{\bm{k}\uparrow}^{\dagger}c_{\bm{k}\uparrow} + \epsilon_{\bm{k}}^{\ast}c_{\bm{k}\downarrow}^{\dagger}c_{\bm{k}\downarrow}\right),
\end{gather}
where the energy dispersion is given as
\begin{equation}
    \epsilon_{\bm{k}} = -2t\sum_{\alpha=1}^d\left(\cos k_{\alpha}+i\gamma\sin k_{\alpha}\right),
    \label{sdBCS_enedis_eq}
\end{equation}
where $d$ is the dimension. 
Note that the presence of spin depairing makes the energy dispersion complex and spin-dependent. 
Throughout this paper, we set $t$ as the unit of energy.

To analyze the ground state for the NH system,
we employ the NH-BCS theory~\cite{yamamoto19}. We start with a path-integral representation of the partition function as 
\begin{gather}
Z = \int \mathcal{D}\bar{c} \mathcal{D}c \exp(-S),\\
S =  \int_{0}^{\beta^\prime}\mathrm{d}\tau\sum_{i}[\bar{c}_{i\sigma}\partial_{\tau}c_{i\sigma} + H(\bar{c}_{i\sigma},c_{i\sigma})],
\end{gather}
where $\beta^\prime$ is a parameter characteristic of the statistical weight of eigenstates, 
and $H(\bar{c}_{i\sigma},c_{i\sigma})$ is obtained by replacing the fermionic operators with Grassman variables. 
We then use the Hubbard-Stratonovich transformation to introduce auxiliary fields $\Delta$ and $\bar{\Delta}$.
Integrating out the fermionic degrees of freeedom and
taking the saddle point of the action with respect to
$\Delta$ and $\bar{\Delta}$, we obtain the NH gap equation as
\begin{equation}
  \frac{1}{U} = \frac{1}{N}\sum_{\bm{k}}\frac{1}{2E_{\bm{k}}}, \label{sdBCS_gap_eq}
\end{equation}
where $N$ is the number of sites, $E_{\bm{k}}=\sqrt{\epsilon_{\bm{k}}^{2}+\Delta\bar{\Delta}}$, and we have taken the limit $\beta^\prime\to \infty$.
Note that the gauges of the order parameters $\Delta$ and $\bar{\Delta}$
can be taken to be real.
This is in contrast to the result for the conventional NH-SF induced
by two-body dissipation,
where the order parameters are intrinsically complex~\cite{yamamoto19}.
In the following, we set $\Delta_{0}=\Delta=\bar{\Delta}\in\mathbb R$.

By using the order parameter $\Delta_0$
that satisfies the NH gap equation~\eqref{sdBCS_gap_eq},
the NH-BCS Hamiltonian can be rewritten as
\begin{align}
H^{\mathrm{BCS}}=
\sum_{\bm{k}}
\begin{pmatrix}
  c_{\bm{k}\uparrow}^{\dagger}  & c_{-\bm{k}\downarrow}
\end{pmatrix}\begin{pmatrix}
  \epsilon_{\bm{k}} & \Delta_{0} \\ \Delta_{0} & -\epsilon_{\bm{k}}
\end{pmatrix} \begin{pmatrix}
  c_{\bm{k}\uparrow} \\ c_{-\bm{k}\downarrow}^{\dagger} 
\end{pmatrix}.
\label{eq_BCSmean}
\end{align}
The ground state energy is given as
  \begin{align}
E_{\mathrm{SF}}=\frac{\Delta_{0}^2}{U}-\sum_{k}[E_{\bm{k}}-\epsilon_{\bm{k}}].
\end{align}
The condensation energy is given by $E_{\mathrm{cond}}=E_{\mathrm{SF}}-E_{\mathrm{N}}$,
where $E_\mathrm{N}$ is the energy for the normal state with $\Delta_0=0$. Here, we remark that we obtain the real condensation energy that gives the same decay rate between the superfluid and normal states.
This is distinct from that for the conventional NH system, 
where the superfluid state usually decays faster than the normal state.
Note that the effective Hamiltonian \eqref{eq_BCSmean} becomes non-diagonalizable at $E_{\bm{k}}=0$,
indicating the presence of the EPs with $\epsilon_k=\pm i\Delta_0$.

\begin{figure}[b]
  \centering
  \includegraphics[width=\linewidth]{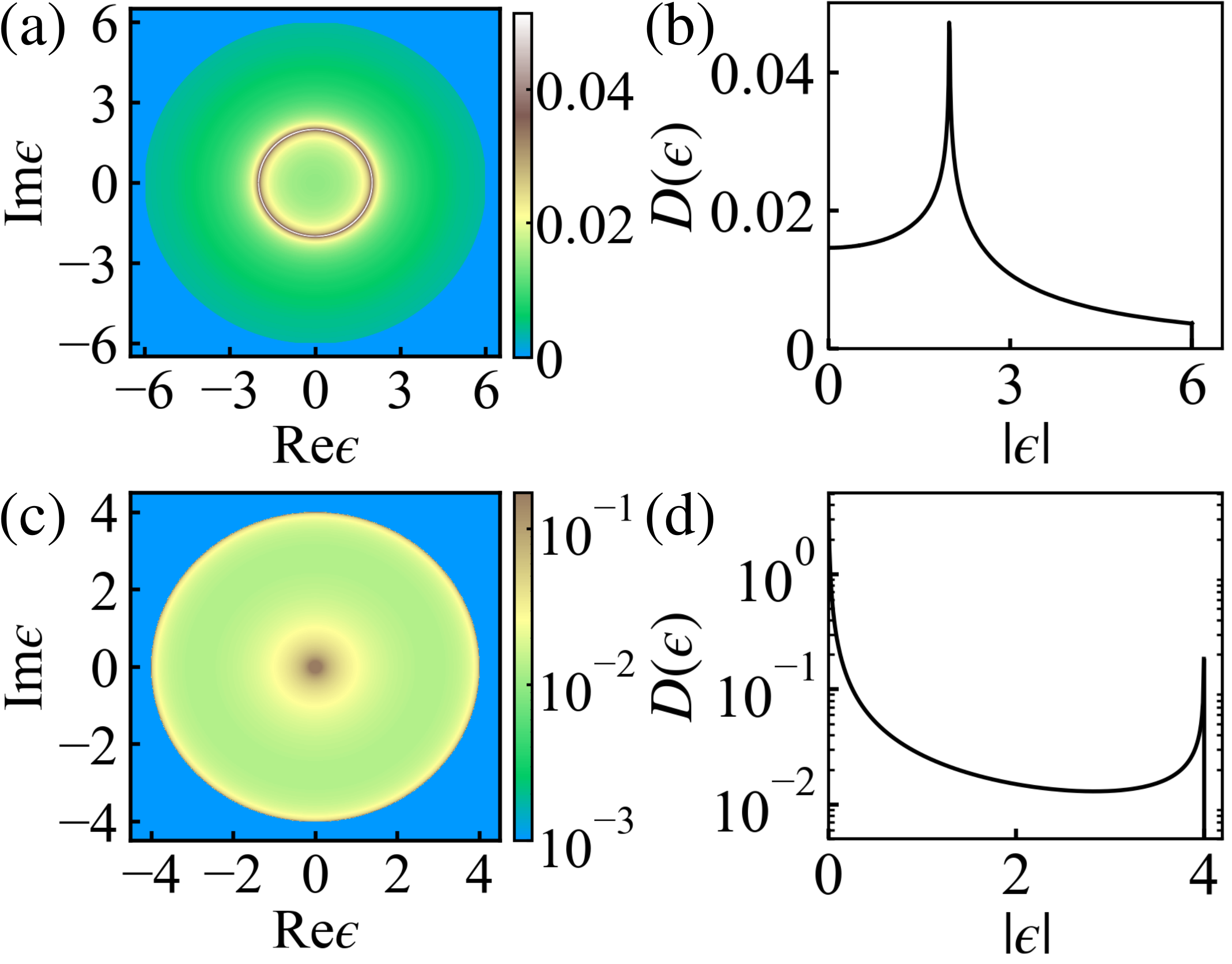}
  \caption{(a) [(c)] 
  Contour plot of the effective DOS~\eqref{sdBCS_eff_DOS_def_eq} on the cubic (square) lattice for $\gamma=1$. 
  Blue region indicates the effective DOS being zero. 
  (b)[(d)] Cross section of the effective DOS. 
  }
  \label{sdBCS_eff_DOS_23D_image}
\end{figure}

To solve the gap equation~\eqref{sdBCS_gap_eq}, 
it is convenient to introduce the effective DOS
of a complex variable $\epsilon$ as
\begin{align}
  D_\gamma(\epsilon) &=\frac{1}{N}\sum_{\bm{k}}
  \delta\left({\rm Re}\,\epsilon-{\rm Re}\,\epsilon_{\bm k}\right)
  \delta({\rm Im}\,\epsilon-{\rm Im}\,\epsilon_{\bm k}),
  \label{sdBCS_eff_DOS_def_eq}
\end{align}
where $\delta(x)$ is the delta function. 
The effective DOS satisfies
$D_\gamma(\epsilon)=D_{\gamma=1}({\rm Re}\,\epsilon+i{\rm Im}\,\epsilon/\gamma)/\gamma$ and
$D_\gamma(\epsilon)=D_\gamma(\epsilon^*)$, and
it is finite at the elliptic region in the complex-$\epsilon$ plane.
Some details are provided in Supplemental Material~\cite{Supple}.
Figure~\ref{sdBCS_eff_DOS_23D_image} shows the effective DOS
in the systems with $\gamma=1$ on the cubic and square lattices.
We find that both lattices have the isotropic effective DOS in the complex-$\epsilon$ plane,
while distinct singularities emerge between them.
The effective DOS logarithmically diverges at $|\epsilon|=2$
in the cubic lattice case, 
while
algebraically diverges
%
at $|\epsilon|=0$ and $|\epsilon|=4$ in the square lattice case,
which is clearly found in the cross section of the effective DOS
shown in Figs.~\ref{sdBCS_eff_DOS_23D_image}(b) and (d), respectively.
The qualitative difference in the effective DOS affects
the ground state phase diagram, which will be discussed later.
The gap equation is then given as
\begin{equation}
    \frac{1}{U} = \int\!\!\!\!\int\mathrm{d}\text{Re}\,\epsilon \:\mathrm{d}\text{Im}\,\epsilon\frac{D_\gamma(\epsilon)}{2\sqrt{\epsilon^{2}+\Delta_0^2}}. \label{sdBCS_gap_effDOS_ref_eq}
\end{equation}

\begin{figure}[t]
    \centering
    \includegraphics[width=8cm]{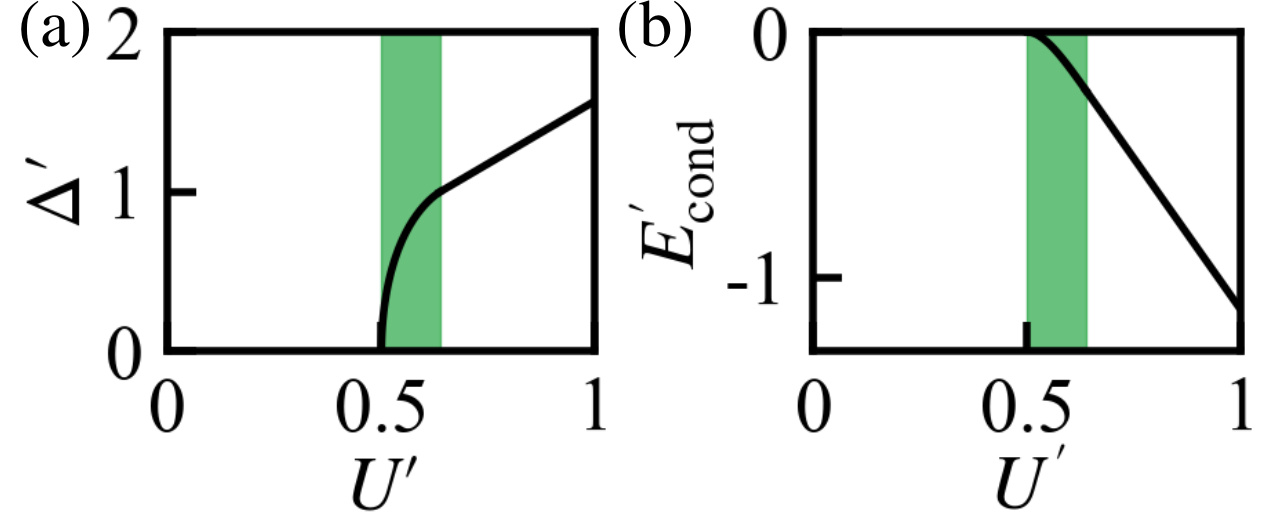}
    \caption{(a) Normalized order parameter and (b) condensation energy
      for the NH system with the constant effective DOS.
      Exceptional SF is realized in the green shaded area.
    }
    \label{sdBCS_App_uniDOS_D_Ec_image}
\end{figure}

\textit{Analytical results for the spin-depairing-induced exceptional SF}--To understand how the attractive interaction induces
the SF unique to the NH system with spin depairing,
we first consider the effective DOS in the absence of singularities.
Specifically, we introduce an effective DOS defined
in the complex-$\epsilon$ plane,
which takes a constant value within a circular domain:
\begin{equation}
  D(\epsilon) = \begin{cases}
    1/(\pi \epsilon_R^{2}), \; & \; (|\epsilon|\le \epsilon_R), \\
    0, \; & \; (|\epsilon|> \epsilon_R),
  \end{cases}
  \label{eq_DOScon}
\end{equation}
where the cutoff $\epsilon_R$ has been introduced
so that the normalization condition $\int\!\!\!\int \mathrm{d}\text{Re}\epsilon\mathrm{d}\text{Im}\epsilon D(\epsilon)=1$
is satisfied.
Integrating over $\epsilon$ in the NH gap equation~\eqref{sdBCS_gap_eq},
we obtain the following equation
\begin{equation}
  \frac{1}{U^\prime} =\begin{cases}
    \frac{\pi}{2\Delta^\prime}, \; & \; (\Delta^\prime>1), \\
    \frac{\pi}{2\Delta^\prime} + \sqrt{1-\Delta^{\prime2}} - \frac{1}{\Delta^\prime}\tan^{-1}(\frac{\sqrt{1-\Delta^{\prime 2}}}{\Delta^\prime}), \; & \; (\Delta^\prime\le 1), \label{sdBCS_cDOS_gapeq_analy_sol_eq}
  \end{cases}
\end{equation}
where we have introduced the parameters
$U^\prime\equiv U/(\pi \epsilon_R)$ and $\Delta^\prime\equiv\Delta_{0}/\epsilon_R$,
for simplicity.
We obtain the condensation energy as 
\begin{align}
  E_{\text{cond}}' = \begin{cases}
    &-\frac{\pi \Delta'}{2} + \frac{4}{3},\;\;\; ( \Delta'>1 ),    \\
    &\rule{0pt}{0.5ex} \\ 
    &-\frac{\pi \Delta'}{2}-\frac{4(1-\Delta'^{2} )^{3/2}}{3} -\Delta'^{2}\sqrt{1-\Delta'^{2}}\\
    &+\Delta'\tan^{-1}(\frac{\sqrt{1-\Delta'^{2}  }}{\Delta'})   +\frac{4}{3}, \; (\Delta'\le 1 ), \label{sdBCS_cDOS_cond_ene_analy_sol}
  \end{cases}
\end{align}
(for the detailed derivation, see Supplemental Material~\cite{Supple}).

The results for this simple case are shown
in Fig.~\ref{sdBCS_App_uniDOS_D_Ec_image}.
When $U'$ is large, the strong attraction stabilizes
the trivial SF state with finite $\Delta'$.
This is consistent with the fact that
the condensation energy is negative.
In this SF phase, $\Delta_0>\epsilon_R$ and
the quasiparticle energy $E_k$ remains finite in the complex energy plane. 
As the attraction decreases,
the order parameter $\Delta_0$ also decreases and
eventually reaches $\Delta_0=\epsilon_{R}$ at $U'=2/\pi\sim 0.637$.
This means that the EPs emerge at the edge of the circular DOS, i.e., $\epsilon=\pm i\Delta_0$,
which is nothing but the emergence of the exceptional SF.
Remarkably, this phase has not been discussed in nonequilibrium systems, to the best of our knowledge.
As the attraction decreases further,
the order parameter continues to decrease and the EPs move toward
the origin in the complex-$\epsilon$ plane.
Finally, $\Delta_0$ vanishes at 
the critical interaction strength $U^\prime=U_c^\prime\equiv1/2$,
as shown in Fig.~\ref{sdBCS_App_uniDOS_D_Ec_image}(a).
Near this point, the order parameter exhibits
mean-field asymptotic behavior
$\Delta^\prime\sim \sqrt{(U^\prime-U_c^\prime)}$.
The key result here is that
the exceptional SF -- characterized by the finite order parameter and the presence of EPs inside the complex-$\epsilon$ plane --
is indeed stabilized within a finite interaction range.
\cred{Our findings of the exceptional SF reveal a qualitatively new connection between EPs and NH many-body physics,
namely,} 
\credsec{the emergence of the fermionic superfluidity accompanied by EPs within}
\cred{a stable finite parameter region.}
\cred{We note that in nonequilibrium systems, a “phase” is commonly defined as one that
can be adiabatically connected to the corresponding conventional phase
through the gradual introduction of dissipation.}
In the following, we examine whether 
such a exceptional SF also arises in the systems with
cubic and square lattices
and discuss the effects of singularities in the effective DOS.

\begin{figure}[t]
  \centering
  \includegraphics[width=\linewidth]{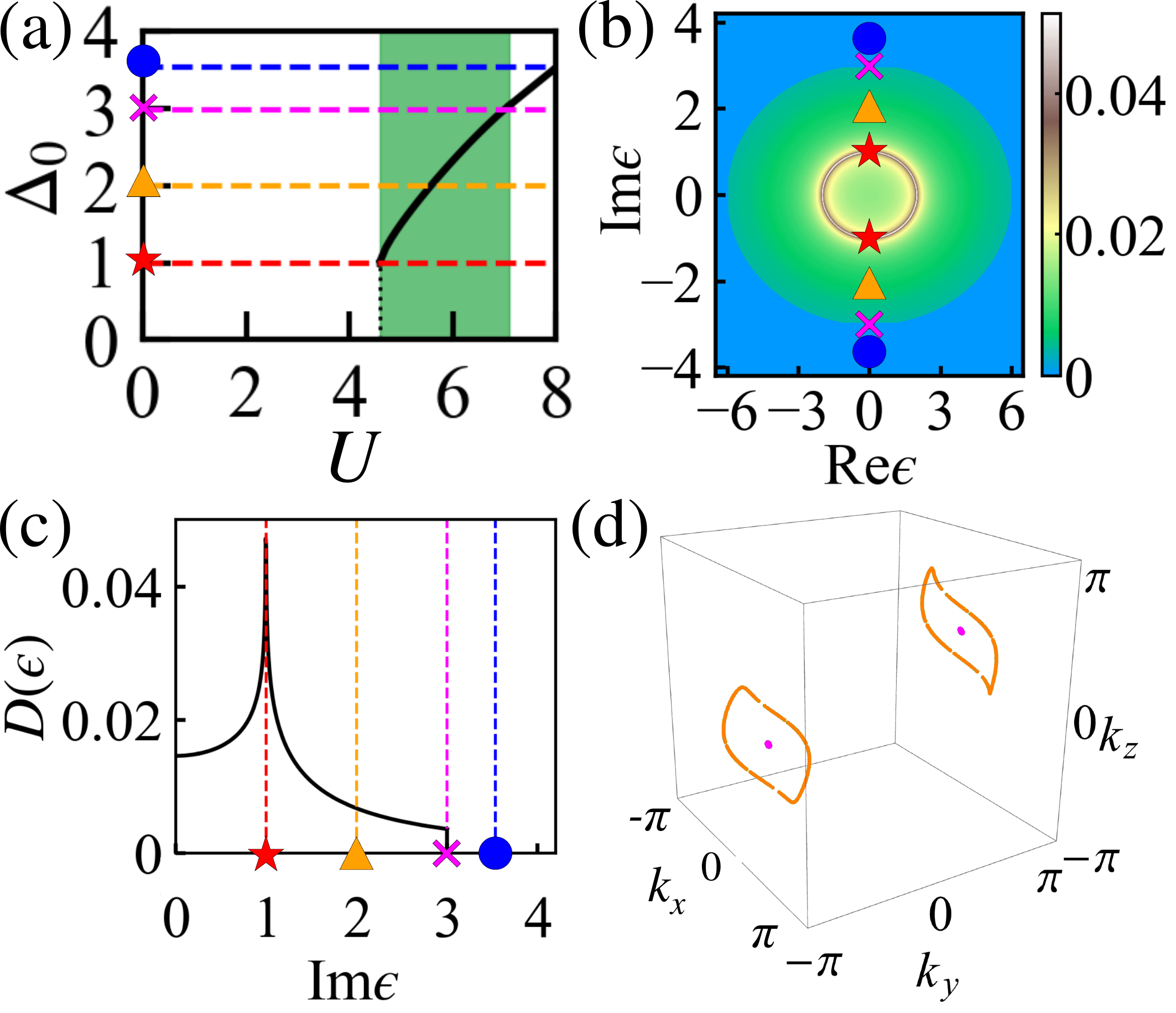}
  \caption{Numerical results for the NH system on the cubic lattice with $\gamma=0.5$.
    (a) Order parameter $\Delta_{0}$ as a function of the attraction $U$.
    Exceptional SF is realized in the green shaded area,
    where EPs appear at $\epsilon=\pm i\Delta_0$.
    (b) Contour plot of the effective DOS~\eqref{sdBCS_eff_DOS_def_eq} and
    (c) its cross section at $\mathrm{Re}\epsilon=0$.
    The marks in (b) and (c) stand for the points
    $\mathrm{Im}\epsilon=i\Delta_0$ at $\mathrm{Re}\epsilon=0$
    by using the corresponding value of $\Delta_0$ shown in (a).
    (d) Exceptional lines in the momentum space when $U=5.57$ and $7.02$.}
  \label{sdBCS_3D_DOS_Delta_image}
\end{figure}

\textit{Exceptional SF on the cubic lattice}--We first consider the NH system on the cubic lattice.
Figure~\ref{sdBCS_3D_DOS_Delta_image}(a) shows the order parameter $\Delta_0$
as a function of the attractive interaction for the system with $\gamma=0.5$.
When $U=8$, a stable SF state is realized with large $\Delta_{0}\sim 3.53$.
In this case, the position $\epsilon=\pm i \Delta_0$ is located
outside the elliptic region with finite $D(\epsilon)$
in the complex-$\epsilon$ plane,
as shown in Figs.~\ref{sdBCS_3D_DOS_Delta_image}(a)-(c) (see the blue circle).
This means the absence of EPs in the superfluid state.
As the attraction decreases,
the order parameter $\Delta_0$ also decreases,
and eventually reaches $\Delta_0=3$ when $U\sim 7.02$.
In this case, EPs emerge at $\epsilon=\pm i \Delta_0$
in the complex-$\epsilon$ plane (see the pink cross mark),
indicating the onset of exceptional SF.
The corresponding EPs in the momemtum space emerges at $\pm(\pi/2,\pi/2,\pi/2)$,
which are shown as the pink circles in Fig.~\ref{sdBCS_3D_DOS_Delta_image}(d). 
When $U\sim 5.57$,
the exceptional SF is realized with $\Delta_0\sim 2$. 
The corresponding EPs
are observed as the lines in the momentum space
in Fig.~\ref{sdBCS_3D_DOS_Delta_image}(d). 
As the attraction further decreases,
the order parameter suddenly vanishes at $\Delta_0=1$.
Since this singularity is distinct from the critical behavior for the system with the constant effective DOS, this behavior can be attributed to the logarithmic divergence
of the effective DOS at that energy (see the red star).
As a result, the exceptional SF becomes unstable and
the normal state is realized instead (see Appendix B in End Matter).
This is in contrast to the Hermitian limit where
the superfluid state remains stable for any attractive interaction $U$.

By performing similar calculations for several $\gamma$,
we obtain the phase diagram, as shown in Fig.~\ref{sdBCS_PD_image}(a).
We find that an exotic exceptional SF, characterized by
the finite order parameter and
the quasiparticle energy exhibiting nodes at EPs,
is stabilized over a finite range of interaction and spin depairing strengths.
This exceptional SF is qualitatively distinct from both metastable and stable
SF states realized in the NH system with two-body dissipation~\cite{yamamoto19}.
In fact, no nodes appear in these states and EPs emerge only
when the metastable solution for the superfluid state vanishes,
as shown in Fig.~\ref{sdBCS_PD_image}(c).

\begin{figure}[tb]
  \centering
  \includegraphics[width=\linewidth]{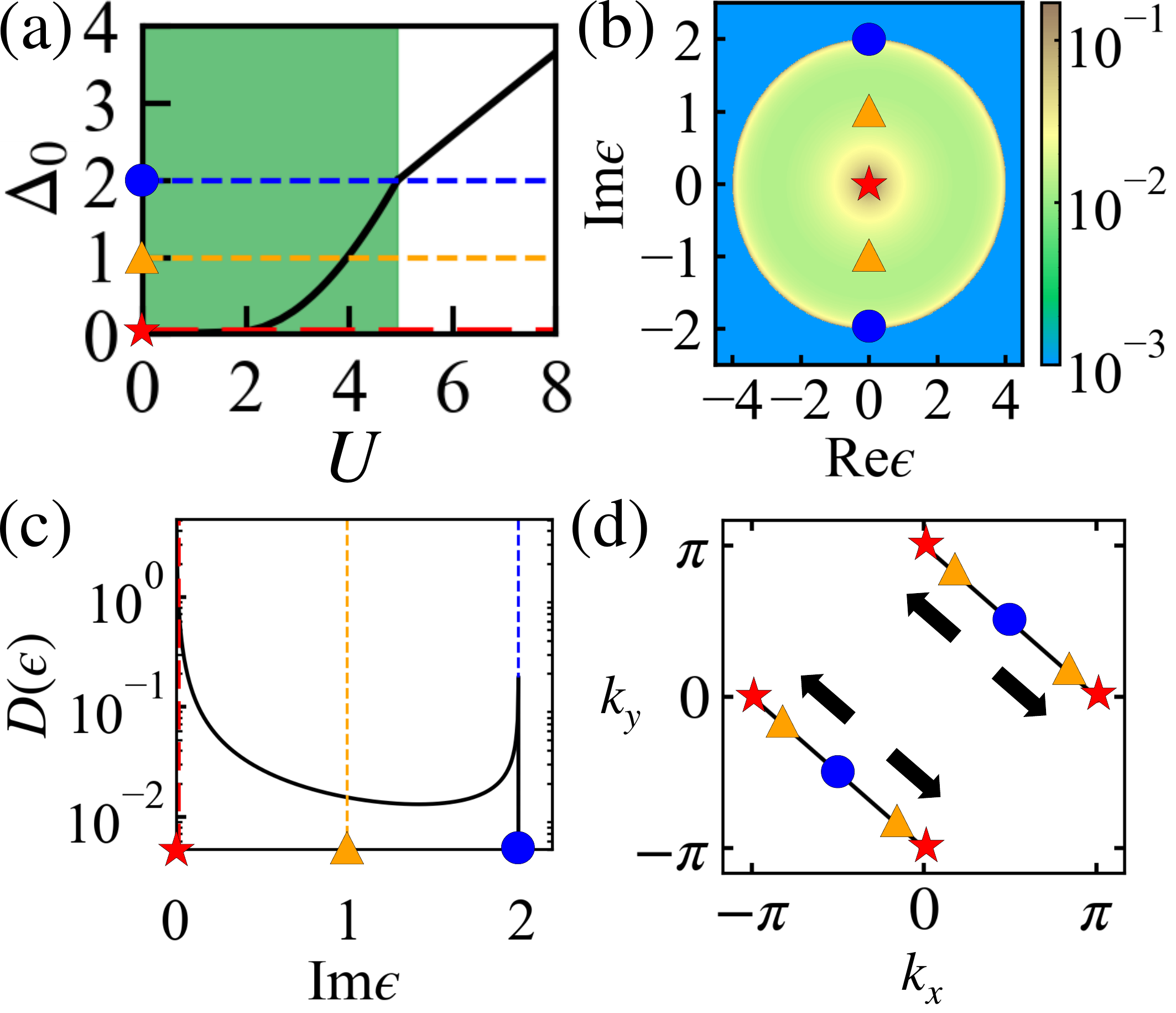}
  \caption{Numerical results for the NH system on the square lattice with $\gamma=0.5$.
    (a) Order parameter $\Delta_{0}$ as a function of the attraction $U$.
    Exceptional SF is realized in the green shaded area.
    (b) Contour plot of the effective DOS~\eqref{sdBCS_eff_DOS_def_eq} and
    (c) its cross section at $\mathrm{Re}\epsilon=0$.
    The marks in (b) and (c) stand for the points $\mathrm{Im}\epsilon=i\Delta_0$ at $\mathrm{Re}\epsilon=0$ by using the corresponding value of $\Delta_0$ shown in (a). (d) EPs in the momentum space in the NH system with $U=0, 3.95$, and $4.92$. }
  \label{sdBCS_2D_DOS_Delta_image}
\end{figure}

\textit{Exceptional SF on the square lattice}--We present the result for the NH system on the square lattice and
highlight the qualitative difference from the results for the cubic lattice.
In Fig.~\ref{sdBCS_2D_DOS_Delta_image}(a),
we show the order parameter $\Delta_{0}$ as a function of $U$ for $\gamma=0.5$.
We find that the order parameter is finite in the whole region,
implying that the spontaneous symmetry breaking always occurs.
Furthermore, we find that when $U\le U_c(\sim 4.92)$,
the order parameter is less than the edge of the elliptic region with finite DOS
{i.e.} $\Delta_0\le 2$,
meaning that the exceptional SF is stabilized.
When $U=U_c$, EPs emerge at $\pm(\pi/2, \pi/2)$ in the momentum space,
as shown in Fig.~\ref{sdBCS_2D_DOS_Delta_image}(d).
Decreasing $U$, each EP split into two EPs and 
move toward $(0,\pm\pi)$ and $(\pm\pi,0)$.
It is intriguing that this exceptional SF is stable
for the weak coupling regime.
This is in contrast to the results for the system on the cubic lattice,
where the phase transition occurs between the exceptional SF and the normal state.
This is attributed to the structure of the effective DOS.
Figures~\ref{sdBCS_2D_DOS_Delta_image}(b) and \ref{sdBCS_2D_DOS_Delta_image}(c)
show that the effective DOS exhibits two singularities at the origin and
the perimeter (see the red star and blue circle).
These singularities are characterized by 
$D(\epsilon)\sim |\epsilon|^{-1}$ for $|\epsilon|\rightarrow +0$ and
$D(\epsilon)\sim (2-|\epsilon|)^{-1/2}$ for $|\epsilon|\rightarrow 2-0$.
When $\Delta_0\sim 2$, the singularity in the effective DOS
leads to a cusp
in the curve of the order parameter,
as shown in Fig.~\ref{sdBCS_2D_DOS_Delta_image}(a).
In contrast, when $\Delta_0\rightarrow 0$,
the integral in Eq.~\eqref{sdBCS_gap_effDOS_ref_eq} tends to diverge due to the strong singularity
in the effective DOS at $|\epsilon|=0$.
By taking this into account correctly,
we find asymptotic behavior $\Delta_{0}\sim \exp(-c_\gamma/U)$ 
with positive constant $c_\gamma$ (see Supplemental Material~\cite{Supple} for details).
Therefore, we can say that in the NH system on the square lattice,
an infinitesimal attractive interaction induce the exceptional SF,
which is clearly found in the phase diagram [see Fig.~\ref{sdBCS_PD_image}(b)].
\credsec{The intrinsic differences in the phase diagrams between the cubic and square lattices
originate from differences in the effective DOS, which arise from the underlying lattice
geometry.}

Finally, we would like to comment on the exceptional SF state.
We have found that the EPs appear in the momentum space 
as exceptional lines in the cubic lattice, and exceptional points in the square lattice. 
This indicates that the dimensionality of EPs is given by $d-2$ in a $d$-dimensional system.
This result contrasts with that of the conventional NH systems with two-body dissipation,
where the dimensionality is typically given by $d-1$~\cite{yamamoto19}.
We also note that
gapless excitations at isolated points in momentum space appear 
in (usually metastable) superfluid state
when equilibrium SF is considered under magnetic fields or with mass imbalance~\cite{sarma63,liu03,sheehy06,sheehy07,barzykin09}. 
However, we emphasize that spin-depairing-induced exceptional SF is always stable 
due to the negative condensation energy and is a unique feature of NH systems characterized by EPs.

\textit{Conclusion}--In this Letter, we have investigated the NH attractive Hubbard model with spin depairing and found the emergence of stable exceptional SF, whose behavior is characterized by the interplay between EPs and the effective DOS. Moreover, we have elucidated that the spin-depairing-induced phase transitions occur on the cubic lattice, while exceptional SF is stable for arbitrary strength of attraction on the square lattice. 
\cred{We have shown that the existence of the exceptional SF is established in the thermodynamic limit, where edge effects are excluded~\cite{yao18,borgnia20}.
Therefore, within the present framework, we are not able to provide a complete numerical analysis
under open boundary conditions (OBCs).}
\credsec{To incorporate edge effects, one would need to reconstruct the NH BCS theory under open boundary conditions by introducing site-dependent and inhomogeneous order parameters, which is highly nontrivial.}
\credsec{We note that, though some works consider NH skin effect under OBCs by assuming bulk order parameters, the order parameters are not self-consistently determined within BCS theory and this approach should be distinguished from our study.}
As we have focused on the analysis of the order parameter,
the NH physical quantities in various setups, such as the susceptibility and the superfluid weight, are worth studying to further understand the exceptional SF. Moreover, it is intriguing how the exceptional SF appears for more complex energy-band structures. 
Finally, since the nonreciprocity in Liouvillian many-body dynamics is actively investigated~\cite{Song19, haga21, yamamoto20, Yang22, Lee23,shu23, Hu23, hanai24}, it is also interesting to study nonreciprocal dissipation effects on the superfluid state in nonequilibrium steady states as well as the exceptional SF in general Markovian dynamics \cite{yamamoto21, mazza23} that may exhibit Liouvillian EPs~\cite{mitnganti19}.

\begin{acknowledgments}
  We thank Masaya Nakagawa, Kohei Kawabata, and Norio Kawakami for fruitful discussions.
  This work was supported by Grant-in-Aid for Scientific Research from JSPS,
  KAKENHI Grants Nos. JP25K17327 (K.Y.), and JP22K03525, JP25H01521, JP25H01398 (A.K.). 
  S.T. was supported by the Sasakawa Scientific Research Grant from the Japan Science Society and JST SPRING, Japan Grant Number JPMJSP2180. K.Y. was also supported by Murata Science and Education Foundation, Hirose Foundation, the Precise Measurement Technology Promotion Foundation, and the Fujikura Foundation.
\end{acknowledgments}


\begin{thebibliography}{128}%
\makeatletter
\providecommand \@ifxundefined [1]{%
 \@ifx{#1\undefined}
}%
\providecommand \@ifnum [1]{%
 \ifnum #1\expandafter \@firstoftwo
 \else \expandafter \@secondoftwo
 \fi
}%
\providecommand \@ifx [1]{%
 \ifx #1\expandafter \@firstoftwo
 \else \expandafter \@secondoftwo
 \fi
}%
\providecommand \natexlab [1]{#1}%
\providecommand \enquote  [1]{``#1''}%
\providecommand \bibnamefont  [1]{#1}%
\providecommand \bibfnamefont [1]{#1}%
\providecommand \citenamefont [1]{#1}%
\providecommand \href@noop [0]{\@secondoftwo}%
\providecommand \href [0]{\begingroup \@sanitize@url \@href}%
\providecommand \@href[1]{\@@startlink{#1}\@@href}%
\providecommand \@@href[1]{\endgroup#1\@@endlink}%
\providecommand \@sanitize@url [0]{\catcode `\\12\catcode `\$12\catcode
  `\&12\catcode `\#12\catcode `\^12\catcode `\_12\catcode `\%12\relax}%
\providecommand \@@startlink[1]{}%
\providecommand \@@endlink[0]{}%
\providecommand \url  [0]{\begingroup\@sanitize@url \@url }%
\providecommand \@url [1]{\endgroup\@href {#1}{\urlprefix }}%
\providecommand \urlprefix  [0]{URL }%
\providecommand \Eprint [0]{\href }%
\providecommand \doibase [0]{https://doi.org/}%
\providecommand \selectlanguage [0]{\@gobble}%
\providecommand \bibinfo  [0]{\@secondoftwo}%
\providecommand \bibfield  [0]{\@secondoftwo}%
\providecommand \translation [1]{[#1]}%
\providecommand \BibitemOpen [0]{}%
\providecommand \bibitemStop [0]{}%
\providecommand \bibitemNoStop [0]{.\EOS\space}%
\providecommand \EOS [0]{\spacefactor3000\relax}%
\providecommand \BibitemShut  [1]{\csname bibitem#1\endcsname}%
\let\auto@bib@innerbib\@empty
\bibitem [{\citenamefont {M{\"u}ller}\ \emph {et~al.}(2012)\citenamefont
  {M{\"u}ller}, \citenamefont {Diehl}, \citenamefont {Pupillo},\ and\
  \citenamefont {Zoller}}]{muller12}%
  \BibitemOpen
  \bibfield  {author} {\bibinfo {author} {\bibfnamefont {M.}~\bibnamefont
  {M{\"u}ller}}, \bibinfo {author} {\bibfnamefont {S.}~\bibnamefont {Diehl}},
  \bibinfo {author} {\bibfnamefont {G.}~\bibnamefont {Pupillo}},\ and\ \bibinfo
  {author} {\bibfnamefont {P.}~\bibnamefont {Zoller}},\ }\bibfield  {title}
  {\bibinfo {title} {Engineered open systems and quantum simulations with atoms
  and ions},\ }\href@noop {} {\bibfield  {journal} {\bibinfo  {journal} {Adv.
  Atom. Mol. Opt. Phys.}\ }\textbf {\bibinfo {volume} {61}},\ \bibinfo {pages}
  {1} (\bibinfo {year} {2012})}\BibitemShut {NoStop}%
\bibitem [{\citenamefont {Daley}(2014)}]{daley14}%
  \BibitemOpen
  \bibfield  {author} {\bibinfo {author} {\bibfnamefont {A.~J.}\ \bibnamefont
  {Daley}},\ }\bibfield  {title} {\bibinfo {title} {{Quantum trajectories and
  open many-body quantum systems}},\ }\href
  {https://doi.org/10.1080/00018732.2014.933502} {\bibfield  {journal}
  {\bibinfo  {journal} {Adv. Phys.}\ }\textbf {\bibinfo {volume} {63}},\
  \bibinfo {pages} {77} (\bibinfo {year} {2014})}\BibitemShut {NoStop}%
\bibitem [{\citenamefont {Sieberer}\ \emph {et~al.}(2016)\citenamefont
  {Sieberer}, \citenamefont {Buchhold},\ and\ \citenamefont
  {Diehl}}]{Sieberer16}%
  \BibitemOpen
  \bibfield  {author} {\bibinfo {author} {\bibfnamefont {L.~M.}\ \bibnamefont
  {Sieberer}}, \bibinfo {author} {\bibfnamefont {M.}~\bibnamefont {Buchhold}},\
  and\ \bibinfo {author} {\bibfnamefont {S.}~\bibnamefont {Diehl}},\ }\bibfield
   {title} {\bibinfo {title} {{Keldysh field theory for driven open quantum
  systems}},\ }\href {https://doi.org/10.1088/0034-4885/79/9/096001} {\bibfield
   {journal} {\bibinfo  {journal} {Rep. Prog. Phys.}\ }\textbf {\bibinfo
  {volume} {79}},\ \bibinfo {pages} {096001} (\bibinfo {year}
  {2016})}\BibitemShut {NoStop}%
\bibitem [{\citenamefont {Fazio}\ \emph {et~al.}(2024)\citenamefont {Fazio},
  \citenamefont {Keeling}, \citenamefont {Mazza},\ and\ \citenamefont
  {Schir\`o}}]{fazio24}%
  \BibitemOpen
  \bibfield  {author} {\bibinfo {author} {\bibfnamefont {R.}~\bibnamefont
  {Fazio}}, \bibinfo {author} {\bibfnamefont {J.}~\bibnamefont {Keeling}},
  \bibinfo {author} {\bibfnamefont {L.}~\bibnamefont {Mazza}},\ and\ \bibinfo
  {author} {\bibfnamefont {M.}~\bibnamefont {Schir\`o}},\ }\bibfield  {title}
  {\bibinfo {title} {Many-body open quantum systems},\ }\href@noop {}
  {\bibfield  {journal} {\bibinfo  {journal} {arXiv:2409.10300}\ } (\bibinfo
  {year} {2024})}\BibitemShut {NoStop}%
\bibitem [{\citenamefont {Syassen}\ \emph {et~al.}(2008)\citenamefont
  {Syassen}, \citenamefont {Bauer}, \citenamefont {Lettner}, \citenamefont
  {Volz}, \citenamefont {Dietze}, \citenamefont {García-Ripoll}, \citenamefont
  {Cirac}, \citenamefont {Rempe},\ and\ \citenamefont {Dürr}}]{syassen08}%
  \BibitemOpen
  \bibfield  {author} {\bibinfo {author} {\bibfnamefont {N.}~\bibnamefont
  {Syassen}}, \bibinfo {author} {\bibfnamefont {D.~M.}\ \bibnamefont {Bauer}},
  \bibinfo {author} {\bibfnamefont {M.}~\bibnamefont {Lettner}}, \bibinfo
  {author} {\bibfnamefont {T.}~\bibnamefont {Volz}}, \bibinfo {author}
  {\bibfnamefont {D.}~\bibnamefont {Dietze}}, \bibinfo {author} {\bibfnamefont
  {J.~J.}\ \bibnamefont {García-Ripoll}}, \bibinfo {author} {\bibfnamefont
  {J.~I.}\ \bibnamefont {Cirac}}, \bibinfo {author} {\bibfnamefont
  {G.}~\bibnamefont {Rempe}},\ and\ \bibinfo {author} {\bibfnamefont
  {S.}~\bibnamefont {Dürr}},\ }\bibfield  {title} {\bibinfo {title} {{Strong
  Dissipation Inhibits Losses and Induces Correlations in Cold Molecular
  Gases}},\ }\href {https://doi.org/10.1126/science.1155309} {\bibfield
  {journal} {\bibinfo  {journal} {Science}\ }\textbf {\bibinfo {volume}
  {320}},\ \bibinfo {pages} {1329} (\bibinfo {year} {2008})}\BibitemShut
  {NoStop}%
\bibitem [{\citenamefont {Mark}\ \emph {et~al.}(2012)\citenamefont {Mark},
  \citenamefont {Haller}, \citenamefont {Lauber}, \citenamefont {Danzl},
  \citenamefont {Janisch}, \citenamefont {B\"uchler}, \citenamefont {Daley},\
  and\ \citenamefont {N\"agerl}}]{mark12}%
  \BibitemOpen
  \bibfield  {author} {\bibinfo {author} {\bibfnamefont {M.~J.}\ \bibnamefont
  {Mark}}, \bibinfo {author} {\bibfnamefont {E.}~\bibnamefont {Haller}},
  \bibinfo {author} {\bibfnamefont {K.}~\bibnamefont {Lauber}}, \bibinfo
  {author} {\bibfnamefont {J.~G.}\ \bibnamefont {Danzl}}, \bibinfo {author}
  {\bibfnamefont {A.}~\bibnamefont {Janisch}}, \bibinfo {author} {\bibfnamefont
  {H.~P.}\ \bibnamefont {B\"uchler}}, \bibinfo {author} {\bibfnamefont {A.~J.}\
  \bibnamefont {Daley}},\ and\ \bibinfo {author} {\bibfnamefont {H.-C.}\
  \bibnamefont {N\"agerl}},\ }\bibfield  {title} {\bibinfo {title}
  {{Preparation and Spectroscopy of a Metastable Mott-Insulator State with
  Attractive Interactions}},\ }\href
  {https://doi.org/10.1103/PhysRevLett.108.215302} {\bibfield  {journal}
  {\bibinfo  {journal} {Phys. Rev. Lett.}\ }\textbf {\bibinfo {volume} {108}},\
  \bibinfo {pages} {215302} (\bibinfo {year} {2012})}\BibitemShut {NoStop}%
\bibitem [{\citenamefont {Yan}\ \emph {et~al.}(2013)\citenamefont {Yan},
  \citenamefont {Moses}, \citenamefont {Gadway}, \citenamefont {Covey},
  \citenamefont {Hazzard}, \citenamefont {Rey}, \citenamefont {Jin},\ and\
  \citenamefont {Ye}}]{yan13}%
  \BibitemOpen
  \bibfield  {author} {\bibinfo {author} {\bibfnamefont {B.}~\bibnamefont
  {Yan}}, \bibinfo {author} {\bibfnamefont {S.~A.}\ \bibnamefont {Moses}},
  \bibinfo {author} {\bibfnamefont {B.}~\bibnamefont {Gadway}}, \bibinfo
  {author} {\bibfnamefont {J.~P.}\ \bibnamefont {Covey}}, \bibinfo {author}
  {\bibfnamefont {K.~R.}\ \bibnamefont {Hazzard}}, \bibinfo {author}
  {\bibfnamefont {A.~M.}\ \bibnamefont {Rey}}, \bibinfo {author} {\bibfnamefont
  {D.~S.}\ \bibnamefont {Jin}},\ and\ \bibinfo {author} {\bibfnamefont
  {J.}~\bibnamefont {Ye}},\ }\bibfield  {title} {\bibinfo {title} {Observation
  of dipolar spin-exchange interactions with lattice-confined polar
  molecules},\ }\href@noop {} {\bibfield  {journal} {\bibinfo  {journal}
  {Nature (London)}\ }\textbf {\bibinfo {volume} {501}},\ \bibinfo {pages}
  {521} (\bibinfo {year} {2013})}\BibitemShut {NoStop}%
\bibitem [{\citenamefont {Barontini}\ \emph {et~al.}(2013)\citenamefont
  {Barontini}, \citenamefont {Labouvie}, \citenamefont {Stubenrauch},
  \citenamefont {Vogler}, \citenamefont {Guarrera},\ and\ \citenamefont
  {Ott}}]{barontini13}%
  \BibitemOpen
  \bibfield  {author} {\bibinfo {author} {\bibfnamefont {G.}~\bibnamefont
  {Barontini}}, \bibinfo {author} {\bibfnamefont {R.}~\bibnamefont {Labouvie}},
  \bibinfo {author} {\bibfnamefont {F.}~\bibnamefont {Stubenrauch}}, \bibinfo
  {author} {\bibfnamefont {A.}~\bibnamefont {Vogler}}, \bibinfo {author}
  {\bibfnamefont {V.}~\bibnamefont {Guarrera}},\ and\ \bibinfo {author}
  {\bibfnamefont {H.}~\bibnamefont {Ott}},\ }\bibfield  {title} {\bibinfo
  {title} {{Controlling the Dynamics of an Open Many-Body Quantum System with
  Localized Dissipation}},\ }\href
  {https://doi.org/10.1103/PhysRevLett.110.035302} {\bibfield  {journal}
  {\bibinfo  {journal} {Phys. Rev. Lett.}\ }\textbf {\bibinfo {volume} {110}},\
  \bibinfo {pages} {035302} (\bibinfo {year} {2013})}\BibitemShut {NoStop}%
\bibitem [{\citenamefont {Zhu}\ \emph {et~al.}(2014)\citenamefont {Zhu},
  \citenamefont {Gadway}, \citenamefont {Foss-Feig}, \citenamefont
  {Schachenmayer}, \citenamefont {Wall}, \citenamefont {Hazzard}, \citenamefont
  {Yan}, \citenamefont {Moses}, \citenamefont {Covey}, \citenamefont {Jin},
  \citenamefont {Ye}, \citenamefont {Holland},\ and\ \citenamefont
  {Rey}}]{zhu14}%
  \BibitemOpen
  \bibfield  {author} {\bibinfo {author} {\bibfnamefont {B.}~\bibnamefont
  {Zhu}}, \bibinfo {author} {\bibfnamefont {B.}~\bibnamefont {Gadway}},
  \bibinfo {author} {\bibfnamefont {M.}~\bibnamefont {Foss-Feig}}, \bibinfo
  {author} {\bibfnamefont {J.}~\bibnamefont {Schachenmayer}}, \bibinfo {author}
  {\bibfnamefont {M.~L.}\ \bibnamefont {Wall}}, \bibinfo {author}
  {\bibfnamefont {K.~R.~A.}\ \bibnamefont {Hazzard}}, \bibinfo {author}
  {\bibfnamefont {B.}~\bibnamefont {Yan}}, \bibinfo {author} {\bibfnamefont
  {S.~A.}\ \bibnamefont {Moses}}, \bibinfo {author} {\bibfnamefont {J.~P.}\
  \bibnamefont {Covey}}, \bibinfo {author} {\bibfnamefont {D.~S.}\ \bibnamefont
  {Jin}}, \bibinfo {author} {\bibfnamefont {J.}~\bibnamefont {Ye}}, \bibinfo
  {author} {\bibfnamefont {M.}~\bibnamefont {Holland}},\ and\ \bibinfo {author}
  {\bibfnamefont {A.~M.}\ \bibnamefont {Rey}},\ }\bibfield  {title} {\bibinfo
  {title} {{Suppressing the Loss of Ultracold Molecules Via the Continuous
  Quantum Zeno Effect}},\ }\href
  {https://doi.org/10.1103/PhysRevLett.112.070404} {\bibfield  {journal}
  {\bibinfo  {journal} {Phys. Rev. Lett.}\ }\textbf {\bibinfo {volume} {112}},\
  \bibinfo {pages} {070404} (\bibinfo {year} {2014})}\BibitemShut {NoStop}%
\bibitem [{\citenamefont {Patil}\ \emph {et~al.}(2015)\citenamefont {Patil},
  \citenamefont {Chakram},\ and\ \citenamefont {Vengalattore}}]{patil15}%
  \BibitemOpen
  \bibfield  {author} {\bibinfo {author} {\bibfnamefont {Y.~S.}\ \bibnamefont
  {Patil}}, \bibinfo {author} {\bibfnamefont {S.}~\bibnamefont {Chakram}},\
  and\ \bibinfo {author} {\bibfnamefont {M.}~\bibnamefont {Vengalattore}},\
  }\bibfield  {title} {\bibinfo {title} {{Measurement-Induced Localization of
  an Ultracold Lattice Gas}},\ }\href
  {https://doi.org/10.1103/PhysRevLett.115.140402} {\bibfield  {journal}
  {\bibinfo  {journal} {Phys. Rev. Lett.}\ }\textbf {\bibinfo {volume} {115}},\
  \bibinfo {pages} {140402} (\bibinfo {year} {2015})}\BibitemShut {NoStop}%
\bibitem [{\citenamefont {Labouvie}\ \emph {et~al.}(2016)\citenamefont
  {Labouvie}, \citenamefont {Santra}, \citenamefont {Heun},\ and\ \citenamefont
  {Ott}}]{labouvie16}%
  \BibitemOpen
  \bibfield  {author} {\bibinfo {author} {\bibfnamefont {R.}~\bibnamefont
  {Labouvie}}, \bibinfo {author} {\bibfnamefont {B.}~\bibnamefont {Santra}},
  \bibinfo {author} {\bibfnamefont {S.}~\bibnamefont {Heun}},\ and\ \bibinfo
  {author} {\bibfnamefont {H.}~\bibnamefont {Ott}},\ }\bibfield  {title}
  {\bibinfo {title} {{Bistability in a Driven-Dissipative Superfluid}},\ }\href
  {https://doi.org/10.1103/PhysRevLett.116.235302} {\bibfield  {journal}
  {\bibinfo  {journal} {Phys. Rev. Lett.}\ }\textbf {\bibinfo {volume} {116}},\
  \bibinfo {pages} {235302} (\bibinfo {year} {2016})}\BibitemShut {NoStop}%
\bibitem [{\citenamefont {L\"uschen}\ \emph {et~al.}(2017)\citenamefont
  {L\"uschen}, \citenamefont {Bordia}, \citenamefont {Hodgman}, \citenamefont
  {Schreiber}, \citenamefont {Sarkar}, \citenamefont {Daley}, \citenamefont
  {Fischer}, \citenamefont {Altman}, \citenamefont {Bloch},\ and\ \citenamefont
  {Schneider}}]{luschen17}%
  \BibitemOpen
  \bibfield  {author} {\bibinfo {author} {\bibfnamefont {H.~P.}\ \bibnamefont
  {L\"uschen}}, \bibinfo {author} {\bibfnamefont {P.}~\bibnamefont {Bordia}},
  \bibinfo {author} {\bibfnamefont {S.~S.}\ \bibnamefont {Hodgman}}, \bibinfo
  {author} {\bibfnamefont {M.}~\bibnamefont {Schreiber}}, \bibinfo {author}
  {\bibfnamefont {S.}~\bibnamefont {Sarkar}}, \bibinfo {author} {\bibfnamefont
  {A.~J.}\ \bibnamefont {Daley}}, \bibinfo {author} {\bibfnamefont {M.~H.}\
  \bibnamefont {Fischer}}, \bibinfo {author} {\bibfnamefont {E.}~\bibnamefont
  {Altman}}, \bibinfo {author} {\bibfnamefont {I.}~\bibnamefont {Bloch}},\ and\
  \bibinfo {author} {\bibfnamefont {U.}~\bibnamefont {Schneider}},\ }\bibfield
  {title} {\bibinfo {title} {{Signatures of Many-Body Localization in a
  Controlled Open Quantum System}},\ }\href
  {https://doi.org/10.1103/PhysRevX.7.011034} {\bibfield  {journal} {\bibinfo
  {journal} {Phys. Rev. X}\ }\textbf {\bibinfo {volume} {7}},\ \bibinfo {pages}
  {011034} (\bibinfo {year} {2017})}\BibitemShut {NoStop}%
\bibitem [{\citenamefont {Sponselee}\ \emph {et~al.}(2018)\citenamefont
  {Sponselee}, \citenamefont {Freystatzky}, \citenamefont {Abeln},
  \citenamefont {Diem}, \citenamefont {Hundt}, \citenamefont {Kochanke},
  \citenamefont {Ponath}, \citenamefont {Santra}, \citenamefont {Mathey},
  \citenamefont {Sengstock},\ and\ \citenamefont {Becker}}]{sponselee18}%
  \BibitemOpen
  \bibfield  {author} {\bibinfo {author} {\bibfnamefont {K.}~\bibnamefont
  {Sponselee}}, \bibinfo {author} {\bibfnamefont {L.}~\bibnamefont
  {Freystatzky}}, \bibinfo {author} {\bibfnamefont {B.}~\bibnamefont {Abeln}},
  \bibinfo {author} {\bibfnamefont {M.}~\bibnamefont {Diem}}, \bibinfo {author}
  {\bibfnamefont {B.}~\bibnamefont {Hundt}}, \bibinfo {author} {\bibfnamefont
  {A.}~\bibnamefont {Kochanke}}, \bibinfo {author} {\bibfnamefont
  {T.}~\bibnamefont {Ponath}}, \bibinfo {author} {\bibfnamefont
  {B.}~\bibnamefont {Santra}}, \bibinfo {author} {\bibfnamefont
  {L.}~\bibnamefont {Mathey}}, \bibinfo {author} {\bibfnamefont
  {K.}~\bibnamefont {Sengstock}},\ and\ \bibinfo {author} {\bibfnamefont
  {C.}~\bibnamefont {Becker}},\ }\bibfield  {title} {\bibinfo {title}
  {{Dynamics of ultracold quantum gases in the dissipative Fermi--Hubbard
  model}},\ }\href {https://doi.org/10.1088/2058-9565/aadccd} {\bibfield
  {journal} {\bibinfo  {journal} {Quantum Sci. Technol.}\ }\textbf {\bibinfo
  {volume} {4}},\ \bibinfo {pages} {014002} (\bibinfo {year}
  {2018})}\BibitemShut {NoStop}%
\bibitem [{\citenamefont {Tomita}\ \emph {et~al.}(2019)\citenamefont {Tomita},
  \citenamefont {Nakajima}, \citenamefont {Takasu},\ and\ \citenamefont
  {Takahashi}}]{tomita19}%
  \BibitemOpen
  \bibfield  {author} {\bibinfo {author} {\bibfnamefont {T.}~\bibnamefont
  {Tomita}}, \bibinfo {author} {\bibfnamefont {S.}~\bibnamefont {Nakajima}},
  \bibinfo {author} {\bibfnamefont {Y.}~\bibnamefont {Takasu}},\ and\ \bibinfo
  {author} {\bibfnamefont {Y.}~\bibnamefont {Takahashi}},\ }\bibfield  {title}
  {\bibinfo {title} {{Dissipative Bose-Hubbard system with intrinsic two-body
  loss}},\ }\href {https://doi.org/10.1103/PhysRevA.99.031601} {\bibfield
  {journal} {\bibinfo  {journal} {Phys. Rev. A}\ }\textbf {\bibinfo {volume}
  {99}},\ \bibinfo {pages} {031601} (\bibinfo {year} {2019})}\BibitemShut
  {NoStop}%
\bibitem [{\citenamefont {Corman}\ \emph {et~al.}(2019)\citenamefont {Corman},
  \citenamefont {Fabritius}, \citenamefont {H\"ausler}, \citenamefont {Mohan},
  \citenamefont {Dogra}, \citenamefont {Husmann}, \citenamefont {Lebrat},\ and\
  \citenamefont {Esslinger}}]{corman19}%
  \BibitemOpen
  \bibfield  {author} {\bibinfo {author} {\bibfnamefont {L.}~\bibnamefont
  {Corman}}, \bibinfo {author} {\bibfnamefont {P.}~\bibnamefont {Fabritius}},
  \bibinfo {author} {\bibfnamefont {S.}~\bibnamefont {H\"ausler}}, \bibinfo
  {author} {\bibfnamefont {J.}~\bibnamefont {Mohan}}, \bibinfo {author}
  {\bibfnamefont {L.~H.}\ \bibnamefont {Dogra}}, \bibinfo {author}
  {\bibfnamefont {D.}~\bibnamefont {Husmann}}, \bibinfo {author} {\bibfnamefont
  {M.}~\bibnamefont {Lebrat}},\ and\ \bibinfo {author} {\bibfnamefont
  {T.}~\bibnamefont {Esslinger}},\ }\bibfield  {title} {\bibinfo {title}
  {{Quantized conductance through a dissipative atomic point contact}},\ }\href
  {https://doi.org/10.1103/PhysRevA.100.053605} {\bibfield  {journal} {\bibinfo
   {journal} {Phys. Rev. A}\ }\textbf {\bibinfo {volume} {100}},\ \bibinfo
  {pages} {053605} (\bibinfo {year} {2019})}\BibitemShut {NoStop}%
\bibitem [{\citenamefont {Mark}\ \emph {et~al.}(2020)\citenamefont {Mark},
  \citenamefont {Flannigan}, \citenamefont {Meinert}, \citenamefont {D'Incao},
  \citenamefont {Daley},\ and\ \citenamefont {N\"agerl}}]{mark20}%
  \BibitemOpen
  \bibfield  {author} {\bibinfo {author} {\bibfnamefont {M.~J.}\ \bibnamefont
  {Mark}}, \bibinfo {author} {\bibfnamefont {S.}~\bibnamefont {Flannigan}},
  \bibinfo {author} {\bibfnamefont {F.}~\bibnamefont {Meinert}}, \bibinfo
  {author} {\bibfnamefont {J.~P.}\ \bibnamefont {D'Incao}}, \bibinfo {author}
  {\bibfnamefont {A.~J.}\ \bibnamefont {Daley}},\ and\ \bibinfo {author}
  {\bibfnamefont {H.-C.}\ \bibnamefont {N\"agerl}},\ }\bibfield  {title}
  {\bibinfo {title} {{Interplay between coherent and dissipative dynamics of
  bosonic doublons in an optical lattice}},\ }\href
  {https://doi.org/10.1103/PhysRevResearch.2.043050} {\bibfield  {journal}
  {\bibinfo  {journal} {Phys. Rev. Res.}\ }\textbf {\bibinfo {volume} {2}},\
  \bibinfo {pages} {043050} (\bibinfo {year} {2020})}\BibitemShut {NoStop}%
\bibitem [{\citenamefont {Takasu}\ \emph {et~al.}(2020)\citenamefont {Takasu},
  \citenamefont {Yagami}, \citenamefont {Ashida}, \citenamefont {Hamazaki},
  \citenamefont {Kuno},\ and\ \citenamefont {Takahashi}}]{takasu20}%
  \BibitemOpen
  \bibfield  {author} {\bibinfo {author} {\bibfnamefont {Y.}~\bibnamefont
  {Takasu}}, \bibinfo {author} {\bibfnamefont {T.}~\bibnamefont {Yagami}},
  \bibinfo {author} {\bibfnamefont {Y.}~\bibnamefont {Ashida}}, \bibinfo
  {author} {\bibfnamefont {R.}~\bibnamefont {Hamazaki}}, \bibinfo {author}
  {\bibfnamefont {Y.}~\bibnamefont {Kuno}},\ and\ \bibinfo {author}
  {\bibfnamefont {Y.}~\bibnamefont {Takahashi}},\ }\bibfield  {title} {\bibinfo
  {title} {{PT-symmetric non-Hermitian quantum many-body system using ultracold
  atoms in an optical lattice with controlled dissipation}},\ }\href
  {https://doi.org/10.1093/ptep/ptaa094} {\bibfield  {journal} {\bibinfo
  {journal} {Prog. Theor. Exp. Phys.}\ }\textbf {\bibinfo {volume} {2020}},\
  \bibinfo {pages} {12A110} (\bibinfo {year} {2020})}\BibitemShut {NoStop}%
\bibitem [{\citenamefont {Öztürk}\ \emph {et~al.}(2021)\citenamefont
  {Öztürk}, \citenamefont {Lappe}, \citenamefont {Hellmann}, \citenamefont
  {Schmitt}, \citenamefont {Klaers}, \citenamefont {Vewinger}, \citenamefont
  {Kroha},\ and\ \citenamefont {Weitz}}]{fahri21}%
  \BibitemOpen
  \bibfield  {author} {\bibinfo {author} {\bibfnamefont {F.~E.}\ \bibnamefont
  {Öztürk}}, \bibinfo {author} {\bibfnamefont {T.}~\bibnamefont {Lappe}},
  \bibinfo {author} {\bibfnamefont {G.}~\bibnamefont {Hellmann}}, \bibinfo
  {author} {\bibfnamefont {J.}~\bibnamefont {Schmitt}}, \bibinfo {author}
  {\bibfnamefont {J.}~\bibnamefont {Klaers}}, \bibinfo {author} {\bibfnamefont
  {F.}~\bibnamefont {Vewinger}}, \bibinfo {author} {\bibfnamefont
  {J.}~\bibnamefont {Kroha}},\ and\ \bibinfo {author} {\bibfnamefont
  {M.}~\bibnamefont {Weitz}},\ }\bibfield  {title} {\bibinfo {title}
  {{Observation of a non-Hermitian phase transition in an optical quantum
  gas}},\ }\href {https://doi.org/10.1126/science.abe9869} {\bibfield
  {journal} {\bibinfo  {journal} {Science}\ }\textbf {\bibinfo {volume}
  {372}},\ \bibinfo {pages} {88} (\bibinfo {year} {2021})}\BibitemShut
  {NoStop}%
\bibitem [{\citenamefont {Benary}\ \emph {et~al.}(2022)\citenamefont {Benary},
  \citenamefont {Baals}, \citenamefont {Bernhart}, \citenamefont {Jiang},
  \citenamefont {Röhrle},\ and\ \citenamefont {Ott}}]{benary22}%
  \BibitemOpen
  \bibfield  {author} {\bibinfo {author} {\bibfnamefont {J.}~\bibnamefont
  {Benary}}, \bibinfo {author} {\bibfnamefont {C.}~\bibnamefont {Baals}},
  \bibinfo {author} {\bibfnamefont {E.}~\bibnamefont {Bernhart}}, \bibinfo
  {author} {\bibfnamefont {J.}~\bibnamefont {Jiang}}, \bibinfo {author}
  {\bibfnamefont {M.}~\bibnamefont {Röhrle}},\ and\ \bibinfo {author}
  {\bibfnamefont {H.}~\bibnamefont {Ott}},\ }\bibfield  {title} {\bibinfo
  {title} {{Experimental observation of a dissipative phase transition in a
  multi-mode many-body quantum system}},\ }\href
  {https://doi.org/10.1088/1367-2630/ac97b6} {\bibfield  {journal} {\bibinfo
  {journal} {New Journal of Physics}\ }\textbf {\bibinfo {volume} {24}},\
  \bibinfo {pages} {103034} (\bibinfo {year} {2022})}\BibitemShut {NoStop}%
\bibitem [{\citenamefont {Ren}\ \emph {et~al.}(2022)\citenamefont {Ren},
  \citenamefont {Liu}, \citenamefont {Zhao}, \citenamefont {He}, \citenamefont
  {Pak}, \citenamefont {Li},\ and\ \citenamefont {Jo}}]{ren22}%
  \BibitemOpen
  \bibfield  {author} {\bibinfo {author} {\bibfnamefont {Z.}~\bibnamefont
  {Ren}}, \bibinfo {author} {\bibfnamefont {D.}~\bibnamefont {Liu}}, \bibinfo
  {author} {\bibfnamefont {E.}~\bibnamefont {Zhao}}, \bibinfo {author}
  {\bibfnamefont {C.}~\bibnamefont {He}}, \bibinfo {author} {\bibfnamefont
  {K.~K.}\ \bibnamefont {Pak}}, \bibinfo {author} {\bibfnamefont
  {J.}~\bibnamefont {Li}},\ and\ \bibinfo {author} {\bibfnamefont {G.-B.}\
  \bibnamefont {Jo}},\ }\bibfield  {title} {\bibinfo {title} {{Chiral control
  of quantum states in non-Hermitian spin--orbit-coupled fermions}},\
  }\href@noop {} {\bibfield  {journal} {\bibinfo  {journal} {Nat. Phys.}\
  }\textbf {\bibinfo {volume} {18}},\ \bibinfo {pages} {385} (\bibinfo {year}
  {2022})}\BibitemShut {NoStop}%
\bibitem [{\citenamefont {Huang}\ \emph {et~al.}(2023)\citenamefont {Huang},
  \citenamefont {Mohan}, \citenamefont {Visuri}, \citenamefont {Fabritius},
  \citenamefont {Talebi}, \citenamefont {Wili}, \citenamefont {Uchino},
  \citenamefont {Giamarchi},\ and\ \citenamefont {Esslinger}}]{huang23}%
  \BibitemOpen
  \bibfield  {author} {\bibinfo {author} {\bibfnamefont {M.-Z.}\ \bibnamefont
  {Huang}}, \bibinfo {author} {\bibfnamefont {J.}~\bibnamefont {Mohan}},
  \bibinfo {author} {\bibfnamefont {A.-M.}\ \bibnamefont {Visuri}}, \bibinfo
  {author} {\bibfnamefont {P.}~\bibnamefont {Fabritius}}, \bibinfo {author}
  {\bibfnamefont {M.}~\bibnamefont {Talebi}}, \bibinfo {author} {\bibfnamefont
  {S.}~\bibnamefont {Wili}}, \bibinfo {author} {\bibfnamefont {S.}~\bibnamefont
  {Uchino}}, \bibinfo {author} {\bibfnamefont {T.}~\bibnamefont {Giamarchi}},\
  and\ \bibinfo {author} {\bibfnamefont {T.}~\bibnamefont {Esslinger}},\
  }\bibfield  {title} {\bibinfo {title} {{Superfluid Signatures in a
  Dissipative Quantum Point Contact}},\ }\href
  {https://doi.org/10.1103/PhysRevLett.130.200404} {\bibfield  {journal}
  {\bibinfo  {journal} {Phys. Rev. Lett.}\ }\textbf {\bibinfo {volume} {130}},\
  \bibinfo {pages} {200404} (\bibinfo {year} {2023})}\BibitemShut {NoStop}%
\bibitem [{\citenamefont {Huang}\ \emph {et~al.}(2025)\citenamefont {Huang},
  \citenamefont {Fabritius}, \citenamefont {Mohan}, \citenamefont {Talebi},
  \citenamefont {Wili},\ and\ \citenamefont {Esslinger}}]{huang24}%
  \BibitemOpen
  \bibfield  {author} {\bibinfo {author} {\bibfnamefont {M.-Z.}\ \bibnamefont
  {Huang}}, \bibinfo {author} {\bibfnamefont {P.}~\bibnamefont {Fabritius}},
  \bibinfo {author} {\bibfnamefont {J.}~\bibnamefont {Mohan}}, \bibinfo
  {author} {\bibfnamefont {M.}~\bibnamefont {Talebi}}, \bibinfo {author}
  {\bibfnamefont {S.}~\bibnamefont {Wili}},\ and\ \bibinfo {author}
  {\bibfnamefont {T.}~\bibnamefont {Esslinger}},\ }\bibfield  {title} {\bibinfo
  {title} {{Saturation of Thermal and Spin Conductances in a Dissipative
  Superfluid Junction}},\ }\href {https://doi.org/10.1103/9ks8-zv9b} {\bibfield
   {journal} {\bibinfo  {journal} {Phys. Rev. Lett.}\ }\textbf {\bibinfo
  {volume} {134}},\ \bibinfo {pages} {253403} (\bibinfo {year}
  {2025})}\BibitemShut {NoStop}%
\bibitem [{\citenamefont {Tsuno}\ \emph {et~al.}()\citenamefont {Tsuno},
  \citenamefont {Taie}, \citenamefont {Takasu}, \citenamefont {Yamashita},
  \citenamefont {Ozawa},\ and\ \citenamefont {Takahashi}}]{tsuno24}%
  \BibitemOpen
  \bibfield  {author} {\bibinfo {author} {\bibfnamefont {T.}~\bibnamefont
  {Tsuno}}, \bibinfo {author} {\bibfnamefont {S.}~\bibnamefont {Taie}},
  \bibinfo {author} {\bibfnamefont {Y.}~\bibnamefont {Takasu}}, \bibinfo
  {author} {\bibfnamefont {K.}~\bibnamefont {Yamashita}}, \bibinfo {author}
  {\bibfnamefont {T.}~\bibnamefont {Ozawa}},\ and\ \bibinfo {author}
  {\bibfnamefont {Y.}~\bibnamefont {Takahashi}},\ }\bibfield  {title} {\bibinfo
  {title} {{Gain engineering and topological atom laser in synthetic
  dimensions}},\ }\href@noop {} {\bibinfo  {journal} {arXiv:2404.13769}\
  }\BibitemShut {NoStop}%
\bibitem [{\citenamefont {Zhao}\ \emph {et~al.}(2025)\citenamefont {Zhao},
  \citenamefont {Wang}, \citenamefont {He}, \citenamefont {Poon}, \citenamefont
  {Pak}, \citenamefont {Liu}, \citenamefont {Ren}, \citenamefont {Liu},\ and\
  \citenamefont {Jo}}]{zhao25}%
  \BibitemOpen
\bibfield  {journal} {  }\bibfield  {author} {\bibinfo {author} {\bibfnamefont
  {E.}~\bibnamefont {Zhao}}, \bibinfo {author} {\bibfnamefont {Z.}~\bibnamefont
  {Wang}}, \bibinfo {author} {\bibfnamefont {C.}~\bibnamefont {He}}, \bibinfo
  {author} {\bibfnamefont {T.~F.~J.}\ \bibnamefont {Poon}}, \bibinfo {author}
  {\bibfnamefont {K.~K.}\ \bibnamefont {Pak}}, \bibinfo {author} {\bibfnamefont
  {Y.-J.}\ \bibnamefont {Liu}}, \bibinfo {author} {\bibfnamefont
  {P.}~\bibnamefont {Ren}}, \bibinfo {author} {\bibfnamefont {X.-J.}\
  \bibnamefont {Liu}},\ and\ \bibinfo {author} {\bibfnamefont {G.-B.}\
  \bibnamefont {Jo}},\ }\bibfield  {title} {\bibinfo {title} {{Two-dimensional
  non-Hermitian skin effect in an ultracold Fermi gas}},\ }\href
  {https://doi.org/10.1038/s41586-024-08347-3} {\bibfield  {journal} {\bibinfo
  {journal} {Nature (London)}\ }\textbf {\bibinfo {volume} {637}},\ \bibinfo
  {pages} {565} (\bibinfo {year} {2025})}\BibitemShut {NoStop}%
\bibitem [{\citenamefont {Tao}\ \emph {et~al.}(2025)\citenamefont {Tao},
  \citenamefont {Mercado-Gutierrez}, \citenamefont {Zhao},\ and\ \citenamefont
  {Spielman}}]{tao25}%
  \BibitemOpen
  \bibfield  {author} {\bibinfo {author} {\bibfnamefont {J.}~\bibnamefont
  {Tao}}, \bibinfo {author} {\bibfnamefont {E.}~\bibnamefont
  {Mercado-Gutierrez}}, \bibinfo {author} {\bibfnamefont {M.}~\bibnamefont
  {Zhao}},\ and\ \bibinfo {author} {\bibfnamefont {I.}~\bibnamefont
  {Spielman}},\ }\bibfield  {title} {\bibinfo {title} {{Imaginary gauge
  potentials in a non-Hermitian spin-orbit coupled quantum gas}},\ }\href@noop
  {} {\bibfield  {journal} {\bibinfo  {journal} {arXiv:2504.08614}\ } (\bibinfo
  {year} {2025})}\BibitemShut {NoStop}%
\bibitem [{\citenamefont {Zhang}\ \emph {et~al.}(2025)\citenamefont {Zhang},
  \citenamefont {Li}, \citenamefont {Wang}, \citenamefont {Liu}, \citenamefont
  {Zhang}, \citenamefont {Zhang}, \citenamefont {Shao}, \citenamefont {Li},
  \citenamefont {Chen}, \citenamefont {Ma} \emph {et~al.}}]{zhang25}%
  \BibitemOpen
  \bibfield  {author} {\bibinfo {author} {\bibfnamefont {J.}~\bibnamefont
  {Zhang}}, \bibinfo {author} {\bibfnamefont {E.-Z.}\ \bibnamefont {Li}},
  \bibinfo {author} {\bibfnamefont {Y.-J.}\ \bibnamefont {Wang}}, \bibinfo
  {author} {\bibfnamefont {B.}~\bibnamefont {Liu}}, \bibinfo {author}
  {\bibfnamefont {L.-H.}\ \bibnamefont {Zhang}}, \bibinfo {author}
  {\bibfnamefont {Z.-Y.}\ \bibnamefont {Zhang}}, \bibinfo {author}
  {\bibfnamefont {S.-Y.}\ \bibnamefont {Shao}}, \bibinfo {author}
  {\bibfnamefont {Q.}~\bibnamefont {Li}}, \bibinfo {author} {\bibfnamefont
  {H.-C.}\ \bibnamefont {Chen}}, \bibinfo {author} {\bibfnamefont
  {Y.}~\bibnamefont {Ma}}, \emph {et~al.},\ }\bibfield  {title} {\bibinfo
  {title} {{Exceptional point and hysteresis trajectories in cold Rydberg
  atomic gases}},\ }\href@noop {} {\bibfield  {journal} {\bibinfo  {journal}
  {Nat. Commun.}\ }\textbf {\bibinfo {volume} {16}},\ \bibinfo {pages} {3511}
  (\bibinfo {year} {2025})}\BibitemShut {NoStop}%
\bibitem [{\citenamefont {Tomita}\ \emph {et~al.}(2017)\citenamefont {Tomita},
  \citenamefont {Nakajima}, \citenamefont {Danshita}, \citenamefont {Takasu},\
  and\ \citenamefont {Takahashi}}]{tomita17}%
  \BibitemOpen
  \bibfield  {author} {\bibinfo {author} {\bibfnamefont {T.}~\bibnamefont
  {Tomita}}, \bibinfo {author} {\bibfnamefont {S.}~\bibnamefont {Nakajima}},
  \bibinfo {author} {\bibfnamefont {I.}~\bibnamefont {Danshita}}, \bibinfo
  {author} {\bibfnamefont {Y.}~\bibnamefont {Takasu}},\ and\ \bibinfo {author}
  {\bibfnamefont {Y.}~\bibnamefont {Takahashi}},\ }\bibfield  {title} {\bibinfo
  {title} {{Observation of the Mott insulator to superfluid crossover of a
  driven-dissipative Bose-Hubbard system}},\ }\href
  {https://doi.org/10.1126/sciadv.1701513} {\bibfield  {journal} {\bibinfo
  {journal} {Sci. Adv.}\ }\textbf {\bibinfo {volume} {3}},\ \bibinfo {pages}
  {e1701513} (\bibinfo {year} {2017})}\BibitemShut {NoStop}%
\bibitem [{\citenamefont {Bouganne}\ \emph {et~al.}(2020)\citenamefont
  {Bouganne}, \citenamefont {Bosch~Aguilera}, \citenamefont {Ghermaoui},
  \citenamefont {Beugnon},\ and\ \citenamefont {Gerbier}}]{bouganne20}%
  \BibitemOpen
  \bibfield  {author} {\bibinfo {author} {\bibfnamefont {R.}~\bibnamefont
  {Bouganne}}, \bibinfo {author} {\bibfnamefont {M.}~\bibnamefont
  {Bosch~Aguilera}}, \bibinfo {author} {\bibfnamefont {A.}~\bibnamefont
  {Ghermaoui}}, \bibinfo {author} {\bibfnamefont {J.}~\bibnamefont {Beugnon}},\
  and\ \bibinfo {author} {\bibfnamefont {F.}~\bibnamefont {Gerbier}},\
  }\bibfield  {title} {\bibinfo {title} {Anomalous decay of coherence in a
  dissipative many-body system},\ }\href@noop {} {\bibfield  {journal}
  {\bibinfo  {journal} {Nat. Phys.}\ }\textbf {\bibinfo {volume} {16}},\
  \bibinfo {pages} {21} (\bibinfo {year} {2020})}\BibitemShut {NoStop}%
\bibitem [{\citenamefont {Honda}\ \emph {et~al.}(2023)\citenamefont {Honda},
  \citenamefont {Taie}, \citenamefont {Takasu}, \citenamefont {Nishizawa},
  \citenamefont {Nakagawa},\ and\ \citenamefont {Takahashi}}]{honda23}%
  \BibitemOpen
  \bibfield  {author} {\bibinfo {author} {\bibfnamefont {K.}~\bibnamefont
  {Honda}}, \bibinfo {author} {\bibfnamefont {S.}~\bibnamefont {Taie}},
  \bibinfo {author} {\bibfnamefont {Y.}~\bibnamefont {Takasu}}, \bibinfo
  {author} {\bibfnamefont {N.}~\bibnamefont {Nishizawa}}, \bibinfo {author}
  {\bibfnamefont {M.}~\bibnamefont {Nakagawa}},\ and\ \bibinfo {author}
  {\bibfnamefont {Y.}~\bibnamefont {Takahashi}},\ }\bibfield  {title} {\bibinfo
  {title} {{Observation of the Sign Reversal of the Magnetic Correlation in a
  Driven-Dissipative Fermi Gas in Double Wells}},\ }\href
  {https://doi.org/10.1103/PhysRevLett.130.063001} {\bibfield  {journal}
  {\bibinfo  {journal} {Phys. Rev. Lett.}\ }\textbf {\bibinfo {volume} {130}},\
  \bibinfo {pages} {063001} (\bibinfo {year} {2023})}\BibitemShut {NoStop}%
\bibitem [{\citenamefont {Ashida}\ \emph {et~al.}(2020)\citenamefont {Ashida},
  \citenamefont {Gong},\ and\ \citenamefont {Ueda}}]{ashida20}%
  \BibitemOpen
  \bibfield  {author} {\bibinfo {author} {\bibfnamefont {Y.}~\bibnamefont
  {Ashida}}, \bibinfo {author} {\bibfnamefont {Z.}~\bibnamefont {Gong}},\ and\
  \bibinfo {author} {\bibfnamefont {M.}~\bibnamefont {Ueda}},\ }\bibfield
  {title} {\bibinfo {title} {{Non-Hermitian physics}},\ }\href
  {https://doi.org/10.1080/00018732.2021.1876991} {\bibfield  {journal}
  {\bibinfo  {journal} {Adv. Phys.}\ }\textbf {\bibinfo {volume} {69}},\
  \bibinfo {pages} {249} (\bibinfo {year} {2020})}\BibitemShut {NoStop}%
\bibitem [{\citenamefont {Ashida}\ \emph {et~al.}(2016)\citenamefont {Ashida},
  \citenamefont {Furukawa},\ and\ \citenamefont {Ueda}}]{ashida16}%
  \BibitemOpen
  \bibfield  {author} {\bibinfo {author} {\bibfnamefont {Y.}~\bibnamefont
  {Ashida}}, \bibinfo {author} {\bibfnamefont {S.}~\bibnamefont {Furukawa}},\
  and\ \bibinfo {author} {\bibfnamefont {M.}~\bibnamefont {Ueda}},\ }\bibfield
  {title} {\bibinfo {title} {Quantum critical behavior influenced by
  measurement backaction in ultracold gases},\ }\href
  {https://doi.org/10.1103/PhysRevA.94.053615} {\bibfield  {journal} {\bibinfo
  {journal} {Phys. Rev. A}\ }\textbf {\bibinfo {volume} {94}},\ \bibinfo
  {pages} {053615} (\bibinfo {year} {2016})}\BibitemShut {NoStop}%
\bibitem [{\citenamefont {Nakagawa}\ \emph {et~al.}(2018)\citenamefont
  {Nakagawa}, \citenamefont {Kawakami},\ and\ \citenamefont
  {Ueda}}]{nakagawa18}%
  \BibitemOpen
  \bibfield  {author} {\bibinfo {author} {\bibfnamefont {M.}~\bibnamefont
  {Nakagawa}}, \bibinfo {author} {\bibfnamefont {N.}~\bibnamefont {Kawakami}},\
  and\ \bibinfo {author} {\bibfnamefont {M.}~\bibnamefont {Ueda}},\ }\bibfield
  {title} {\bibinfo {title} {{Non-Hermitian Kondo Effect in Ultracold
  Alkaline-Earth Atoms}},\ }\href
  {https://doi.org/10.1103/PhysRevLett.121.203001} {\bibfield  {journal}
  {\bibinfo  {journal} {Phys. Rev. Lett.}\ }\textbf {\bibinfo {volume} {121}},\
  \bibinfo {pages} {203001} (\bibinfo {year} {2018})}\BibitemShut {NoStop}%
\bibitem [{\citenamefont {Res\'endiz-V\'azquez}\ \emph
  {et~al.}(2020)\citenamefont {Res\'endiz-V\'azquez}, \citenamefont
  {Tschernig}, \citenamefont {Perez-Leija}, \citenamefont {Busch},\ and\
  \citenamefont {Le\'on-Montiel}}]{resendiz20}%
  \BibitemOpen
  \bibfield  {author} {\bibinfo {author} {\bibfnamefont {P.}~\bibnamefont
  {Res\'endiz-V\'azquez}}, \bibinfo {author} {\bibfnamefont {K.}~\bibnamefont
  {Tschernig}}, \bibinfo {author} {\bibfnamefont {A.}~\bibnamefont
  {Perez-Leija}}, \bibinfo {author} {\bibfnamefont {K.}~\bibnamefont {Busch}},\
  and\ \bibinfo {author} {\bibfnamefont {R.~d.~J.}\ \bibnamefont
  {Le\'on-Montiel}},\ }\bibfield  {title} {\bibinfo {title} {{Topological
  protection in non-Hermitian Haldane honeycomb lattices}},\ }\href
  {https://doi.org/10.1103/PhysRevResearch.2.013387} {\bibfield  {journal}
  {\bibinfo  {journal} {Phys. Rev. Res.}\ }\textbf {\bibinfo {volume} {2}},\
  \bibinfo {pages} {013387} (\bibinfo {year} {2020})}\BibitemShut {NoStop}%
\bibitem [{\citenamefont {Nakagawa}\ \emph {et~al.}(2020)\citenamefont
  {Nakagawa}, \citenamefont {Tsuji}, \citenamefont {Kawakami},\ and\
  \citenamefont {Ueda}}]{nakagawa20}%
  \BibitemOpen
  \bibfield  {author} {\bibinfo {author} {\bibfnamefont {M.}~\bibnamefont
  {Nakagawa}}, \bibinfo {author} {\bibfnamefont {N.}~\bibnamefont {Tsuji}},
  \bibinfo {author} {\bibfnamefont {N.}~\bibnamefont {Kawakami}},\ and\
  \bibinfo {author} {\bibfnamefont {M.}~\bibnamefont {Ueda}},\ }\bibfield
  {title} {\bibinfo {title} {{Dynamical Sign Reversal of Magnetic Correlations
  in Dissipative Hubbard Models}},\ }\href
  {https://doi.org/10.1103/PhysRevLett.124.147203} {\bibfield  {journal}
  {\bibinfo  {journal} {Phys. Rev. Lett.}\ }\textbf {\bibinfo {volume} {124}},\
  \bibinfo {pages} {147203} (\bibinfo {year} {2020})}\BibitemShut {NoStop}%
\bibitem [{\citenamefont {Xu}\ and\ \citenamefont {Chen}(2020)}]{xu20}%
  \BibitemOpen
  \bibfield  {author} {\bibinfo {author} {\bibfnamefont {Z.}~\bibnamefont
  {Xu}}\ and\ \bibinfo {author} {\bibfnamefont {S.}~\bibnamefont {Chen}},\
  }\bibfield  {title} {\bibinfo {title} {{Topological Bose-Mott insulators in
  one-dimensional non-Hermitian superlattices}},\ }\href
  {https://doi.org/10.1103/PhysRevB.102.035153} {\bibfield  {journal} {\bibinfo
   {journal} {Phys. Rev. B}\ }\textbf {\bibinfo {volume} {102}},\ \bibinfo
  {pages} {035153} (\bibinfo {year} {2020})}\BibitemShut {NoStop}%
\bibitem [{\citenamefont {Matsumoto}\ \emph {et~al.}(2020)\citenamefont
  {Matsumoto}, \citenamefont {Kawabata}, \citenamefont {Ashida}, \citenamefont
  {Furukawa},\ and\ \citenamefont {Ueda}}]{matsumoto20}%
  \BibitemOpen
  \bibfield  {author} {\bibinfo {author} {\bibfnamefont {N.}~\bibnamefont
  {Matsumoto}}, \bibinfo {author} {\bibfnamefont {K.}~\bibnamefont {Kawabata}},
  \bibinfo {author} {\bibfnamefont {Y.}~\bibnamefont {Ashida}}, \bibinfo
  {author} {\bibfnamefont {S.}~\bibnamefont {Furukawa}},\ and\ \bibinfo
  {author} {\bibfnamefont {M.}~\bibnamefont {Ueda}},\ }\bibfield  {title}
  {\bibinfo {title} {{Continuous Phase Transition without Gap Closing in
  Non-Hermitian Quantum Many-Body Systems}},\ }\href
  {https://doi.org/10.1103/PhysRevLett.125.260601} {\bibfield  {journal}
  {\bibinfo  {journal} {Phys. Rev. Lett.}\ }\textbf {\bibinfo {volume} {125}},\
  \bibinfo {pages} {260601} (\bibinfo {year} {2020})}\BibitemShut {NoStop}%
\bibitem [{\citenamefont {Zhang}\ and\ \citenamefont
  {Song}(2021)}]{zhang21eta}%
  \BibitemOpen
  \bibfield  {author} {\bibinfo {author} {\bibfnamefont {X.~Z.}\ \bibnamefont
  {Zhang}}\ and\ \bibinfo {author} {\bibfnamefont {Z.}~\bibnamefont {Song}},\
  }\bibfield  {title} {\bibinfo {title} {{$\ensuremath{\eta}$-pairing ground
  states in the non-Hermitian Hubbard model}},\ }\href
  {https://doi.org/10.1103/PhysRevB.103.235153} {\bibfield  {journal} {\bibinfo
   {journal} {Phys. Rev. B}\ }\textbf {\bibinfo {volume} {103}},\ \bibinfo
  {pages} {235153} (\bibinfo {year} {2021})}\BibitemShut {NoStop}%
\bibitem [{\citenamefont {Tajima}\ and\ \citenamefont {Iida}(2021)}]{tajima21}%
  \BibitemOpen
  \bibfield  {author} {\bibinfo {author} {\bibfnamefont {H.}~\bibnamefont
  {Tajima}}\ and\ \bibinfo {author} {\bibfnamefont {K.}~\bibnamefont {Iida}},\
  }\bibfield  {title} {\bibinfo {title} {{Non-Hermitian Ferromagnetism in an
  Ultracold Fermi Gas}},\ }\href {https://doi.org/10.7566/JPSJ.90.024004}
  {\bibfield  {journal} {\bibinfo  {journal} {Journal of the Physical Society
  of Japan}\ }\textbf {\bibinfo {volume} {90}},\ \bibinfo {pages} {024004}
  (\bibinfo {year} {2021})}\BibitemShut {NoStop}%
\bibitem [{\citenamefont {Yamamoto}\ \emph {et~al.}(2022)\citenamefont
  {Yamamoto}, \citenamefont {Nakagawa}, \citenamefont {Tezuka}, \citenamefont
  {Ueda},\ and\ \citenamefont {Kawakami}}]{yamamoto22}%
  \BibitemOpen
  \bibfield  {author} {\bibinfo {author} {\bibfnamefont {K.}~\bibnamefont
  {Yamamoto}}, \bibinfo {author} {\bibfnamefont {M.}~\bibnamefont {Nakagawa}},
  \bibinfo {author} {\bibfnamefont {M.}~\bibnamefont {Tezuka}}, \bibinfo
  {author} {\bibfnamefont {M.}~\bibnamefont {Ueda}},\ and\ \bibinfo {author}
  {\bibfnamefont {N.}~\bibnamefont {Kawakami}},\ }\bibfield  {title} {\bibinfo
  {title} {{Universal properties of dissipative Tomonaga-Luttinger liquids:
  Case study of a non-Hermitian XXZ spin chain}},\ }\href
  {https://doi.org/10.1103/PhysRevB.105.205125} {\bibfield  {journal} {\bibinfo
   {journal} {Phys. Rev. B}\ }\textbf {\bibinfo {volume} {105}},\ \bibinfo
  {pages} {205125} (\bibinfo {year} {2022})}\BibitemShut {NoStop}%
\bibitem [{\citenamefont {Yamamoto}\ and\ \citenamefont
  {Kawakami}(2023)}]{yamamoto23sun}%
  \BibitemOpen
  \bibfield  {author} {\bibinfo {author} {\bibfnamefont {K.}~\bibnamefont
  {Yamamoto}}\ and\ \bibinfo {author} {\bibfnamefont {N.}~\bibnamefont
  {Kawakami}},\ }\bibfield  {title} {\bibinfo {title} {{Universal description
  of dissipative Tomonaga-Luttinger liquids with $\mathrm{SU}(N)$ spin
  symmetry: Exact spectrum and critical exponents}},\ }\href
  {https://doi.org/10.1103/PhysRevB.107.045110} {\bibfield  {journal} {\bibinfo
   {journal} {Phys. Rev. B}\ }\textbf {\bibinfo {volume} {107}},\ \bibinfo
  {pages} {045110} (\bibinfo {year} {2023})}\BibitemShut {NoStop}%
\bibitem [{\citenamefont {Han}\ \emph {et~al.}(2023)\citenamefont {Han},
  \citenamefont {Schultz},\ and\ \citenamefont {Kim}}]{han23}%
  \BibitemOpen
  \bibfield  {author} {\bibinfo {author} {\bibfnamefont {S.~E.}\ \bibnamefont
  {Han}}, \bibinfo {author} {\bibfnamefont {D.~J.}\ \bibnamefont {Schultz}},\
  and\ \bibinfo {author} {\bibfnamefont {Y.~B.}\ \bibnamefont {Kim}},\
  }\bibfield  {title} {\bibinfo {title} {{Complex fixed points of the
  non-Hermitian Kondo model in a Luttinger liquid}},\ }\href
  {https://doi.org/10.1103/PhysRevB.107.235153} {\bibfield  {journal} {\bibinfo
   {journal} {Phys. Rev. B}\ }\textbf {\bibinfo {volume} {107}},\ \bibinfo
  {pages} {235153} (\bibinfo {year} {2023})}\BibitemShut {NoStop}%
\bibitem [{\citenamefont {Wang}\ \emph {et~al.}(2023)\citenamefont {Wang},
  \citenamefont {Yi}, \citenamefont {Li},\ and\ \citenamefont
  {Mondaini}}]{wang23Hal}%
  \BibitemOpen
  \bibfield  {author} {\bibinfo {author} {\bibfnamefont {C.}~\bibnamefont
  {Wang}}, \bibinfo {author} {\bibfnamefont {T.-C.}\ \bibnamefont {Yi}},
  \bibinfo {author} {\bibfnamefont {J.}~\bibnamefont {Li}},\ and\ \bibinfo
  {author} {\bibfnamefont {R.}~\bibnamefont {Mondaini}},\ }\bibfield  {title}
  {\bibinfo {title} {{Non-Hermitian Haldane-Hubbard model: Effective
  description of one- and two-body dissipation}},\ }\href
  {https://doi.org/10.1103/PhysRevB.108.085134} {\bibfield  {journal} {\bibinfo
   {journal} {Phys. Rev. B}\ }\textbf {\bibinfo {volume} {108}},\ \bibinfo
  {pages} {085134} (\bibinfo {year} {2023})}\BibitemShut {NoStop}%
\bibitem [{\citenamefont {Yang}(2024)}]{yang24}%
  \BibitemOpen
  \bibfield  {author} {\bibinfo {author} {\bibfnamefont {L.}~\bibnamefont
  {Yang}},\ }\bibfield  {title} {\bibinfo {title}
  {{Dissipative-interaction-induced polaron-molecule transition in
  three-dimensional Fermi polarons}},\ }\href
  {https://doi.org/10.1103/PhysRevA.109.063305} {\bibfield  {journal} {\bibinfo
   {journal} {Phys. Rev. A}\ }\textbf {\bibinfo {volume} {109}},\ \bibinfo
  {pages} {063305} (\bibinfo {year} {2024})}\BibitemShut {NoStop}%
\bibitem [{\citenamefont {Yamamoto}\ \emph {et~al.}(2025)\citenamefont
  {Yamamoto}, \citenamefont {Nakagawa},\ and\ \citenamefont
  {Kawakami}}]{yamamoto24}%
  \BibitemOpen
  \bibfield  {author} {\bibinfo {author} {\bibfnamefont {K.}~\bibnamefont
  {Yamamoto}}, \bibinfo {author} {\bibfnamefont {M.}~\bibnamefont {Nakagawa}},\
  and\ \bibinfo {author} {\bibfnamefont {N.}~\bibnamefont {Kawakami}},\
  }\bibfield  {title} {\bibinfo {title} {{Correlation versus dissipation in a
  non-Hermitian Anderson impurity model}},\ }\href
  {https://doi.org/10.1103/PhysRevB.111.125157} {\bibfield  {journal} {\bibinfo
   {journal} {Phys. Rev. B}\ }\textbf {\bibinfo {volume} {111}},\ \bibinfo
  {pages} {125157} (\bibinfo {year} {2025})}\BibitemShut {NoStop}%
\bibitem [{\citenamefont {Hatano}\ and\ \citenamefont
  {Nelson}(1996)}]{hatano96}%
  \BibitemOpen
  \bibfield  {author} {\bibinfo {author} {\bibfnamefont {N.}~\bibnamefont
  {Hatano}}\ and\ \bibinfo {author} {\bibfnamefont {D.~R.}\ \bibnamefont
  {Nelson}},\ }\bibfield  {title} {\bibinfo {title} {{Localization Transitions
  in Non-Hermitian Quantum Mechanics}},\ }\href
  {https://doi.org/10.1103/PhysRevLett.77.570} {\bibfield  {journal} {\bibinfo
  {journal} {Phys. Rev. Lett.}\ }\textbf {\bibinfo {volume} {77}},\ \bibinfo
  {pages} {570} (\bibinfo {year} {1996})}\BibitemShut {NoStop}%
\bibitem [{\citenamefont {Hatano}\ and\ \citenamefont
  {Nelson}(1997)}]{hatano97}%
  \BibitemOpen
  \bibfield  {author} {\bibinfo {author} {\bibfnamefont {N.}~\bibnamefont
  {Hatano}}\ and\ \bibinfo {author} {\bibfnamefont {D.~R.}\ \bibnamefont
  {Nelson}},\ }\bibfield  {title} {\bibinfo {title} {{Vortex pinning and
  non-Hermitian quantum mechanics}},\ }\href
  {https://doi.org/10.1103/PhysRevB.56.8651} {\bibfield  {journal} {\bibinfo
  {journal} {Phys. Rev. B}\ }\textbf {\bibinfo {volume} {56}},\ \bibinfo
  {pages} {8651} (\bibinfo {year} {1997})}\BibitemShut {NoStop}%
\bibitem [{\citenamefont {Hatano}\ and\ \citenamefont
  {Nelson}(1998)}]{hatano98}%
  \BibitemOpen
  \bibfield  {author} {\bibinfo {author} {\bibfnamefont {N.}~\bibnamefont
  {Hatano}}\ and\ \bibinfo {author} {\bibfnamefont {D.~R.}\ \bibnamefont
  {Nelson}},\ }\bibfield  {title} {\bibinfo {title} {{Non-Hermitian
  delocalization and eigenfunctions}},\ }\href
  {https://doi.org/10.1103/PhysRevB.58.8384} {\bibfield  {journal} {\bibinfo
  {journal} {Phys. Rev. B}\ }\textbf {\bibinfo {volume} {58}},\ \bibinfo
  {pages} {8384} (\bibinfo {year} {1998})}\BibitemShut {NoStop}%
\bibitem [{\citenamefont {Gong}\ \emph {et~al.}(2018)\citenamefont {Gong},
  \citenamefont {Ashida}, \citenamefont {Kawabata}, \citenamefont {Takasan},
  \citenamefont {Higashikawa},\ and\ \citenamefont {Ueda}}]{gong18}%
  \BibitemOpen
  \bibfield  {author} {\bibinfo {author} {\bibfnamefont {Z.}~\bibnamefont
  {Gong}}, \bibinfo {author} {\bibfnamefont {Y.}~\bibnamefont {Ashida}},
  \bibinfo {author} {\bibfnamefont {K.}~\bibnamefont {Kawabata}}, \bibinfo
  {author} {\bibfnamefont {K.}~\bibnamefont {Takasan}}, \bibinfo {author}
  {\bibfnamefont {S.}~\bibnamefont {Higashikawa}},\ and\ \bibinfo {author}
  {\bibfnamefont {M.}~\bibnamefont {Ueda}},\ }\bibfield  {title} {\bibinfo
  {title} {{Topological Phases of Non-Hermitian Systems}},\ }\href
  {https://doi.org/10.1103/PhysRevX.8.031079} {\bibfield  {journal} {\bibinfo
  {journal} {Phys. Rev. X}\ }\textbf {\bibinfo {volume} {8}},\ \bibinfo {pages}
  {031079} (\bibinfo {year} {2018})}\BibitemShut {NoStop}%
\bibitem [{\citenamefont {Liu}\ \emph {et~al.}(2019)\citenamefont {Liu},
  \citenamefont {Zhang}, \citenamefont {Ai}, \citenamefont {Gong},
  \citenamefont {Kawabata}, \citenamefont {Ueda},\ and\ \citenamefont
  {Nori}}]{liu19}%
  \BibitemOpen
  \bibfield  {author} {\bibinfo {author} {\bibfnamefont {T.}~\bibnamefont
  {Liu}}, \bibinfo {author} {\bibfnamefont {Y.-R.}\ \bibnamefont {Zhang}},
  \bibinfo {author} {\bibfnamefont {Q.}~\bibnamefont {Ai}}, \bibinfo {author}
  {\bibfnamefont {Z.}~\bibnamefont {Gong}}, \bibinfo {author} {\bibfnamefont
  {K.}~\bibnamefont {Kawabata}}, \bibinfo {author} {\bibfnamefont
  {M.}~\bibnamefont {Ueda}},\ and\ \bibinfo {author} {\bibfnamefont
  {F.}~\bibnamefont {Nori}},\ }\bibfield  {title} {\bibinfo {title}
  {{Second-Order Topological Phases in Non-Hermitian Systems}},\ }\href
  {https://doi.org/10.1103/PhysRevLett.122.076801} {\bibfield  {journal}
  {\bibinfo  {journal} {Phys. Rev. Lett.}\ }\textbf {\bibinfo {volume} {122}},\
  \bibinfo {pages} {076801} (\bibinfo {year} {2019})}\BibitemShut {NoStop}%
\bibitem [{\citenamefont {Yao}\ and\ \citenamefont {Wang}(2018)}]{yao18}%
  \BibitemOpen
  \bibfield  {author} {\bibinfo {author} {\bibfnamefont {S.}~\bibnamefont
  {Yao}}\ and\ \bibinfo {author} {\bibfnamefont {Z.}~\bibnamefont {Wang}},\
  }\bibfield  {title} {\bibinfo {title} {{Edge States and Topological
  Invariants of Non-Hermitian Systems}},\ }\href
  {https://doi.org/10.1103/PhysRevLett.121.086803} {\bibfield  {journal}
  {\bibinfo  {journal} {Phys. Rev. Lett.}\ }\textbf {\bibinfo {volume} {121}},\
  \bibinfo {pages} {086803} (\bibinfo {year} {2018})}\BibitemShut {NoStop}%
\bibitem [{\citenamefont {Borgnia}\ \emph {et~al.}(2020)\citenamefont
  {Borgnia}, \citenamefont {Kruchkov},\ and\ \citenamefont
  {Slager}}]{borgnia20}%
  \BibitemOpen
  \bibfield  {author} {\bibinfo {author} {\bibfnamefont {D.~S.}\ \bibnamefont
  {Borgnia}}, \bibinfo {author} {\bibfnamefont {A.~J.}\ \bibnamefont
  {Kruchkov}},\ and\ \bibinfo {author} {\bibfnamefont {R.-J.}\ \bibnamefont
  {Slager}},\ }\bibfield  {title} {\bibinfo {title} {Non-hermitian boundary
  modes and topology},\ }\href {https://doi.org/10.1103/PhysRevLett.124.056802}
  {\bibfield  {journal} {\bibinfo  {journal} {Phys. Rev. Lett.}\ }\textbf
  {\bibinfo {volume} {124}},\ \bibinfo {pages} {056802} (\bibinfo {year}
  {2020})}\BibitemShut {NoStop}%
\bibitem [{\citenamefont {Fukui}\ and\ \citenamefont
  {Kawakami}(1998{\natexlab{a}})}]{fukui98B}%
  \BibitemOpen
  \bibfield  {author} {\bibinfo {author} {\bibfnamefont {T.}~\bibnamefont
  {Fukui}}\ and\ \bibinfo {author} {\bibfnamefont {N.}~\bibnamefont
  {Kawakami}},\ }\bibfield  {title} {\bibinfo {title} {Spectral flow of
  non-hermitian heisenberg spin chain with complex twist},\ }\href@noop {}
  {\bibfield  {journal} {\bibinfo  {journal} {Nucl. Phys. B}\ }\textbf
  {\bibinfo {volume} {519}},\ \bibinfo {pages} {715} (\bibinfo {year}
  {1998}{\natexlab{a}})}\BibitemShut {NoStop}%
\bibitem [{\citenamefont {Hamazaki}\ \emph {et~al.}(2019)\citenamefont
  {Hamazaki}, \citenamefont {Kawabata},\ and\ \citenamefont
  {Ueda}}]{hamazaki19}%
  \BibitemOpen
  \bibfield  {author} {\bibinfo {author} {\bibfnamefont {R.}~\bibnamefont
  {Hamazaki}}, \bibinfo {author} {\bibfnamefont {K.}~\bibnamefont {Kawabata}},\
  and\ \bibinfo {author} {\bibfnamefont {M.}~\bibnamefont {Ueda}},\ }\bibfield
  {title} {\bibinfo {title} {{Non-Hermitian Many-Body Localization}},\ }\href
  {https://doi.org/10.1103/PhysRevLett.123.090603} {\bibfield  {journal}
  {\bibinfo  {journal} {Phys. Rev. Lett.}\ }\textbf {\bibinfo {volume} {123}},\
  \bibinfo {pages} {090603} (\bibinfo {year} {2019})}\BibitemShut {NoStop}%
\bibitem [{\citenamefont {Zhang}\ \emph {et~al.}(2020)\citenamefont {Zhang},
  \citenamefont {Chen}, \citenamefont {Zhang}, \citenamefont {Lang},
  \citenamefont {Li},\ and\ \citenamefont {Zhu}}]{zhang20}%
  \BibitemOpen
  \bibfield  {author} {\bibinfo {author} {\bibfnamefont {D.-W.}\ \bibnamefont
  {Zhang}}, \bibinfo {author} {\bibfnamefont {Y.-L.}\ \bibnamefont {Chen}},
  \bibinfo {author} {\bibfnamefont {G.-Q.}\ \bibnamefont {Zhang}}, \bibinfo
  {author} {\bibfnamefont {L.-J.}\ \bibnamefont {Lang}}, \bibinfo {author}
  {\bibfnamefont {Z.}~\bibnamefont {Li}},\ and\ \bibinfo {author}
  {\bibfnamefont {S.-L.}\ \bibnamefont {Zhu}},\ }\bibfield  {title} {\bibinfo
  {title} {{Skin superfluid, topological Mott insulators, and asymmetric
  dynamics in an interacting non-Hermitian Aubry-Andr\'e-Harper model}},\
  }\href {https://doi.org/10.1103/PhysRevB.101.235150} {\bibfield  {journal}
  {\bibinfo  {journal} {Phys. Rev. B}\ }\textbf {\bibinfo {volume} {101}},\
  \bibinfo {pages} {235150} (\bibinfo {year} {2020})}\BibitemShut {NoStop}%
\bibitem [{\citenamefont {Suthar}\ \emph {et~al.}(2022)\citenamefont {Suthar},
  \citenamefont {Wang}, \citenamefont {Huang}, \citenamefont {Jen},\ and\
  \citenamefont {You}}]{suthar22}%
  \BibitemOpen
  \bibfield  {author} {\bibinfo {author} {\bibfnamefont {K.}~\bibnamefont
  {Suthar}}, \bibinfo {author} {\bibfnamefont {Y.-C.}\ \bibnamefont {Wang}},
  \bibinfo {author} {\bibfnamefont {Y.-P.}\ \bibnamefont {Huang}}, \bibinfo
  {author} {\bibfnamefont {H.~H.}\ \bibnamefont {Jen}},\ and\ \bibinfo {author}
  {\bibfnamefont {J.-S.}\ \bibnamefont {You}},\ }\bibfield  {title} {\bibinfo
  {title} {{Non-Hermitian many-body localization with open boundaries}},\
  }\href {https://doi.org/10.1103/PhysRevB.106.064208} {\bibfield  {journal}
  {\bibinfo  {journal} {Phys. Rev. B}\ }\textbf {\bibinfo {volume} {106}},\
  \bibinfo {pages} {064208} (\bibinfo {year} {2022})}\BibitemShut {NoStop}%
\bibitem [{\citenamefont {Kawabata}\ \emph {et~al.}(2022)\citenamefont
  {Kawabata}, \citenamefont {Shiozaki},\ and\ \citenamefont
  {Ryu}}]{kawabata22}%
  \BibitemOpen
  \bibfield  {author} {\bibinfo {author} {\bibfnamefont {K.}~\bibnamefont
  {Kawabata}}, \bibinfo {author} {\bibfnamefont {K.}~\bibnamefont {Shiozaki}},\
  and\ \bibinfo {author} {\bibfnamefont {S.}~\bibnamefont {Ryu}},\ }\bibfield
  {title} {\bibinfo {title} {{Many-body topology of non-Hermitian systems}},\
  }\href {https://doi.org/10.1103/PhysRevB.105.165137} {\bibfield  {journal}
  {\bibinfo  {journal} {Phys. Rev. B}\ }\textbf {\bibinfo {volume} {105}},\
  \bibinfo {pages} {165137} (\bibinfo {year} {2022})}\BibitemShut {NoStop}%
\bibitem [{\citenamefont {Orito}\ and\ \citenamefont
  {Imura}(2022)}]{orito22dis}%
  \BibitemOpen
  \bibfield  {author} {\bibinfo {author} {\bibfnamefont {T.}~\bibnamefont
  {Orito}}\ and\ \bibinfo {author} {\bibfnamefont {K.-I.}\ \bibnamefont
  {Imura}},\ }\bibfield  {title} {\bibinfo {title} {{Unusual wave-packet
  spreading and entanglement dynamics in non-Hermitian disordered many-body
  systems}},\ }\href {https://doi.org/10.1103/PhysRevB.105.024303} {\bibfield
  {journal} {\bibinfo  {journal} {Phys. Rev. B}\ }\textbf {\bibinfo {volume}
  {105}},\ \bibinfo {pages} {024303} (\bibinfo {year} {2022})}\BibitemShut
  {NoStop}%
\bibitem [{\citenamefont {Faugno}\ and\ \citenamefont
  {Ozawa}(2022)}]{faugno22}%
  \BibitemOpen
  \bibfield  {author} {\bibinfo {author} {\bibfnamefont {W.~N.}\ \bibnamefont
  {Faugno}}\ and\ \bibinfo {author} {\bibfnamefont {T.}~\bibnamefont {Ozawa}},\
  }\bibfield  {title} {\bibinfo {title} {{Interaction-Induced Non-Hermitian
  Topological Phases from a Dynamical Gauge Field}},\ }\href
  {https://doi.org/10.1103/PhysRevLett.129.180401} {\bibfield  {journal}
  {\bibinfo  {journal} {Phys. Rev. Lett.}\ }\textbf {\bibinfo {volume} {129}},\
  \bibinfo {pages} {180401} (\bibinfo {year} {2022})}\BibitemShut {NoStop}%
\bibitem [{\citenamefont {Li}\ \emph {et~al.}(2023{\natexlab{a}})\citenamefont
  {Li}, \citenamefont {Yu},\ and\ \citenamefont {Zhong}}]{li23stark}%
  \BibitemOpen
  \bibfield  {author} {\bibinfo {author} {\bibfnamefont {H.-Z.}\ \bibnamefont
  {Li}}, \bibinfo {author} {\bibfnamefont {X.-J.}\ \bibnamefont {Yu}},\ and\
  \bibinfo {author} {\bibfnamefont {J.-X.}\ \bibnamefont {Zhong}},\ }\bibfield
  {title} {\bibinfo {title} {{Non-Hermitian Stark many-body localization}},\
  }\href {https://doi.org/10.1103/PhysRevA.108.043301} {\bibfield  {journal}
  {\bibinfo  {journal} {Phys. Rev. A}\ }\textbf {\bibinfo {volume} {108}},\
  \bibinfo {pages} {043301} (\bibinfo {year} {2023}{\natexlab{a}})}\BibitemShut
  {NoStop}%
\bibitem [{\citenamefont {D\'ora}\ and\ \citenamefont {Moca}(2024)}]{dora24}%
  \BibitemOpen
  \bibfield  {author} {\bibinfo {author} {\bibfnamefont {B.}~\bibnamefont
  {D\'ora}}\ and\ \bibinfo {author} {\bibfnamefont {C.~u. u. u. u. P. m.~c.}\
  \bibnamefont {Moca}},\ }\bibfield  {title} {\bibinfo {title} {Work statistics
  and generalized loschmidt echo for the hatano-nelson model},\ }\href
  {https://doi.org/10.1103/PhysRevB.110.L121116} {\bibfield  {journal}
  {\bibinfo  {journal} {Phys. Rev. B}\ }\textbf {\bibinfo {volume} {110}},\
  \bibinfo {pages} {L121116} (\bibinfo {year} {2024})}\BibitemShut {NoStop}%
\bibitem [{\citenamefont {M{\'a}k}\ \emph {et~al.}(2024)\citenamefont
  {M{\'a}k}, \citenamefont {Bhaseen},\ and\ \citenamefont {Pal}}]{mak24}%
  \BibitemOpen
  \bibfield  {author} {\bibinfo {author} {\bibfnamefont {J.}~\bibnamefont
  {M{\'a}k}}, \bibinfo {author} {\bibfnamefont {M.}~\bibnamefont {Bhaseen}},\
  and\ \bibinfo {author} {\bibfnamefont {A.}~\bibnamefont {Pal}},\ }\bibfield
  {title} {\bibinfo {title} {{Statics and dynamics of non-Hermitian many-body
  localization}},\ }\href@noop {} {\bibfield  {journal} {\bibinfo  {journal}
  {Commun. Phys.}\ }\textbf {\bibinfo {volume} {7}},\ \bibinfo {pages} {92}
  (\bibinfo {year} {2024})}\BibitemShut {NoStop}%
\bibitem [{\citenamefont {Yoshida}\ \emph {et~al.}(2024)\citenamefont
  {Yoshida}, \citenamefont {Zhang}, \citenamefont {Neupert},\ and\
  \citenamefont {Kawakami}}]{yoshida24}%
  \BibitemOpen
  \bibfield  {author} {\bibinfo {author} {\bibfnamefont {T.}~\bibnamefont
  {Yoshida}}, \bibinfo {author} {\bibfnamefont {S.-B.}\ \bibnamefont {Zhang}},
  \bibinfo {author} {\bibfnamefont {T.}~\bibnamefont {Neupert}},\ and\ \bibinfo
  {author} {\bibfnamefont {N.}~\bibnamefont {Kawakami}},\ }\bibfield  {title}
  {\bibinfo {title} {{Non-Hermitian Mott Skin Effect}},\ }\href
  {https://doi.org/10.1103/PhysRevLett.133.076502} {\bibfield  {journal}
  {\bibinfo  {journal} {Phys. Rev. Lett.}\ }\textbf {\bibinfo {volume} {133}},\
  \bibinfo {pages} {076502} (\bibinfo {year} {2024})}\BibitemShut {NoStop}%
\bibitem [{\citenamefont {Dupays}\ \emph {et~al.}(2025)\citenamefont {Dupays},
  \citenamefont {del Campo},\ and\ \citenamefont {D\'ora}}]{dupays25}%
  \BibitemOpen
  \bibfield  {author} {\bibinfo {author} {\bibfnamefont {L.}~\bibnamefont
  {Dupays}}, \bibinfo {author} {\bibfnamefont {A.}~\bibnamefont {del Campo}},\
  and\ \bibinfo {author} {\bibfnamefont {B.}~\bibnamefont {D\'ora}},\
  }\bibfield  {title} {\bibinfo {title} {{Slow approach to adiabaticity in
  many-body non-Hermitian systems: The Hatano-Nelson model}},\ }\href
  {https://doi.org/10.1103/PhysRevB.111.045130} {\bibfield  {journal} {\bibinfo
   {journal} {Phys. Rev. B}\ }\textbf {\bibinfo {volume} {111}},\ \bibinfo
  {pages} {045130} (\bibinfo {year} {2025})}\BibitemShut {NoStop}%
\bibitem [{\citenamefont {Lee}\ \emph {et~al.}(2020)\citenamefont {Lee},
  \citenamefont {Lee},\ and\ \citenamefont {Yang}}]{LeeE20}%
  \BibitemOpen
  \bibfield  {author} {\bibinfo {author} {\bibfnamefont {E.}~\bibnamefont
  {Lee}}, \bibinfo {author} {\bibfnamefont {H.}~\bibnamefont {Lee}},\ and\
  \bibinfo {author} {\bibfnamefont {B.-J.}\ \bibnamefont {Yang}},\ }\bibfield
  {title} {\bibinfo {title} {{Many-body approach to non-Hermitian physics in
  fermionic systems}},\ }\href {https://doi.org/10.1103/PhysRevB.101.121109}
  {\bibfield  {journal} {\bibinfo  {journal} {Phys. Rev. B}\ }\textbf {\bibinfo
  {volume} {101}},\ \bibinfo {pages} {121109} (\bibinfo {year}
  {2020})}\BibitemShut {NoStop}%
\bibitem [{\citenamefont {Liu}\ \emph {et~al.}(2020)\citenamefont {Liu},
  \citenamefont {He}, \citenamefont {Yoshida}, \citenamefont {Xiang},\ and\
  \citenamefont {Nori}}]{liu20}%
  \BibitemOpen
  \bibfield  {author} {\bibinfo {author} {\bibfnamefont {T.}~\bibnamefont
  {Liu}}, \bibinfo {author} {\bibfnamefont {J.~J.}\ \bibnamefont {He}},
  \bibinfo {author} {\bibfnamefont {T.}~\bibnamefont {Yoshida}}, \bibinfo
  {author} {\bibfnamefont {Z.-L.}\ \bibnamefont {Xiang}},\ and\ \bibinfo
  {author} {\bibfnamefont {F.}~\bibnamefont {Nori}},\ }\bibfield  {title}
  {\bibinfo {title} {{Non-Hermitian topological Mott insulators in
  one-dimensional fermionic superlattices}},\ }\href
  {https://doi.org/10.1103/PhysRevB.102.235151} {\bibfield  {journal} {\bibinfo
   {journal} {Phys. Rev. B}\ }\textbf {\bibinfo {volume} {102}},\ \bibinfo
  {pages} {235151} (\bibinfo {year} {2020})}\BibitemShut {NoStop}%
\bibitem [{\citenamefont {Fukui}\ and\ \citenamefont
  {Kawakami}(1998{\natexlab{b}})}]{fukui98}%
  \BibitemOpen
  \bibfield  {author} {\bibinfo {author} {\bibfnamefont {T.}~\bibnamefont
  {Fukui}}\ and\ \bibinfo {author} {\bibfnamefont {N.}~\bibnamefont
  {Kawakami}},\ }\bibfield  {title} {\bibinfo {title} {{Breakdown of the Mott
  insulator: Exact solution of an asymmetric Hubbard model}},\ }\href
  {https://doi.org/10.1103/PhysRevB.58.16051} {\bibfield  {journal} {\bibinfo
  {journal} {Phys. Rev. B}\ }\textbf {\bibinfo {volume} {58}},\ \bibinfo
  {pages} {16051} (\bibinfo {year} {1998}{\natexlab{b}})}\BibitemShut {NoStop}%
\bibitem [{\citenamefont {Uchino}\ and\ \citenamefont
  {Kawakami}(2012)}]{uchino12}%
  \BibitemOpen
  \bibfield  {author} {\bibinfo {author} {\bibfnamefont {S.}~\bibnamefont
  {Uchino}}\ and\ \bibinfo {author} {\bibfnamefont {N.}~\bibnamefont
  {Kawakami}},\ }\bibfield  {title} {\bibinfo {title} {{Spin-depairing
  transition of attractive Fermi gases on a ring driven by synthetic gauge
  fields}},\ }\href {https://doi.org/10.1103/PhysRevA.85.013610} {\bibfield
  {journal} {\bibinfo  {journal} {Phys. Rev. A}\ }\textbf {\bibinfo {volume}
  {85}},\ \bibinfo {pages} {013610} (\bibinfo {year} {2012})}\BibitemShut
  {NoStop}%
\bibitem [{\citenamefont {Hayata}\ and\ \citenamefont
  {Yamamoto}(2021)}]{hayata21sign}%
  \BibitemOpen
  \bibfield  {author} {\bibinfo {author} {\bibfnamefont {T.}~\bibnamefont
  {Hayata}}\ and\ \bibinfo {author} {\bibfnamefont {A.}~\bibnamefont
  {Yamamoto}},\ }\bibfield  {title} {\bibinfo {title} {{Non-Hermitian Hubbard
  model without the sign problem}},\ }\href@noop {} {\bibfield  {journal}
  {\bibinfo  {journal} {Phys. Rev. B}\ }\textbf {\bibinfo {volume} {104}},\
  \bibinfo {pages} {125102} (\bibinfo {year} {2021})}\BibitemShut {NoStop}%
\bibitem [{\citenamefont {Yu}\ \emph {et~al.}(2024)\citenamefont {Yu},
  \citenamefont {Pan}, \citenamefont {Xu},\ and\ \citenamefont {Li}}]{yu24}%
  \BibitemOpen
  \bibfield  {author} {\bibinfo {author} {\bibfnamefont {X.-J.}\ \bibnamefont
  {Yu}}, \bibinfo {author} {\bibfnamefont {Z.}~\bibnamefont {Pan}}, \bibinfo
  {author} {\bibfnamefont {L.}~\bibnamefont {Xu}},\ and\ \bibinfo {author}
  {\bibfnamefont {Z.-X.}\ \bibnamefont {Li}},\ }\bibfield  {title} {\bibinfo
  {title} {{Non-Hermitian Strongly Interacting Dirac Fermions}},\ }\href
  {https://doi.org/10.1103/PhysRevLett.132.116503} {\bibfield  {journal}
  {\bibinfo  {journal} {Phys. Rev. Lett.}\ }\textbf {\bibinfo {volume} {132}},\
  \bibinfo {pages} {116503} (\bibinfo {year} {2024})}\BibitemShut {NoStop}%
\bibitem [{\citenamefont {Kawabata}\ \emph
  {et~al.}(2019{\natexlab{a}})\citenamefont {Kawabata}, \citenamefont
  {Higashikawa}, \citenamefont {Gong}, \citenamefont {Ashida},\ and\
  \citenamefont {Ueda}}]{kawabata19}%
  \BibitemOpen
  \bibfield  {author} {\bibinfo {author} {\bibfnamefont {K.}~\bibnamefont
  {Kawabata}}, \bibinfo {author} {\bibfnamefont {S.}~\bibnamefont
  {Higashikawa}}, \bibinfo {author} {\bibfnamefont {Z.}~\bibnamefont {Gong}},
  \bibinfo {author} {\bibfnamefont {Y.}~\bibnamefont {Ashida}},\ and\ \bibinfo
  {author} {\bibfnamefont {M.}~\bibnamefont {Ueda}},\ }\bibfield  {title}
  {\bibinfo {title} {{Topological unification of time-reversal and
  particle-hole symmetries in non-Hermitian physics}},\ }\href@noop {}
  {\bibfield  {journal} {\bibinfo  {journal} {Nat. Commun.}\ }\textbf {\bibinfo
  {volume} {10}},\ \bibinfo {pages} {297} (\bibinfo {year}
  {2019}{\natexlab{a}})}\BibitemShut {NoStop}%
\bibitem [{\citenamefont {Str\'ansk\'y}\ \emph {et~al.}(2018)\citenamefont
  {Str\'ansk\'y}, \citenamefont {Dvo\ifmmode~\check{r}\else \v{r}\fi{}\'ak},\
  and\ \citenamefont {Cejnar}}]{stranky18}%
  \BibitemOpen
  \bibfield  {author} {\bibinfo {author} {\bibfnamefont {P.}~\bibnamefont
  {Str\'ansk\'y}}, \bibinfo {author} {\bibfnamefont {M.}~\bibnamefont
  {Dvo\ifmmode~\check{r}\else \v{r}\fi{}\'ak}},\ and\ \bibinfo {author}
  {\bibfnamefont {P.}~\bibnamefont {Cejnar}},\ }\bibfield  {title} {\bibinfo
  {title} {Exceptional points near first- and second-order quantum phase
  transitions},\ }\href {https://doi.org/10.1103/PhysRevE.97.012112} {\bibfield
   {journal} {\bibinfo  {journal} {Phys. Rev. E}\ }\textbf {\bibinfo {volume}
  {97}},\ \bibinfo {pages} {012112} (\bibinfo {year} {2018})}\BibitemShut
  {NoStop}%
\bibitem [{\citenamefont {Kawabata}\ \emph
  {et~al.}(2019{\natexlab{b}})\citenamefont {Kawabata}, \citenamefont
  {Shiozaki}, \citenamefont {Ueda},\ and\ \citenamefont {Sato}}]{kawabata19X}%
  \BibitemOpen
  \bibfield  {author} {\bibinfo {author} {\bibfnamefont {K.}~\bibnamefont
  {Kawabata}}, \bibinfo {author} {\bibfnamefont {K.}~\bibnamefont {Shiozaki}},
  \bibinfo {author} {\bibfnamefont {M.}~\bibnamefont {Ueda}},\ and\ \bibinfo
  {author} {\bibfnamefont {M.}~\bibnamefont {Sato}},\ }\bibfield  {title}
  {\bibinfo {title} {{Symmetry and Topology in Non-Hermitian Physics}},\ }\href
  {https://doi.org/10.1103/PhysRevX.9.041015} {\bibfield  {journal} {\bibinfo
  {journal} {Phys. Rev. X}\ }\textbf {\bibinfo {volume} {9}},\ \bibinfo {pages}
  {041015} (\bibinfo {year} {2019}{\natexlab{b}})}\BibitemShut {NoStop}%
\bibitem [{\citenamefont {Yang}\ \emph {et~al.}(2021)\citenamefont {Yang},
  \citenamefont {Morampudi},\ and\ \citenamefont {Bergholtz}}]{yang21}%
  \BibitemOpen
  \bibfield  {author} {\bibinfo {author} {\bibfnamefont {K.}~\bibnamefont
  {Yang}}, \bibinfo {author} {\bibfnamefont {S.~C.}\ \bibnamefont
  {Morampudi}},\ and\ \bibinfo {author} {\bibfnamefont {E.~J.}\ \bibnamefont
  {Bergholtz}},\ }\bibfield  {title} {\bibinfo {title} {{Exceptional Spin
  Liquids from Couplings to the Environment}},\ }\href
  {https://doi.org/10.1103/PhysRevLett.126.077201} {\bibfield  {journal}
  {\bibinfo  {journal} {Phys. Rev. Lett.}\ }\textbf {\bibinfo {volume} {126}},\
  \bibinfo {pages} {077201} (\bibinfo {year} {2021})}\BibitemShut {NoStop}%
\bibitem [{\citenamefont {Miri}\ and\ \citenamefont {Alu}(2019)}]{miri19}%
  \BibitemOpen
  \bibfield  {author} {\bibinfo {author} {\bibfnamefont {M.-A.}\ \bibnamefont
  {Miri}}\ and\ \bibinfo {author} {\bibfnamefont {A.}~\bibnamefont {Alu}},\
  }\bibfield  {title} {\bibinfo {title} {Exceptional points in optics and
  photonics},\ }\href@noop {} {\bibfield  {journal} {\bibinfo  {journal}
  {Science}\ }\textbf {\bibinfo {volume} {363}},\ \bibinfo {pages} {eaar7709}
  (\bibinfo {year} {2019})}\BibitemShut {NoStop}%
\bibitem [{\citenamefont {Kozii}\ and\ \citenamefont {Fu}(2024)}]{Kozii24}%
  \BibitemOpen
  \bibfield  {author} {\bibinfo {author} {\bibfnamefont {V.}~\bibnamefont
  {Kozii}}\ and\ \bibinfo {author} {\bibfnamefont {L.}~\bibnamefont {Fu}},\
  }\bibfield  {title} {\bibinfo {title} {{Non-Hermitian topological theory of
  finite-lifetime quasiparticles: Prediction of bulk Fermi arc due to
  exceptional point}},\ }\href {https://doi.org/10.1103/PhysRevB.109.235139}
  {\bibfield  {journal} {\bibinfo  {journal} {Phys. Rev. B}\ }\textbf {\bibinfo
  {volume} {109}},\ \bibinfo {pages} {235139} (\bibinfo {year}
  {2024})}\BibitemShut {NoStop}%
\bibitem [{\citenamefont {Yoshida}\ \emph {et~al.}(2018)\citenamefont
  {Yoshida}, \citenamefont {Peters},\ and\ \citenamefont
  {Kawakami}}]{yoshida18}%
  \BibitemOpen
  \bibfield  {author} {\bibinfo {author} {\bibfnamefont {T.}~\bibnamefont
  {Yoshida}}, \bibinfo {author} {\bibfnamefont {R.}~\bibnamefont {Peters}},\
  and\ \bibinfo {author} {\bibfnamefont {N.}~\bibnamefont {Kawakami}},\
  }\bibfield  {title} {\bibinfo {title} {{Non-Hermitian perspective of the band
  structure in heavy-fermion systems}},\ }\href
  {https://doi.org/10.1103/PhysRevB.98.035141} {\bibfield  {journal} {\bibinfo
  {journal} {Phys. Rev. B}\ }\textbf {\bibinfo {volume} {98}},\ \bibinfo
  {pages} {035141} (\bibinfo {year} {2018})}\BibitemShut {NoStop}%
\bibitem [{\citenamefont {Nagai}\ \emph {et~al.}(2020)\citenamefont {Nagai},
  \citenamefont {Qi}, \citenamefont {Isobe}, \citenamefont {Kozii},\ and\
  \citenamefont {Fu}}]{Nagai20}%
  \BibitemOpen
  \bibfield  {author} {\bibinfo {author} {\bibfnamefont {Y.}~\bibnamefont
  {Nagai}}, \bibinfo {author} {\bibfnamefont {Y.}~\bibnamefont {Qi}}, \bibinfo
  {author} {\bibfnamefont {H.}~\bibnamefont {Isobe}}, \bibinfo {author}
  {\bibfnamefont {V.}~\bibnamefont {Kozii}},\ and\ \bibinfo {author}
  {\bibfnamefont {L.}~\bibnamefont {Fu}},\ }\bibfield  {title} {\bibinfo
  {title} {{DMFT Reveals the Non-Hermitian Topology and Fermi Arcs in
  Heavy-Fermion Systems}},\ }\href
  {https://doi.org/10.1103/PhysRevLett.125.227204} {\bibfield  {journal}
  {\bibinfo  {journal} {Phys. Rev. Lett.}\ }\textbf {\bibinfo {volume} {125}},\
  \bibinfo {pages} {227204} (\bibinfo {year} {2020})}\BibitemShut {NoStop}%
\bibitem [{\citenamefont {Budich}\ \emph {et~al.}(2019)\citenamefont {Budich},
  \citenamefont {Carlstr\"om}, \citenamefont {Kunst},\ and\ \citenamefont
  {Bergholtz}}]{Budich19}%
  \BibitemOpen
  \bibfield  {author} {\bibinfo {author} {\bibfnamefont {J.~C.}\ \bibnamefont
  {Budich}}, \bibinfo {author} {\bibfnamefont {J.}~\bibnamefont {Carlstr\"om}},
  \bibinfo {author} {\bibfnamefont {F.~K.}\ \bibnamefont {Kunst}},\ and\
  \bibinfo {author} {\bibfnamefont {E.~J.}\ \bibnamefont {Bergholtz}},\
  }\bibfield  {title} {\bibinfo {title} {{Symmetry-protected nodal phases in
  non-Hermitian systems}},\ }\href {https://doi.org/10.1103/PhysRevB.99.041406}
  {\bibfield  {journal} {\bibinfo  {journal} {Phys. Rev. B}\ }\textbf {\bibinfo
  {volume} {99}},\ \bibinfo {pages} {041406} (\bibinfo {year}
  {2019})}\BibitemShut {NoStop}%
\bibitem [{\citenamefont {Okugawa}\ and\ \citenamefont
  {Yokoyama}(2019)}]{Okugawa19}%
  \BibitemOpen
  \bibfield  {author} {\bibinfo {author} {\bibfnamefont {R.}~\bibnamefont
  {Okugawa}}\ and\ \bibinfo {author} {\bibfnamefont {T.}~\bibnamefont
  {Yokoyama}},\ }\bibfield  {title} {\bibinfo {title} {{Topological exceptional
  surfaces in non-Hermitian systems with parity-time and parity-particle-hole
  symmetries}},\ }\href {https://doi.org/10.1103/PhysRevB.99.041202} {\bibfield
   {journal} {\bibinfo  {journal} {Phys. Rev. B}\ }\textbf {\bibinfo {volume}
  {99}},\ \bibinfo {pages} {041202} (\bibinfo {year} {2019})}\BibitemShut
  {NoStop}%
\bibitem [{\citenamefont {Yoshida}\ \emph {et~al.}(2019)\citenamefont
  {Yoshida}, \citenamefont {Peters}, \citenamefont {Kawakami},\ and\
  \citenamefont {Hatsugai}}]{YoshidaT19}%
  \BibitemOpen
  \bibfield  {author} {\bibinfo {author} {\bibfnamefont {T.}~\bibnamefont
  {Yoshida}}, \bibinfo {author} {\bibfnamefont {R.}~\bibnamefont {Peters}},
  \bibinfo {author} {\bibfnamefont {N.}~\bibnamefont {Kawakami}},\ and\
  \bibinfo {author} {\bibfnamefont {Y.}~\bibnamefont {Hatsugai}},\ }\bibfield
  {title} {\bibinfo {title} {Symmetry-protected exceptional rings in
  two-dimensional correlated systems with chiral symmetry},\ }\href
  {https://doi.org/10.1103/PhysRevB.99.121101} {\bibfield  {journal} {\bibinfo
  {journal} {Phys. Rev. B}\ }\textbf {\bibinfo {volume} {99}},\ \bibinfo
  {pages} {121101} (\bibinfo {year} {2019})}\BibitemShut {NoStop}%
\bibitem [{\citenamefont {Zhou}\ \emph {et~al.}(2019)\citenamefont {Zhou},
  \citenamefont {Lee}, \citenamefont {Liu},\ and\ \citenamefont
  {Zhen}}]{ZhouH19}%
  \BibitemOpen
  \bibfield  {author} {\bibinfo {author} {\bibfnamefont {H.}~\bibnamefont
  {Zhou}}, \bibinfo {author} {\bibfnamefont {J.~Y.}\ \bibnamefont {Lee}},
  \bibinfo {author} {\bibfnamefont {S.}~\bibnamefont {Liu}},\ and\ \bibinfo
  {author} {\bibfnamefont {B.}~\bibnamefont {Zhen}},\ }\bibfield  {title}
  {\bibinfo {title} {{Exceptional surfaces in PT-symmetric non-Hermitian
  photonic systems}},\ }\href@noop {} {\bibfield  {journal} {\bibinfo
  {journal} {Optica}\ }\textbf {\bibinfo {volume} {6}},\ \bibinfo {pages} {190}
  (\bibinfo {year} {2019})}\BibitemShut {NoStop}%
\bibitem [{\citenamefont {Sch\"afer}\ \emph {et~al.}(2022)\citenamefont
  {Sch\"afer}, \citenamefont {Budich},\ and\ \citenamefont
  {Luitz}}]{schafer22}%
  \BibitemOpen
  \bibfield  {author} {\bibinfo {author} {\bibfnamefont {R.}~\bibnamefont
  {Sch\"afer}}, \bibinfo {author} {\bibfnamefont {J.~C.}\ \bibnamefont
  {Budich}},\ and\ \bibinfo {author} {\bibfnamefont {D.~J.}\ \bibnamefont
  {Luitz}},\ }\bibfield  {title} {\bibinfo {title} {Symmetry protected
  exceptional points of interacting fermions},\ }\href
  {https://doi.org/10.1103/PhysRevResearch.4.033181} {\bibfield  {journal}
  {\bibinfo  {journal} {Phys. Rev. Res.}\ }\textbf {\bibinfo {volume} {4}},\
  \bibinfo {pages} {033181} (\bibinfo {year} {2022})}\BibitemShut {NoStop}%
\bibitem [{\citenamefont {Wei}\ and\ \citenamefont {Jin}(2017)}]{wei17}%
  \BibitemOpen
  \bibfield  {author} {\bibinfo {author} {\bibfnamefont {B.-B.}\ \bibnamefont
  {Wei}}\ and\ \bibinfo {author} {\bibfnamefont {L.}~\bibnamefont {Jin}},\
  }\bibfield  {title} {\bibinfo {title} {{Universal critical behaviours in
  non-Hermitian phase transitions}},\ }\href@noop {} {\bibfield  {journal}
  {\bibinfo  {journal} {Scientific reports}\ }\textbf {\bibinfo {volume} {7}},\
  \bibinfo {pages} {7165} (\bibinfo {year} {2017})}\BibitemShut {NoStop}%
\bibitem [{\citenamefont {Yang}\ and\ \citenamefont {Song}(2020)}]{yang20}%
  \BibitemOpen
  \bibfield  {author} {\bibinfo {author} {\bibfnamefont {X.~M.}\ \bibnamefont
  {Yang}}\ and\ \bibinfo {author} {\bibfnamefont {Z.}~\bibnamefont {Song}},\
  }\bibfield  {title} {\bibinfo {title} {{Resonant generation of a $p$-wave
  Cooper pair in a non-Hermitian Kitaev chain at the exceptional point}},\
  }\href {https://doi.org/10.1103/PhysRevA.102.022219} {\bibfield  {journal}
  {\bibinfo  {journal} {Phys. Rev. A}\ }\textbf {\bibinfo {volume} {102}},\
  \bibinfo {pages} {022219} (\bibinfo {year} {2020})}\BibitemShut {NoStop}%
\bibitem [{\citenamefont {Fruchart}\ \emph {et~al.}(2021)\citenamefont
  {Fruchart}, \citenamefont {Hanai}, \citenamefont {Littlewood},\ and\
  \citenamefont {Vitelli}}]{fruchart21}%
  \BibitemOpen
  \bibfield  {author} {\bibinfo {author} {\bibfnamefont {M.}~\bibnamefont
  {Fruchart}}, \bibinfo {author} {\bibfnamefont {R.}~\bibnamefont {Hanai}},
  \bibinfo {author} {\bibfnamefont {P.~B.}\ \bibnamefont {Littlewood}},\ and\
  \bibinfo {author} {\bibfnamefont {V.}~\bibnamefont {Vitelli}},\ }\bibfield
  {title} {\bibinfo {title} {Non-reciprocal phase transitions},\ }\href@noop {}
  {\bibfield  {journal} {\bibinfo  {journal} {Nature (London)}\ }\textbf
  {\bibinfo {volume} {592}},\ \bibinfo {pages} {363} (\bibinfo {year}
  {2021})}\BibitemShut {NoStop}%
\bibitem [{\citenamefont {Kim}\ \emph {et~al.}(2024)\citenamefont {Kim},
  \citenamefont {Han},\ and\ \citenamefont {Park}}]{kim24}%
  \BibitemOpen
  \bibfield  {author} {\bibinfo {author} {\bibfnamefont {B.~H.}\ \bibnamefont
  {Kim}}, \bibinfo {author} {\bibfnamefont {J.-H.}\ \bibnamefont {Han}},\ and\
  \bibinfo {author} {\bibfnamefont {M.~J.}\ \bibnamefont {Park}},\ }\bibfield
  {title} {\bibinfo {title} {{Collective non-Hermitian skin effect: point-gap
  topology and the doublon-holon excitations in non-reciprocal many-body
  systems}},\ }\href@noop {} {\bibfield  {journal} {\bibinfo  {journal}
  {Commun. Phys.}\ }\textbf {\bibinfo {volume} {7}},\ \bibinfo {pages} {73}
  (\bibinfo {year} {2024})}\BibitemShut {NoStop}%
\bibitem [{\citenamefont {Ashida}\ \emph {et~al.}(2017)\citenamefont {Ashida},
  \citenamefont {Furukawa},\ and\ \citenamefont {Ueda}}]{ashida17}%
  \BibitemOpen
  \bibfield  {author} {\bibinfo {author} {\bibfnamefont {Y.}~\bibnamefont
  {Ashida}}, \bibinfo {author} {\bibfnamefont {S.}~\bibnamefont {Furukawa}},\
  and\ \bibinfo {author} {\bibfnamefont {M.}~\bibnamefont {Ueda}},\ }\bibfield
  {title} {\bibinfo {title} {Parity-time-symmetric quantum critical
  phenomena},\ }\href {https://doi.org/10.1038/ncomms15791} {\bibfield
  {journal} {\bibinfo  {journal} {Nat. Commun.}\ }\textbf {\bibinfo {volume}
  {8}},\ \bibinfo {pages} {15791} (\bibinfo {year} {2017})}\BibitemShut
  {NoStop}%
\bibitem [{\citenamefont {Zhang}\ \emph {et~al.}(2022)\citenamefont {Zhang},
  \citenamefont {Denner}, \citenamefont {Bzdu\ifmmode~\check{s}\else
  \v{s}\fi{}ek}, \citenamefont {Sentef},\ and\ \citenamefont
  {Neupert}}]{zhang22}%
  \BibitemOpen
  \bibfield  {author} {\bibinfo {author} {\bibfnamefont {S.-B.}\ \bibnamefont
  {Zhang}}, \bibinfo {author} {\bibfnamefont {M.~M.}\ \bibnamefont {Denner}},
  \bibinfo {author} {\bibfnamefont {T.~c.~v.}\ \bibnamefont
  {Bzdu\ifmmode~\check{s}\else \v{s}\fi{}ek}}, \bibinfo {author} {\bibfnamefont
  {M.~A.}\ \bibnamefont {Sentef}},\ and\ \bibinfo {author} {\bibfnamefont
  {T.}~\bibnamefont {Neupert}},\ }\bibfield  {title} {\bibinfo {title}
  {{Symmetry breaking and spectral structure of the interacting Hatano-Nelson
  model}},\ }\href {https://doi.org/10.1103/PhysRevB.106.L121102} {\bibfield
  {journal} {\bibinfo  {journal} {Phys. Rev. B}\ }\textbf {\bibinfo {volume}
  {106}},\ \bibinfo {pages} {L121102} (\bibinfo {year} {2022})}\BibitemShut
  {NoStop}%
\bibitem [{\citenamefont {Louren\ifmmode~\mbox{\c{c}}\else \c{c}\fi{}o}\ \emph
  {et~al.}(2018)\citenamefont {Louren\ifmmode~\mbox{\c{c}}\else \c{c}\fi{}o},
  \citenamefont {Eneias},\ and\ \citenamefont {Pereira}}]{lourencco18}%
  \BibitemOpen
  \bibfield  {author} {\bibinfo {author} {\bibfnamefont {J.~A.~S.}\
  \bibnamefont {Louren\ifmmode~\mbox{\c{c}}\else \c{c}\fi{}o}}, \bibinfo
  {author} {\bibfnamefont {R.~L.}\ \bibnamefont {Eneias}},\ and\ \bibinfo
  {author} {\bibfnamefont {R.~G.}\ \bibnamefont {Pereira}},\ }\bibfield
  {title} {\bibinfo {title} {{Kondo effect in a $\mathcal{PT}$-symmetric
  non-Hermitian Hamiltonian}},\ }\href
  {https://doi.org/10.1103/PhysRevB.98.085126} {\bibfield  {journal} {\bibinfo
  {journal} {Phys. Rev. B}\ }\textbf {\bibinfo {volume} {98}},\ \bibinfo
  {pages} {085126} (\bibinfo {year} {2018})}\BibitemShut {NoStop}%
\bibitem [{\citenamefont {Hanai}\ \emph {et~al.}(2019)\citenamefont {Hanai},
  \citenamefont {Edelman}, \citenamefont {Ohashi},\ and\ \citenamefont
  {Littlewood}}]{Hanai19}%
  \BibitemOpen
  \bibfield  {author} {\bibinfo {author} {\bibfnamefont {R.}~\bibnamefont
  {Hanai}}, \bibinfo {author} {\bibfnamefont {A.}~\bibnamefont {Edelman}},
  \bibinfo {author} {\bibfnamefont {Y.}~\bibnamefont {Ohashi}},\ and\ \bibinfo
  {author} {\bibfnamefont {P.~B.}\ \bibnamefont {Littlewood}},\ }\bibfield
  {title} {\bibinfo {title} {{Non-Hermitian Phase Transition from a Polariton
  Bose-Einstein Condensate to a Photon Laser}},\ }\href
  {https://doi.org/10.1103/PhysRevLett.122.185301} {\bibfield  {journal}
  {\bibinfo  {journal} {Phys. Rev. Lett.}\ }\textbf {\bibinfo {volume} {122}},\
  \bibinfo {pages} {185301} (\bibinfo {year} {2019})}\BibitemShut {NoStop}%
\bibitem [{\citenamefont {Hanai}\ and\ \citenamefont
  {Littlewood}(2020)}]{hanai20}%
  \BibitemOpen
  \bibfield  {author} {\bibinfo {author} {\bibfnamefont {R.}~\bibnamefont
  {Hanai}}\ and\ \bibinfo {author} {\bibfnamefont {P.~B.}\ \bibnamefont
  {Littlewood}},\ }\bibfield  {title} {\bibinfo {title} {Critical fluctuations
  at a many-body exceptional point},\ }\href
  {https://doi.org/10.1103/PhysRevResearch.2.033018} {\bibfield  {journal}
  {\bibinfo  {journal} {Phys. Rev. Res.}\ }\textbf {\bibinfo {volume} {2}},\
  \bibinfo {pages} {033018} (\bibinfo {year} {2020})}\BibitemShut {NoStop}%
\bibitem [{\citenamefont {Nakagawa}\ \emph {et~al.}(2021)\citenamefont
  {Nakagawa}, \citenamefont {Kawakami},\ and\ \citenamefont
  {Ueda}}]{nakagawa21}%
  \BibitemOpen
  \bibfield  {author} {\bibinfo {author} {\bibfnamefont {M.}~\bibnamefont
  {Nakagawa}}, \bibinfo {author} {\bibfnamefont {N.}~\bibnamefont {Kawakami}},\
  and\ \bibinfo {author} {\bibfnamefont {M.}~\bibnamefont {Ueda}},\ }\bibfield
  {title} {\bibinfo {title} {{Exact Liouvillian Spectrum of a One-Dimensional
  Dissipative Hubbard Model}},\ }\href
  {https://doi.org/10.1103/PhysRevLett.126.110404} {\bibfield  {journal}
  {\bibinfo  {journal} {Phys. Rev. Lett.}\ }\textbf {\bibinfo {volume} {126}},\
  \bibinfo {pages} {110404} (\bibinfo {year} {2021})}\BibitemShut {NoStop}%
\bibitem [{\citenamefont {Yamamoto}\ \emph {et~al.}(2019)\citenamefont
  {Yamamoto}, \citenamefont {Nakagawa}, \citenamefont {Adachi}, \citenamefont
  {Takasan}, \citenamefont {Ueda},\ and\ \citenamefont
  {Kawakami}}]{yamamoto19}%
  \BibitemOpen
  \bibfield  {author} {\bibinfo {author} {\bibfnamefont {K.}~\bibnamefont
  {Yamamoto}}, \bibinfo {author} {\bibfnamefont {M.}~\bibnamefont {Nakagawa}},
  \bibinfo {author} {\bibfnamefont {K.}~\bibnamefont {Adachi}}, \bibinfo
  {author} {\bibfnamefont {K.}~\bibnamefont {Takasan}}, \bibinfo {author}
  {\bibfnamefont {M.}~\bibnamefont {Ueda}},\ and\ \bibinfo {author}
  {\bibfnamefont {N.}~\bibnamefont {Kawakami}},\ }\bibfield  {title} {\bibinfo
  {title} {{Theory of Non-Hermitian Fermionic Superfluidity with a
  Complex-Valued Interaction}},\ }\href
  {https://doi.org/10.1103/PhysRevLett.123.123601} {\bibfield  {journal}
  {\bibinfo  {journal} {Phys. Rev. Lett.}\ }\textbf {\bibinfo {volume} {123}},\
  \bibinfo {pages} {123601} (\bibinfo {year} {2019})}\BibitemShut {NoStop}%
\bibitem [{\citenamefont {Ghatak}\ and\ \citenamefont {Das}(2018)}]{gahtak18}%
  \BibitemOpen
  \bibfield  {author} {\bibinfo {author} {\bibfnamefont {A.}~\bibnamefont
  {Ghatak}}\ and\ \bibinfo {author} {\bibfnamefont {T.}~\bibnamefont {Das}},\
  }\bibfield  {title} {\bibinfo {title} {{Theory of superconductivity with
  non-Hermitian and parity-time reversal symmetric Cooper pairing symmetry}},\
  }\href {https://doi.org/10.1103/PhysRevB.97.014512} {\bibfield  {journal}
  {\bibinfo  {journal} {Phys. Rev. B}\ }\textbf {\bibinfo {volume} {97}},\
  \bibinfo {pages} {014512} (\bibinfo {year} {2018})}\BibitemShut {NoStop}%
\bibitem [{\citenamefont {Kanazawa}(2021)}]{kanazawa21}%
  \BibitemOpen
  \bibfield  {author} {\bibinfo {author} {\bibfnamefont {T.}~\bibnamefont
  {Kanazawa}},\ }\bibfield  {title} {\bibinfo {title} {{Non-Hermitian BCS-BEC
  crossover of Dirac fermions}},\ }\href@noop {} {\bibfield  {journal}
  {\bibinfo  {journal} {J. High Energy Phys.}\ }\textbf {\bibinfo {volume}
  {03}}\bibinfo  {number} { (2021)},\ \bibinfo {pages} {121}}\BibitemShut
  {NoStop}%
\bibitem [{\citenamefont {Iskin}(2021)}]{iskin21}%
  \BibitemOpen
\bibfield  {number} {  }\bibfield  {author} {\bibinfo {author} {\bibfnamefont
  {M.}~\bibnamefont {Iskin}},\ }\bibfield  {title} {\bibinfo {title}
  {{Non-Hermitian BCS-BEC evolution with a complex scattering length}},\ }\href
  {https://doi.org/10.1103/PhysRevA.103.013724} {\bibfield  {journal} {\bibinfo
   {journal} {Phys. Rev. A}\ }\textbf {\bibinfo {volume} {103}},\ \bibinfo
  {pages} {013724} (\bibinfo {year} {2021})}\BibitemShut {NoStop}%
\bibitem [{\citenamefont {He}\ \emph {et~al.}(2021)\citenamefont {He},
  \citenamefont {Ding},\ and\ \citenamefont {Zhu}}]{he21}%
  \BibitemOpen
  \bibfield  {author} {\bibinfo {author} {\bibfnamefont {P.}~\bibnamefont
  {He}}, \bibinfo {author} {\bibfnamefont {H.-T.}\ \bibnamefont {Ding}},\ and\
  \bibinfo {author} {\bibfnamefont {S.-L.}\ \bibnamefont {Zhu}},\ }\bibfield
  {title} {\bibinfo {title} {{Geometry and superfluidity of the flat band in a
  non-Hermitian optical lattice}},\ }\href
  {https://doi.org/10.1103/PhysRevA.103.043329} {\bibfield  {journal} {\bibinfo
   {journal} {Phys. Rev. A}\ }\textbf {\bibinfo {volume} {103}},\ \bibinfo
  {pages} {043329} (\bibinfo {year} {2021})}\BibitemShut {NoStop}%
\bibitem [{\citenamefont {Tajima}\ \emph {et~al.}(2023)\citenamefont {Tajima},
  \citenamefont {Sekino}, \citenamefont {Inotani}, \citenamefont {Dohi},
  \citenamefont {Nagataki},\ and\ \citenamefont {Hayata}}]{tajima23topo}%
  \BibitemOpen
  \bibfield  {author} {\bibinfo {author} {\bibfnamefont {H.}~\bibnamefont
  {Tajima}}, \bibinfo {author} {\bibfnamefont {Y.}~\bibnamefont {Sekino}},
  \bibinfo {author} {\bibfnamefont {D.}~\bibnamefont {Inotani}}, \bibinfo
  {author} {\bibfnamefont {A.}~\bibnamefont {Dohi}}, \bibinfo {author}
  {\bibfnamefont {S.}~\bibnamefont {Nagataki}},\ and\ \bibinfo {author}
  {\bibfnamefont {T.}~\bibnamefont {Hayata}},\ }\bibfield  {title} {\bibinfo
  {title} {{Non-Hermitian topological Fermi superfluid near the $p$-wave
  unitary limit}},\ }\href {https://doi.org/10.1103/PhysRevA.107.033331}
  {\bibfield  {journal} {\bibinfo  {journal} {Phys. Rev. A}\ }\textbf {\bibinfo
  {volume} {107}},\ \bibinfo {pages} {033331} (\bibinfo {year}
  {2023})}\BibitemShut {NoStop}%
\bibitem [{\citenamefont {Li}\ \emph {et~al.}(2023{\natexlab{b}})\citenamefont
  {Li}, \citenamefont {Yu}, \citenamefont {Nakagawa},\ and\ \citenamefont
  {Ueda}}]{li23yang}%
  \BibitemOpen
  \bibfield  {author} {\bibinfo {author} {\bibfnamefont {H.}~\bibnamefont
  {Li}}, \bibinfo {author} {\bibfnamefont {X.-H.}\ \bibnamefont {Yu}}, \bibinfo
  {author} {\bibfnamefont {M.}~\bibnamefont {Nakagawa}},\ and\ \bibinfo
  {author} {\bibfnamefont {M.}~\bibnamefont {Ueda}},\ }\bibfield  {title}
  {\bibinfo {title} {{Yang-Lee Zeros, Semicircle Theorem, and Nonunitary
  Criticality in Bardeen-Cooper-Schrieffer Superconductivity}},\ }\href
  {https://doi.org/10.1103/PhysRevLett.131.216001} {\bibfield  {journal}
  {\bibinfo  {journal} {Phys. Rev. Lett.}\ }\textbf {\bibinfo {volume} {131}},\
  \bibinfo {pages} {216001} (\bibinfo {year} {2023}{\natexlab{b}})}\BibitemShut
  {NoStop}%
\bibitem [{\citenamefont {Tajima}\ \emph {et~al.}(2024)\citenamefont {Tajima},
  \citenamefont {Sekino}, \citenamefont {Inotani}, \citenamefont {Dohi},
  \citenamefont {Nagataki},\ and\ \citenamefont {Hayata}}]{tajima24}%
  \BibitemOpen
  \bibfield  {author} {\bibinfo {author} {\bibfnamefont {H.}~\bibnamefont
  {Tajima}}, \bibinfo {author} {\bibfnamefont {Y.}~\bibnamefont {Sekino}},
  \bibinfo {author} {\bibfnamefont {D.}~\bibnamefont {Inotani}}, \bibinfo
  {author} {\bibfnamefont {A.}~\bibnamefont {Dohi}}, \bibinfo {author}
  {\bibfnamefont {S.}~\bibnamefont {Nagataki}},\ and\ \bibinfo {author}
  {\bibfnamefont {T.}~\bibnamefont {Hayata}},\ }\bibfield  {title} {\bibinfo
  {title} {{Non-Hermitian $p$-wave superfluid and effects of the inelastic
  three-body loss in a one-dimensional spin-polarized Fermi gas}},\ }\href
  {https://doi.org/10.1103/PhysRevResearch.6.023060} {\bibfield  {journal}
  {\bibinfo  {journal} {Phys. Rev. Res.}\ }\textbf {\bibinfo {volume} {6}},\
  \bibinfo {pages} {023060} (\bibinfo {year} {2024})}\BibitemShut {NoStop}%
\bibitem [{\citenamefont {Shi}\ \emph {et~al.}(2024)\citenamefont {Shi},
  \citenamefont {Wang}, \citenamefont {Zheng},\ and\ \citenamefont
  {Zhang}}]{shi24}%
  \BibitemOpen
  \bibfield  {author} {\bibinfo {author} {\bibfnamefont {T.}~\bibnamefont
  {Shi}}, \bibinfo {author} {\bibfnamefont {S.}~\bibnamefont {Wang}}, \bibinfo
  {author} {\bibfnamefont {Z.}~\bibnamefont {Zheng}},\ and\ \bibinfo {author}
  {\bibfnamefont {W.}~\bibnamefont {Zhang}},\ }\bibfield  {title} {\bibinfo
  {title} {{Two-dimensional non-Hermitian fermionic superfluidity with spin
  imbalance}},\ }\href {https://doi.org/10.1103/PhysRevA.109.063306} {\bibfield
   {journal} {\bibinfo  {journal} {Phys. Rev. A}\ }\textbf {\bibinfo {volume}
  {109}},\ \bibinfo {pages} {063306} (\bibinfo {year} {2024})}\BibitemShut
  {NoStop}%
\bibitem [{\citenamefont {Takemori}\ \emph
  {et~al.}(2024{\natexlab{a}})\citenamefont {Takemori}, \citenamefont
  {Yamamoto},\ and\ \citenamefont {Koga}}]{takemori24honey}%
  \BibitemOpen
  \bibfield  {author} {\bibinfo {author} {\bibfnamefont {S.}~\bibnamefont
  {Takemori}}, \bibinfo {author} {\bibfnamefont {K.}~\bibnamefont {Yamamoto}},\
  and\ \bibinfo {author} {\bibfnamefont {A.}~\bibnamefont {Koga}},\ }\bibfield
  {title} {\bibinfo {title} {{Theory of non-Hermitian fermionic superfluidity
  on a honeycomb lattice: Interplay between exceptional manifolds and Van Hove
  singularity}},\ }\href {https://doi.org/10.1103/PhysRevB.109.L060501}
  {\bibfield  {journal} {\bibinfo  {journal} {Phys. Rev. B}\ }\textbf {\bibinfo
  {volume} {109}},\ \bibinfo {pages} {L060501} (\bibinfo {year}
  {2024}{\natexlab{a}})}\BibitemShut {NoStop}%
\bibitem [{\citenamefont {Takemori}\ \emph
  {et~al.}(2024{\natexlab{b}})\citenamefont {Takemori}, \citenamefont
  {Yamamoto},\ and\ \citenamefont {Koga}}]{takemori24asym}%
  \BibitemOpen
  \bibfield  {author} {\bibinfo {author} {\bibfnamefont {S.}~\bibnamefont
  {Takemori}}, \bibinfo {author} {\bibfnamefont {K.}~\bibnamefont {Yamamoto}},\
  and\ \bibinfo {author} {\bibfnamefont {A.}~\bibnamefont {Koga}},\ }\bibfield
  {title} {\bibinfo {title} {{Phase diagram of non-Hermitian BCS superfluids in
  a dissipative asymmetric Hubbard model}},\ }\href
  {https://doi.org/10.1103/PhysRevB.110.184518} {\bibfield  {journal} {\bibinfo
   {journal} {Phys. Rev. B}\ }\textbf {\bibinfo {volume} {110}},\ \bibinfo
  {pages} {184518} (\bibinfo {year} {2024}{\natexlab{b}})}\BibitemShut
  {NoStop}%
\bibitem [{foo()}]{footnote}%
  \BibitemOpen
  \href@noop {} {}\bibinfo {note} {\cred{Here, the NH attractive Hubbard model
  with two-body dissipation is given by $H=-t\sum_{\langle
  i,j\rangle,\sigma}(c_{i\sigma}^{\dagger}c_{j\sigma} + \text{H.c.})
  -(U+i\gamma_{2}/2)\sum_{i}(n_{i\uparrow}-1/2)(n_{i\downarrow}-1/2)$
  \cite{yamamoto19, takemori24asym}.}}\BibitemShut {Stop}%
\bibitem [{\citenamefont {Lindblad}(1976)}]{lindblad76}%
  \BibitemOpen
  \bibfield  {author} {\bibinfo {author} {\bibfnamefont {G.}~\bibnamefont
  {Lindblad}},\ }\bibfield  {title} {\bibinfo {title} {On the generators of
  quantum dynamical semigroups},\ }\href {https://doi.org/10.1007/BF01608499}
  {\bibfield  {journal} {\bibinfo  {journal} {Commun. Math. Phys.}\ }\textbf
  {\bibinfo {volume} {48}},\ \bibinfo {pages} {119} (\bibinfo {year}
  {1976})}\BibitemShut {NoStop}%
\bibitem [{Sup()}]{Supple}%
  \BibitemOpen
  \href@noop {} {}\bibinfo {note} {\cred{See Supplemental Material, which
  includes Ref. [93], for the detailed drivation of the effective DOS on the
  square lattice, the detailed calculation of the superfluid gap and the
  condensation energy for constant $D(\epsilon)$, and the evaluation of the
  superfluid gap at $U\to0$ on the square lattice.}}\BibitemShut {Stop}%
\bibitem [{\citenamefont {Sarma}(1963)}]{sarma63}%
  \BibitemOpen
  \bibfield  {author} {\bibinfo {author} {\bibfnamefont {G.}~\bibnamefont
  {Sarma}},\ }\bibfield  {title} {\bibinfo {title} {{On the influence of a
  uniform exchange field acting on the spins of the conduction electrons in a
  superconductor}},\ }\href
  {https://doi.org/https://doi.org/10.1016/0022-3697(63)90007-6} {\bibfield
  {journal} {\bibinfo  {journal} {J. Phys. Chem. Solids}\ }\textbf {\bibinfo
  {volume} {24}},\ \bibinfo {pages} {1029} (\bibinfo {year}
  {1963})}\BibitemShut {NoStop}%
\bibitem [{\citenamefont {Liu}\ and\ \citenamefont {Wilczek}(2003)}]{liu03}%
  \BibitemOpen
  \bibfield  {author} {\bibinfo {author} {\bibfnamefont {W.~V.}\ \bibnamefont
  {Liu}}\ and\ \bibinfo {author} {\bibfnamefont {F.}~\bibnamefont {Wilczek}},\
  }\bibfield  {title} {\bibinfo {title} {{Interior Gap Superfluidity}},\ }\href
  {https://doi.org/10.1103/PhysRevLett.90.047002} {\bibfield  {journal}
  {\bibinfo  {journal} {Phys. Rev. Lett.}\ }\textbf {\bibinfo {volume} {90}},\
  \bibinfo {pages} {047002} (\bibinfo {year} {2003})}\BibitemShut {NoStop}%
\bibitem [{\citenamefont {Sheehy}\ and\ \citenamefont
  {Radzihovsky}(2006)}]{sheehy06}%
  \BibitemOpen
  \bibfield  {author} {\bibinfo {author} {\bibfnamefont {D.~E.}\ \bibnamefont
  {Sheehy}}\ and\ \bibinfo {author} {\bibfnamefont {L.}~\bibnamefont
  {Radzihovsky}},\ }\bibfield  {title} {\bibinfo {title} {{BEC-BCS Crossover in
  ``Magnetized'' Feshbach-Resonantly Paired Superfluids}},\ }\href
  {https://doi.org/10.1103/PhysRevLett.96.060401} {\bibfield  {journal}
  {\bibinfo  {journal} {Phys. Rev. Lett.}\ }\textbf {\bibinfo {volume} {96}},\
  \bibinfo {pages} {060401} (\bibinfo {year} {2006})}\BibitemShut {NoStop}%
\bibitem [{\citenamefont {Sheehy}\ and\ \citenamefont
  {Radzihovsky}(2007)}]{sheehy07}%
  \BibitemOpen
  \bibfield  {author} {\bibinfo {author} {\bibfnamefont {D.~E.}\ \bibnamefont
  {Sheehy}}\ and\ \bibinfo {author} {\bibfnamefont {L.}~\bibnamefont
  {Radzihovsky}},\ }\bibfield  {title} {\bibinfo {title} {{BEC–BCS crossover,
  phase transitions and phase separation in polarized resonantly-paired
  superfluids}},\ }\href
  {https://doi.org/https://doi.org/10.1016/j.aop.2006.09.009} {\bibfield
  {journal} {\bibinfo  {journal} {Ann. Phys.}\ }\textbf {\bibinfo {volume}
  {322}},\ \bibinfo {pages} {1790} (\bibinfo {year} {2007})}\BibitemShut
  {NoStop}%
\bibitem [{\citenamefont {Barzykin}(2009)}]{barzykin09}%
  \BibitemOpen
  \bibfield  {author} {\bibinfo {author} {\bibfnamefont {V.}~\bibnamefont
  {Barzykin}},\ }\bibfield  {title} {\bibinfo {title} {Magnetic-field-induced
  gapless state in multiband superconductors},\ }\href
  {https://doi.org/10.1103/PhysRevB.79.134517} {\bibfield  {journal} {\bibinfo
  {journal} {Phys. Rev. B}\ }\textbf {\bibinfo {volume} {79}},\ \bibinfo
  {pages} {134517} (\bibinfo {year} {2009})}\BibitemShut {NoStop}%
\bibitem [{\citenamefont {Song}\ \emph {et~al.}(2019)\citenamefont {Song},
  \citenamefont {Yao},\ and\ \citenamefont {Wang}}]{Song19}%
  \BibitemOpen
  \bibfield  {author} {\bibinfo {author} {\bibfnamefont {F.}~\bibnamefont
  {Song}}, \bibinfo {author} {\bibfnamefont {S.}~\bibnamefont {Yao}},\ and\
  \bibinfo {author} {\bibfnamefont {Z.}~\bibnamefont {Wang}},\ }\bibfield
  {title} {\bibinfo {title} {{Non-Hermitian Skin Effect and Chiral Damping in
  Open Quantum Systems}},\ }\href
  {https://doi.org/10.1103/PhysRevLett.123.170401} {\bibfield  {journal}
  {\bibinfo  {journal} {Phys. Rev. Lett.}\ }\textbf {\bibinfo {volume} {123}},\
  \bibinfo {pages} {170401} (\bibinfo {year} {2019})}\BibitemShut {NoStop}%
\bibitem [{\citenamefont {Haga}\ \emph {et~al.}(2021)\citenamefont {Haga},
  \citenamefont {Nakagawa}, \citenamefont {Hamazaki},\ and\ \citenamefont
  {Ueda}}]{haga21}%
  \BibitemOpen
  \bibfield  {author} {\bibinfo {author} {\bibfnamefont {T.}~\bibnamefont
  {Haga}}, \bibinfo {author} {\bibfnamefont {M.}~\bibnamefont {Nakagawa}},
  \bibinfo {author} {\bibfnamefont {R.}~\bibnamefont {Hamazaki}},\ and\
  \bibinfo {author} {\bibfnamefont {M.}~\bibnamefont {Ueda}},\ }\bibfield
  {title} {\bibinfo {title} {{Liouvillian Skin Effect: Slowing Down of
  Relaxation Processes without Gap Closing}},\ }\href
  {https://doi.org/10.1103/PhysRevLett.127.070402} {\bibfield  {journal}
  {\bibinfo  {journal} {Phys. Rev. Lett.}\ }\textbf {\bibinfo {volume} {127}},\
  \bibinfo {pages} {070402} (\bibinfo {year} {2021})}\BibitemShut {NoStop}%
\bibitem [{\citenamefont {Yamamoto}\ \emph {et~al.}(2020)\citenamefont
  {Yamamoto}, \citenamefont {Ashida},\ and\ \citenamefont
  {Kawakami}}]{yamamoto20}%
  \BibitemOpen
  \bibfield  {author} {\bibinfo {author} {\bibfnamefont {K.}~\bibnamefont
  {Yamamoto}}, \bibinfo {author} {\bibfnamefont {Y.}~\bibnamefont {Ashida}},\
  and\ \bibinfo {author} {\bibfnamefont {N.}~\bibnamefont {Kawakami}},\
  }\bibfield  {title} {\bibinfo {title} {Rectification in nonequilibrium steady
  states of open many-body systems},\ }\href
  {https://doi.org/10.1103/PhysRevResearch.2.043343} {\bibfield  {journal}
  {\bibinfo  {journal} {Phys. Rev. Res.}\ }\textbf {\bibinfo {volume} {2}},\
  \bibinfo {pages} {043343} (\bibinfo {year} {2020})}\BibitemShut {NoStop}%
\bibitem [{\citenamefont {Yang}\ \emph {et~al.}(2022)\citenamefont {Yang},
  \citenamefont {Jiang},\ and\ \citenamefont {Bergholtz}}]{Yang22}%
  \BibitemOpen
  \bibfield  {author} {\bibinfo {author} {\bibfnamefont {F.}~\bibnamefont
  {Yang}}, \bibinfo {author} {\bibfnamefont {Q.-D.}\ \bibnamefont {Jiang}},\
  and\ \bibinfo {author} {\bibfnamefont {E.~J.}\ \bibnamefont {Bergholtz}},\
  }\bibfield  {title} {\bibinfo {title} {Liouvillian skin effect in an exactly
  solvable model},\ }\href {https://doi.org/10.1103/PhysRevResearch.4.023160}
  {\bibfield  {journal} {\bibinfo  {journal} {Phys. Rev. Res.}\ }\textbf
  {\bibinfo {volume} {4}},\ \bibinfo {pages} {023160} (\bibinfo {year}
  {2022})}\BibitemShut {NoStop}%
\bibitem [{\citenamefont {Lee}\ \emph {et~al.}(2023)\citenamefont {Lee},
  \citenamefont {McDonald},\ and\ \citenamefont {Clerk}}]{Lee23}%
  \BibitemOpen
  \bibfield  {author} {\bibinfo {author} {\bibfnamefont {G.}~\bibnamefont
  {Lee}}, \bibinfo {author} {\bibfnamefont {A.}~\bibnamefont {McDonald}},\ and\
  \bibinfo {author} {\bibfnamefont {A.}~\bibnamefont {Clerk}},\ }\bibfield
  {title} {\bibinfo {title} {{Anomalously large relaxation times in dissipative
  lattice models beyond the non-Hermitian skin effect}},\ }\href
  {https://doi.org/10.1103/PhysRevB.108.064311} {\bibfield  {journal} {\bibinfo
   {journal} {Phys. Rev. B}\ }\textbf {\bibinfo {volume} {108}},\ \bibinfo
  {pages} {064311} (\bibinfo {year} {2023})}\BibitemShut {NoStop}%
\bibitem [{\citenamefont {Hamanaka}\ \emph {et~al.}(2023)\citenamefont
  {Hamanaka}, \citenamefont {Yamamoto},\ and\ \citenamefont {Yoshida}}]{shu23}%
  \BibitemOpen
  \bibfield  {author} {\bibinfo {author} {\bibfnamefont {S.}~\bibnamefont
  {Hamanaka}}, \bibinfo {author} {\bibfnamefont {K.}~\bibnamefont {Yamamoto}},\
  and\ \bibinfo {author} {\bibfnamefont {T.}~\bibnamefont {Yoshida}},\
  }\bibfield  {title} {\bibinfo {title} {{Interaction-induced Liouvillian skin
  effect in a fermionic chain with a two-body loss}},\ }\href
  {https://doi.org/10.1103/PhysRevB.108.155114} {\bibfield  {journal} {\bibinfo
   {journal} {Phys. Rev. B}\ }\textbf {\bibinfo {volume} {108}},\ \bibinfo
  {pages} {155114} (\bibinfo {year} {2023})}\BibitemShut {NoStop}%
\bibitem [{\citenamefont {Hu}\ \emph {et~al.}(2023)\citenamefont {Hu},
  \citenamefont {Xue}, \citenamefont {Song},\ and\ \citenamefont
  {Wang}}]{Hu23}%
  \BibitemOpen
  \bibfield  {author} {\bibinfo {author} {\bibfnamefont {Y.-M.}\ \bibnamefont
  {Hu}}, \bibinfo {author} {\bibfnamefont {W.-T.}\ \bibnamefont {Xue}},
  \bibinfo {author} {\bibfnamefont {F.}~\bibnamefont {Song}},\ and\ \bibinfo
  {author} {\bibfnamefont {Z.}~\bibnamefont {Wang}},\ }\bibfield  {title}
  {\bibinfo {title} {{Steady-state edge burst: From free-particle systems to
  interaction-induced phenomena}},\ }\href
  {https://doi.org/10.1103/PhysRevB.108.235422} {\bibfield  {journal} {\bibinfo
   {journal} {Phys. Rev. B}\ }\textbf {\bibinfo {volume} {108}},\ \bibinfo
  {pages} {235422} (\bibinfo {year} {2023})}\BibitemShut {NoStop}%
\bibitem [{\citenamefont {Begg}\ and\ \citenamefont {Hanai}(2024)}]{hanai24}%
  \BibitemOpen
  \bibfield  {author} {\bibinfo {author} {\bibfnamefont {S.~E.}\ \bibnamefont
  {Begg}}\ and\ \bibinfo {author} {\bibfnamefont {R.}~\bibnamefont {Hanai}},\
  }\bibfield  {title} {\bibinfo {title} {{Quantum Criticality in Open Quantum
  Spin Chains with Nonreciprocity}},\ }\href
  {https://doi.org/10.1103/PhysRevLett.132.120401} {\bibfield  {journal}
  {\bibinfo  {journal} {Phys. Rev. Lett.}\ }\textbf {\bibinfo {volume} {132}},\
  \bibinfo {pages} {120401} (\bibinfo {year} {2024})}\BibitemShut {NoStop}%
\bibitem [{\citenamefont {Yamamoto}\ \emph {et~al.}(2021)\citenamefont
  {Yamamoto}, \citenamefont {Nakagawa}, \citenamefont {Tsuji}, \citenamefont
  {Ueda},\ and\ \citenamefont {Kawakami}}]{yamamoto21}%
  \BibitemOpen
  \bibfield  {author} {\bibinfo {author} {\bibfnamefont {K.}~\bibnamefont
  {Yamamoto}}, \bibinfo {author} {\bibfnamefont {M.}~\bibnamefont {Nakagawa}},
  \bibinfo {author} {\bibfnamefont {N.}~\bibnamefont {Tsuji}}, \bibinfo
  {author} {\bibfnamefont {M.}~\bibnamefont {Ueda}},\ and\ \bibinfo {author}
  {\bibfnamefont {N.}~\bibnamefont {Kawakami}},\ }\bibfield  {title} {\bibinfo
  {title} {{Collective Excitations and Nonequilibrium Phase Transition in
  Dissipative Fermionic Superfluids}},\ }\href
  {https://doi.org/10.1103/PhysRevLett.127.055301} {\bibfield  {journal}
  {\bibinfo  {journal} {Phys. Rev. Lett.}\ }\textbf {\bibinfo {volume} {127}},\
  \bibinfo {pages} {055301} (\bibinfo {year} {2021})}\BibitemShut {NoStop}%
\bibitem [{\citenamefont {Mazza}\ and\ \citenamefont
  {Schir\`o}(2023)}]{mazza23}%
  \BibitemOpen
  \bibfield  {author} {\bibinfo {author} {\bibfnamefont {G.}~\bibnamefont
  {Mazza}}\ and\ \bibinfo {author} {\bibfnamefont {M.}~\bibnamefont
  {Schir\`o}},\ }\bibfield  {title} {\bibinfo {title} {{Dissipative dynamics of
  a fermionic superfluid with two-body losses}},\ }\href
  {https://doi.org/10.1103/PhysRevA.107.L051301} {\bibfield  {journal}
  {\bibinfo  {journal} {Phys. Rev. A}\ }\textbf {\bibinfo {volume} {107}},\
  \bibinfo {pages} {L051301} (\bibinfo {year} {2023})}\BibitemShut {NoStop}%
\bibitem [{\citenamefont {Minganti}\ \emph {et~al.}(2019)\citenamefont
  {Minganti}, \citenamefont {Miranowicz}, \citenamefont {Chhajlany},\ and\
  \citenamefont {Nori}}]{mitnganti19}%
  \BibitemOpen
  \bibfield  {author} {\bibinfo {author} {\bibfnamefont {F.}~\bibnamefont
  {Minganti}}, \bibinfo {author} {\bibfnamefont {A.}~\bibnamefont
  {Miranowicz}}, \bibinfo {author} {\bibfnamefont {R.~W.}\ \bibnamefont
  {Chhajlany}},\ and\ \bibinfo {author} {\bibfnamefont {F.}~\bibnamefont
  {Nori}},\ }\bibfield  {title} {\bibinfo {title} {{Quantum exceptional points
  of non-Hermitian Hamiltonians and Liouvillians: The effects of quantum
  jumps}},\ }\href {https://doi.org/10.1103/PhysRevA.100.062131} {\bibfield
  {journal} {\bibinfo  {journal} {Phys. Rev. A}\ }\textbf {\bibinfo {volume}
  {100}},\ \bibinfo {pages} {062131} (\bibinfo {year} {2019})}\BibitemShut
  {NoStop}%
\bibitem [{\citenamefont {Yamamoto}\ and\ \citenamefont
  {Hamazaki}(2023)}]{yamamoto23local}%
  \BibitemOpen
  \bibfield  {author} {\bibinfo {author} {\bibfnamefont {K.}~\bibnamefont
  {Yamamoto}}\ and\ \bibinfo {author} {\bibfnamefont {R.}~\bibnamefont
  {Hamazaki}},\ }\bibfield  {title} {\bibinfo {title} {Localization properties
  in disordered quantum many-body dynamics under continuous measurement},\
  }\href {https://doi.org/10.1103/PhysRevB.107.L220201} {\bibfield  {journal}
  {\bibinfo  {journal} {Phys. Rev. B}\ }\textbf {\bibinfo {volume} {107}},\
  \bibinfo {pages} {L220201} (\bibinfo {year} {2023})}\BibitemShut {NoStop}%
\bibitem [{\citenamefont {Yamamoto}\ and\ \citenamefont
  {Hamazaki}(2025)}]{yamamoto25}%
  \BibitemOpen
  \bibfield  {author} {\bibinfo {author} {\bibfnamefont {K.}~\bibnamefont
  {Yamamoto}}\ and\ \bibinfo {author} {\bibfnamefont {R.}~\bibnamefont
  {Hamazaki}},\ }\bibfield  {title} {\bibinfo {title} {{Measurement-Induced
  Crossover of Quantum Jump Statistics in Postselection-Free Many-Body
  Dynamics}},\ }\href@noop {} {\bibfield  {journal} {\bibinfo  {journal}
  {arXiv:2503.02418}\ } (\bibinfo {year} {2025})}\BibitemShut {NoStop}%
\bibitem [{\citenamefont {Livi}\ \emph {et~al.}(2016)\citenamefont {Livi},
  \citenamefont {Cappellini}, \citenamefont {Diem}, \citenamefont {Franchi},
  \citenamefont {Clivati}, \citenamefont {Frittelli}, \citenamefont {Levi},
  \citenamefont {Calonico}, \citenamefont {Catani}, \citenamefont {Inguscio},\
  and\ \citenamefont {Fallani}}]{livi16}%
  \BibitemOpen
  \bibfield  {author} {\bibinfo {author} {\bibfnamefont {L.~F.}\ \bibnamefont
  {Livi}}, \bibinfo {author} {\bibfnamefont {G.}~\bibnamefont {Cappellini}},
  \bibinfo {author} {\bibfnamefont {M.}~\bibnamefont {Diem}}, \bibinfo {author}
  {\bibfnamefont {L.}~\bibnamefont {Franchi}}, \bibinfo {author} {\bibfnamefont
  {C.}~\bibnamefont {Clivati}}, \bibinfo {author} {\bibfnamefont
  {M.}~\bibnamefont {Frittelli}}, \bibinfo {author} {\bibfnamefont
  {F.}~\bibnamefont {Levi}}, \bibinfo {author} {\bibfnamefont {D.}~\bibnamefont
  {Calonico}}, \bibinfo {author} {\bibfnamefont {J.}~\bibnamefont {Catani}},
  \bibinfo {author} {\bibfnamefont {M.}~\bibnamefont {Inguscio}},\ and\
  \bibinfo {author} {\bibfnamefont {L.}~\bibnamefont {Fallani}},\ }\bibfield
  {title} {\bibinfo {title} {{Synthetic Dimensions and Spin-Orbit Coupling with
  an Optical Clock Transition}},\ }\href
  {https://doi.org/10.1103/PhysRevLett.117.220401} {\bibfield  {journal}
  {\bibinfo  {journal} {Phys. Rev. Lett.}\ }\textbf {\bibinfo {volume} {117}},\
  \bibinfo {pages} {220401} (\bibinfo {year} {2016})}\BibitemShut {NoStop}%
\bibitem [{\citenamefont {Chin}\ \emph {et~al.}(2004)\citenamefont {Chin},
  \citenamefont {Bartenstein}, \citenamefont {Altmeyer}, \citenamefont {Riedl},
  \citenamefont {Jochim}, \citenamefont {Denschlag},\ and\ \citenamefont
  {Grimm}}]{chin04}%
  \BibitemOpen
  \bibfield  {author} {\bibinfo {author} {\bibfnamefont {C.}~\bibnamefont
  {Chin}}, \bibinfo {author} {\bibfnamefont {M.}~\bibnamefont {Bartenstein}},
  \bibinfo {author} {\bibfnamefont {A.}~\bibnamefont {Altmeyer}}, \bibinfo
  {author} {\bibfnamefont {S.}~\bibnamefont {Riedl}}, \bibinfo {author}
  {\bibfnamefont {S.}~\bibnamefont {Jochim}}, \bibinfo {author} {\bibfnamefont
  {J.~H.}\ \bibnamefont {Denschlag}},\ and\ \bibinfo {author} {\bibfnamefont
  {R.}~\bibnamefont {Grimm}},\ }\bibfield  {title} {\bibinfo {title}
  {{Observation of the Pairing Gap in a Strongly Interacting Fermi Gas}},\
  }\href {https://doi.org/10.1126/science.1100818} {\bibfield  {journal}
  {\bibinfo  {journal} {Science}\ }\textbf {\bibinfo {volume} {305}},\ \bibinfo
  {pages} {1128} (\bibinfo {year} {2004})}\BibitemShut {NoStop}%
\bibitem [{\citenamefont {Schunck}\ \emph {et~al.}(2008)\citenamefont
  {Schunck}, \citenamefont {Shin}, \citenamefont {Schirotzek},\ and\
  \citenamefont {Ketterle}}]{schunck08}%
  \BibitemOpen
  \bibfield  {author} {\bibinfo {author} {\bibfnamefont {C.~H.}\ \bibnamefont
  {Schunck}}, \bibinfo {author} {\bibfnamefont {Y.-i.}\ \bibnamefont {Shin}},
  \bibinfo {author} {\bibfnamefont {A.}~\bibnamefont {Schirotzek}},\ and\
  \bibinfo {author} {\bibfnamefont {W.}~\bibnamefont {Ketterle}},\ }\bibfield
  {title} {\bibinfo {title} {Determination of the fermion pair size in a
  resonantly interacting superfluid},\ }\href@noop {} {\bibfield  {journal}
  {\bibinfo  {journal} {Nature (London)}\ }\textbf {\bibinfo {volume} {454}},\
  \bibinfo {pages} {739} (\bibinfo {year} {2008})}\BibitemShut {NoStop}%
\bibitem [{\citenamefont {Schirotzek}\ \emph {et~al.}(2008)\citenamefont
  {Schirotzek}, \citenamefont {Shin}, \citenamefont {Schunck},\ and\
  \citenamefont {Ketterle}}]{schirotzek08}%
  \BibitemOpen
  \bibfield  {author} {\bibinfo {author} {\bibfnamefont {A.}~\bibnamefont
  {Schirotzek}}, \bibinfo {author} {\bibfnamefont {Y.-i.}\ \bibnamefont
  {Shin}}, \bibinfo {author} {\bibfnamefont {C.~H.}\ \bibnamefont {Schunck}},\
  and\ \bibinfo {author} {\bibfnamefont {W.}~\bibnamefont {Ketterle}},\
  }\bibfield  {title} {\bibinfo {title} {Determination of the superfluid gap in
  atomic fermi gases by quasiparticle spectroscopy},\ }\href
  {https://doi.org/10.1103/PhysRevLett.101.140403} {\bibfield  {journal}
  {\bibinfo  {journal} {Phys. Rev. Lett.}\ }\textbf {\bibinfo {volume} {101}},\
  \bibinfo {pages} {140403} (\bibinfo {year} {2008})}\BibitemShut {NoStop}%
\end{thebibliography}

%

\onecolumngrid
\begin{center}
   \textbf{End Matter}
 \end{center}

\twocolumngrid

\cred{\textit{Appendix A: Experimental implementation}--}
\cred{In this section, we explain the experimental setup of our system by following the studies~\cite{gong18,takemori24asym}. For the sake of simplicity, we consider the one-dimensional system, and its extension for higher dimensional systems follows similarly. As illustrated in Fig.~\ref{sdBCSrepa_primary_auxiliary_optlat_image}, we consider ultracold fermionic atoms with four hyperfine states $|g,\sigma\rangle,|e,\sigma\rangle\; (\sigma=\uparrow,\downarrow)$ in an optical lattice. The potential minima of the auxiliary lattice for $|e,\sigma\rangle$ are positioned at the middle of those for the primary lattice. The auxiliary lattice undergoes the spin-dependent one-body loss with rate $\kappa_{\sigma}$. Furthermore, we introduce two running-wave lasers; one propagates from the left and couples the ground state with $|e,\uparrow\rangle$ with the strength $\Omega_{\uparrow}$, while the other propagates from the right and couples the ground state with $|e,\downarrow\rangle$ with the strength $\Omega_{\downarrow}$.}

\cred{The dissipative dynamics is described by the following Lindblad equation~\cite{daley14,yamamoto23local,yamamoto25}
\begin{equation}
  \frac{\partial \rho}{\partial t} = -i[H,\rho] + \sum_{i,\sigma}\mathcal{D}[L_{i,\sigma}]\rho,
\end{equation}
where $\rho$ is the density matrix. We introduce the system Hamiltonian as $H=H_{g}+H_{e}+V$ with
\begin{align}
  H_{g} = &-t \sum_{i,\sigma} (c_{i\sigma g}^{\dagger}c_{i+1\sigma g} + \text{H.c.})\notag\\ &-U\sum_{i}(n_{i\uparrow g}-\frac{1}{2})(n_{i\downarrow g}-\frac{1}{2}),  \\
  H_{e} = &-t \sum_{i,\sigma} (c_{i\sigma e}^{\dagger}c_{i+1\sigma e}  + \text{H.c.}),                                 \\
  V  = &\frac{\Omega_{\sigma}}{2}\sum_{i,\sigma} [c_{i\sigma e}^{\dagger}(c_{i\sigma g}+i\sigma c_{i+1\sigma g}) + \text{H.c.}],
\end{align}
where $t$ is the hopping amplitude, and $U$ is the on-site attraction strength in the primary lattice. The dissipator for the Lindblad operator $L_{m}$ is given by
\begin{equation}
  \mathcal{D}[L_{m}]\rho = L_{m}\rho {L_{m}}^{\dagger} -\frac{1}{2}\{{L_{m}}^{\dagger}L_{m},\rho\},
\end{equation}
where $L_{i,\sigma}=\sqrt{\kappa_{\sigma}}c_{i\sigma e}$ describes the spin-dependent one-body loss in the auxiliary lattice with rate $\kappa_{\sigma}$.}
\cred{When $\kappa_{\sigma}\gg \Omega_{\sigma}$, we can employ the adiabatically elimination of the auxiliary degree of freedom~\cite{gong18}. This scheme effectively transforms the local spin-dependent one-body loss in the auxiliary lattice into the collective spin-dependent loss in the primary lattice. After the adiabatically elimination, we obtain the effective Lindblad equation as
\begin{equation}
  \frac{\partial\rho}{\partial t} = -i[H_{g},\rho] + \sum_{i,\sigma}\mathcal{D}[L_{\text{eff},i\sigma}]\rho.
\end{equation}
Here, the Lindblad operator is given by
\begin{equation}
  L_{\text{eff},i\sigma} = \sqrt{\Gamma}(c_{i\sigma g} + i \sigma c_{i+1\sigma g}),
\end{equation}
where we have introduced $\Gamma=\Omega^{2}/\kappa$, $\Omega=\Omega_\uparrow=\Omega_\downarrow$, and $\kappa=\kappa_\uparrow=\kappa_\downarrow$. Then, we obtain the NH Hamiltonian given in Eq.~\eqref{sdBCS_effNHHamiltonian_eq} by postselecting the null measurement outcome with the use of a quantum-gas microscopes~\cite{ashida16}.}
\begin{figure}[tb]
  \centering
  \includegraphics[width = \linewidth]{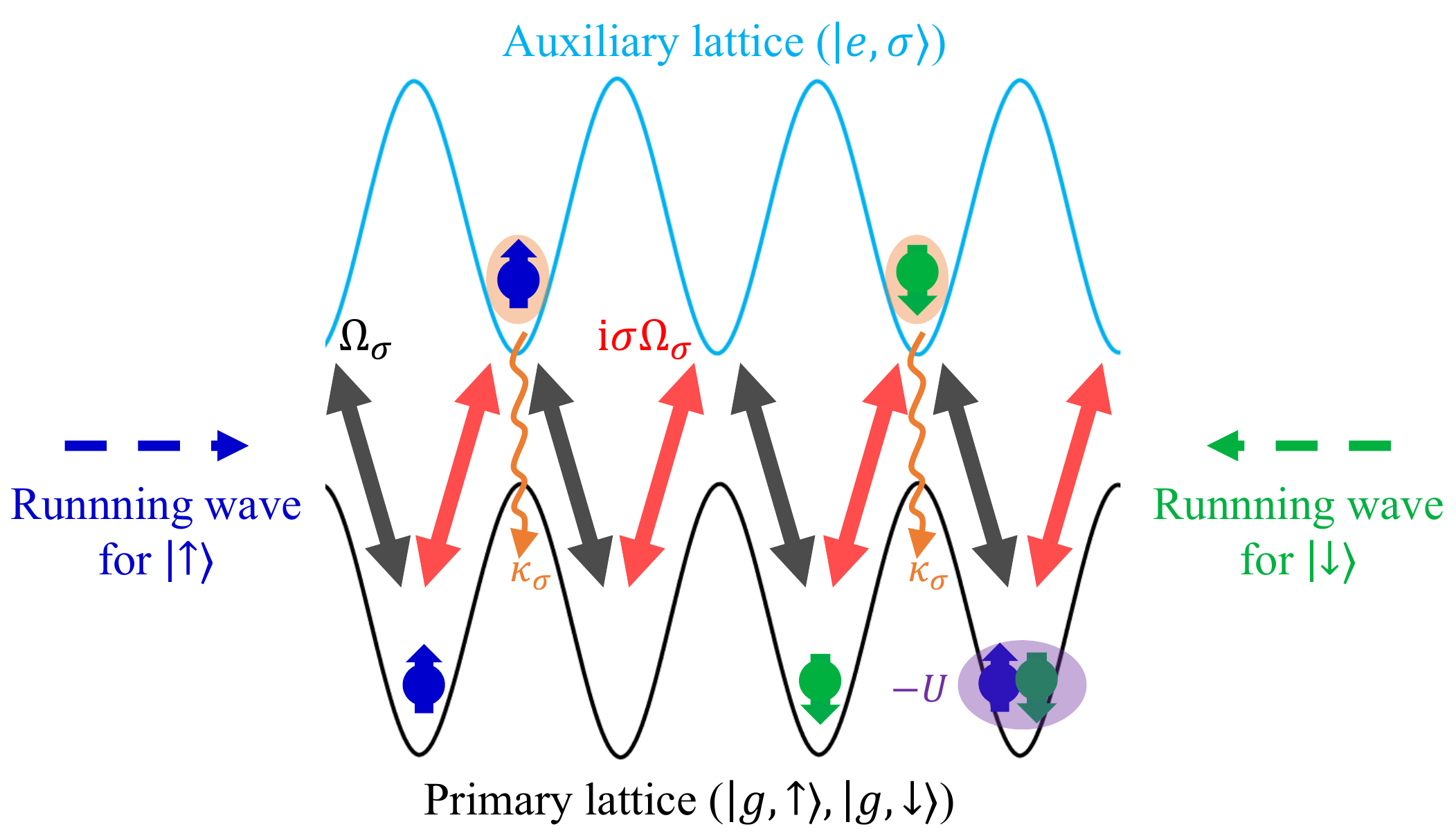}
  \caption{\cred{Schematic image of the experimental setup. We consider the dissipative dynamics of fermionic atoms in the primary (black) and auxiliary (light blue) lattices. Rabi coupling between these two lattices is introduced in a spin-dependent manner: $\Omega_{\uparrow}$ ($\Omega_{\downarrow}$) is introduced by using a running wave from left (right), and the rate is changed by $i$ ($-i$) compared to the left nearest sites. The auxiliary lattice is subject to the spin-dependent one-body loss with rate $\kappa_{\sigma}$. On-site attractive interaction $-U$ is also introduced in the primary lattice.}}
 \label{sdBCSrepa_primary_auxiliary_optlat_image}
\end{figure}

\cred{Next, 
we provide the estimate of the realistic parameters for the experimental setup. Our system can be implemented by using hyperfine states in, e.g., $^{6}\mathrm{Li}$, $^{40}\mathrm{K}$, and $^{173}\mathrm{Yb}$ atoms in an optical lattice. In the following, we estimate experimental parameters for $^{173}\mathrm{Yb}$. 
First, the hopping amplitude $t$ can be tunable by adjusting the lattice depth. The attractive interaction $U$ is controlled by using the Feshbach resonance. For the repulsive Fermi-Hubbard model, the hopping amplitude and attractive interaction strength are estimated to be several $\mathrm{kHz}$~\cite{honda23}.
Second, the collective one-body loss rate $\Gamma$ and the strength of spin-depairing $\gamma$ can be controlled by changing the Rabi coupling $\Omega$ and the one-body loss rate $\kappa$. In Ref.~\cite{livi16}, the Rabi coupling between two hyperfine states is estimated to be several $\mathrm{kHz}$. The one-body loss can be introduced by using a near-resonant optical beam with loss rate of several $\mathrm{kHz}$~\cite{ren22}. To realize our model, enhancing the loss rate by using a high power beam may be necessary. We estimate the required one-body loss rate $\kappa$ for the spin-depairing $\gamma=0.4$. If we use $\Omega\sim 2\pi \times 0.59 \mathrm{kHz}$ and $t\sim 2\pi\times 0.32 \mathrm{kHz}$ given in Refs.~\cite{honda23,livi16}, the one-body loss rate $\kappa= \Omega^{2}/(2t\gamma)\sim 2\pi \times 8.5 \mathrm{kHz}$ should be necessary. Thus, we expect that our model can be accessed in the experiment.
}

\cred{Finally, we note that exceptional SF is derived for the bulk system in the thermodynamic limit and large experimental platforms in ultracold atoms (say order of $L=10^2$) are useful to obtain theoretical predictions of the order parameter. In experiments, we expect that the superfluid gap and the excitation spectrum can be observed by using the rf spectroscopy~\cite{chin04,schunck08,schirotzek08}. 
}

\cred{\textit{Appendix B: Discontinuous jump of the order parameter on a cubic lattice}--}

\cred{
The gap equation [Eq. (10) in the main text] is rewritten as,
\begin{align}
  U&=F(\Delta_0), \\
  F(\Delta_{0}) &= \left(\int\!\!\!\!\int\frac{D_\gamma(\epsilon)}{2\sqrt{\epsilon^{2}+\Delta_0^2}} \; \mathrm{d}\text{Re}\,\epsilon \:\mathrm{d}\text{Im}\,\epsilon   \right)^{-1}. \label{sdBCSrepa_FR}
\end{align}
In Fig.~\ref{sdBCSrepa_gapeqRLHS_image}, we show $F(\Delta_0)$ for the systems on the cubic and square lattices.
In the large-$\Delta_0$ region,
$F(\Delta_0)$ increases monotonically, 
indicating the existence of a trivial physical solution
and hence the realization of the superfluid state.
In the cubic lattice, $F(\Delta_0)$ exhibits a minimum,
implying that no physical solution exists for $U<U_c (U_c\sim 4.59)$,
which leads to the discontinuous jump in the order parameter $\Delta_0$.
This nonmonotonic behavior arises from 
the interplay between the singularity of the effective DOS and EPs.
When $\Delta_{0}>1$, EPs appear at $\epsilon=i\Delta_0$ in the complex $\epsilon$ plane.
This means that the integrand in $F$ diverges at $\epsilon=i\Delta_0$, and
$D_{\gamma}(\epsilon\sim i\Delta_0)$ dominates $F$.
Since the singularity of the effective DOS is located at $\epsilon=i$,
this yields the minimum in $F$, as shown in Fig.~\ref{sdBCSrepa_gapeqRLHS_image}(a).
We note that the solution for $\Delta_{0}< 1$,
which is shown as the dashed line in Fig.~\ref{sdBCSrepa_gapeqRLHS_image},
is not physical and should be omitted
because they does not adiabatically connect to the solution in the Hermitian limit $\gamma\to 0$.
In the square lattice case, $F(\Delta_0)$ approaches zero as $\Delta_0\rightarrow 0$,
so the superfluid solution always exists
although the order parameter is extremely small for small $U$.
}
\begin{figure}[htb]
  \centering
  \includegraphics[width=\linewidth]{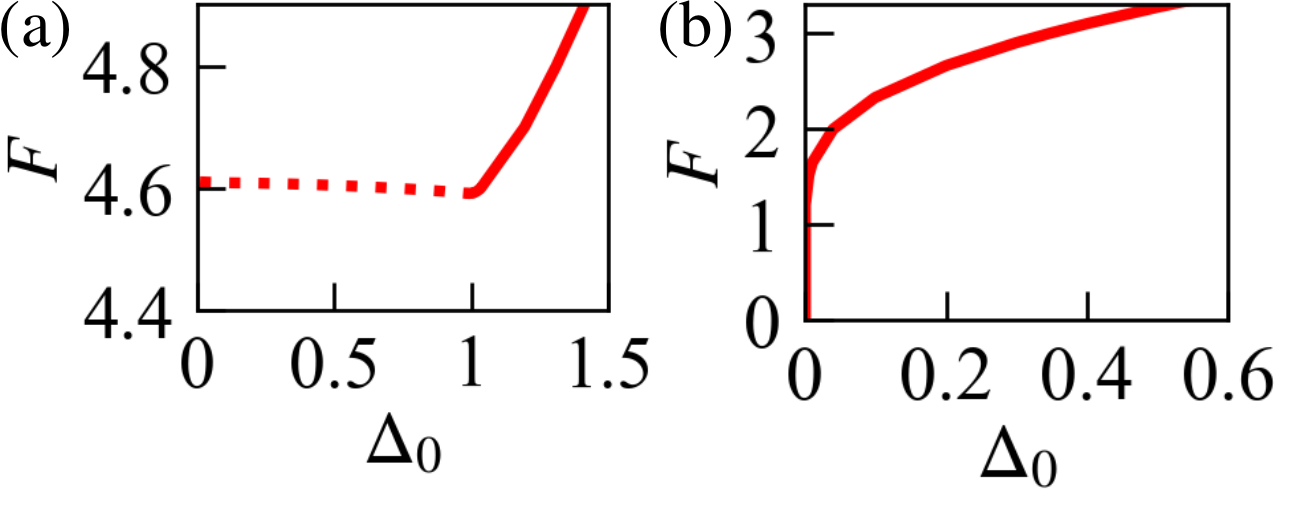}
  \caption{\cred{Lines represent $F(\Delta_0)$, defined in Eq.~\eqref{sdBCSrepa_FR},
    on (a) cubic and (b) square lattices.
    In~(a), solid (dashed) line corresponds to the physical (unphysical) solution of
    the self-consistency equation $U=F(\Delta_0)$.}}
  \label{sdBCSrepa_gapeqRLHS_image}
\end{figure}

\clearpage

\renewcommand{\thesection}{S\arabic{section}}
\renewcommand{\theequation}{S\arabic{equation}}
\setcounter{equation}{0}
\renewcommand{\thefigure}{S\arabic{figure}}
\setcounter{figure}{0}

\onecolumngrid
\appendix
\begin{center}
\large{Supplemental Material for}\\
\textbf{''Spin-Depairing-Induced Exceptional Fermionic Superfluidity"}
\end{center}

\maketitle

\section{Detailed derivation of the effective density of states for the square lattice}
We explain the detailed derivation of the effective density of states (DOS) for the square lattice. First, we show the relation between the effective DOS for $\gamma=1$ and that for $\gamma=\gamma_{0}\neq 1$. We can rewrite the effective DOS given in Eq. (9) in the main text as
\begin{equation}
    D_{\gamma}(\epsilon=x+iy) = \frac{1}{N}\sum_{\bm{k}}\delta(x-\text{Re}\epsilon_{\bm{k}})\delta(y-\text{Im}\epsilon_{\bm{k}}).
\end{equation}
Using the relation $\text{Im}\epsilon_{\bm{k}}|_{\gamma=\gamma_{0}}=\gamma_{0}\text{Im}\epsilon_{\bm{k}}|_{\gamma=1}$ and $\delta(ax)=\delta(x)/a$ ($a>0$), we obtain the following relation:
\begin{equation}
    D_{\gamma_{0}}(x+iy) = \frac{1}{\gamma_{0}}D_{\gamma=1}(x+iY),
\end{equation}
where $Y=y/\gamma_{0}$.
This relation indicates that the effective DOS for $\gamma=1$ contains all the information of $D_{\gamma}(\epsilon)$.

Next, we show the explicit form of the effective DOS on the square lattice for $\gamma=1$. Since the effective DOS for $\gamma=1$ only depends on the absolute value of the energy $\epsilon$, we get
\begin{equation}
    D_{\gamma=1}(\epsilon=re^{i\theta})=D(r) = \frac{1}{2\pi rN}\sum_{\bm{k}}\delta(r-|\epsilon_{\bm{k}}|) = \frac{1}{2\pi r}R(r),
\end{equation}
where $R(r)=\frac{1}{N}\sum_{\bm{k}}\delta(r-|\epsilon_{\bm{k}}|)$. We can rewrite $R(r)$ as
\begin{align}
    R(r)=\int_{-\pi}^{\pi}\mathrm{d}k_{y}\int_{-\pi}^{\pi}\mathrm{d}k_{x}\delta\left(r-4\left|\cos\left(\frac{k_{x}-k_{y}}{2}\right)\right|\right).
\end{align}
Using the relation $\cos(-x)=\cos(x)$ and introducing $p=k_{x}-k_{y}$, we obtain 
\begin{equation}
    R(r) = 2\int_{-\pi}^{\pi}\mathrm{d}k_{y}\int_{0}^{\pi-k_{y}}\mathrm{d}p\:\delta\left(r-4\left|\cos\frac{p}{2}\right|\right).
\end{equation}
Then, using the following property of the delta function:
\begin{equation}
    \delta(f(x)) = \sum_{i}\frac{1}{|\mathrm{d}f(a_{i})/\mathrm{d}x|}\delta(x-a_{i}),
\end{equation}
where $a_{i}$ is the zeros of the function $f(x)$, we can carry out the integration over $p$ and $k_{y}$ and arrive at
\begin{equation}
    R(r) = \Theta(4-r)\frac{2}{\pi}\frac{1}{\sqrt{16-r^{2}}}.
\end{equation}
Here, $\Theta(r)$ is the step function. Finally, we obtain the effective DOS for the square lattice as
\begin{equation}
    D(r) = \frac{\Theta(4-r)}{\pi^{2}}\frac{1}{r\sqrt{16-r^{2}}}.
\end{equation}

\section{Detailed calculation of the order parameter and the condensation energy for the constant effective density of states}
We show the detailed calculation of the order parameter and the condensation energy for the constant effective DOS given in Eq. (11) in the main text. First, we show the analytical solution of the NH gap equation (10) in the main text by taking into account the fact that the physical order parameter should be real. We can rewrite the NH gap equation for the constant effective DOS as
\begin{align}
    \frac{1}{U} &= \frac{1}{\pi \epsilon_{R}^{2}}\int\mathrm{d}\text{Re}\epsilon\mathrm{d}\text{Im}\epsilon \frac{1}{2\sqrt{\epsilon^{2}+\Delta_{0}^{2}}} \notag \\
    &= \frac{1}{\pi \epsilon_{R}^{2}}\int_{0}^{\epsilon_{R}}\dd{r}\int_{-\pi}^{\pi}\dd{\theta}\frac{r}{2\sqrt{r^{2}e^{2i\theta}+\Delta_{0}^{2}}} \notag \\
    &= \frac{1}{\pi \epsilon_{R}^{2}}\int_{0}^{\epsilon_{R}}\dd{r} 2\pi r R_{1}(r), \label{sdBCS_App_gapeq_analy_eq1}
\end{align}
where
\begin{align}
    R_{1}(r) &= \frac{1}{4\pi r}\int_{-\pi}^{\pi}\dd{\theta}\frac{r}{\sqrt{r^{2}e^{2i\theta}+\Delta_{0}^{2}}} \notag \\
    &=-i\frac{1}{4\pi r} \oint_{|z|=1}\dd{z} \frac{1}{z\sqrt{z^{2}+(\Delta_{0}/r)^{2}}},
\end{align}
and we have introduced $z=e^{i\theta}$. By performing the contour integral with the use of the residue theorem, we get
\begin{align}
      R_{1}(r) &= \begin{cases}
            \frac{1}{2\Delta_{0}},                                                                                & \quad (\Delta_{0}>r ),   \\
            \frac{1}{2\Delta_{0}}-\frac{1}{\pi\Delta_{0} }\arctan(\frac{\sqrt{r^{2}-\Delta_{0}^{2}}}{\Delta_{0}}), & \quad (\Delta_{0}\le r ).
          \end{cases} \label{sdBCS_App_gapeq_analy_R1_eq}
\end{align}
We note that this calculation is also applicable for the square and cubic lattices with $\delta=1$. Here, we have to pay attention to the fact that the pole of the integrand lies inside the contour for $\Delta_{0}\le r$. Then, by perfoming the integration in Eq.~\eqref{sdBCS_App_gapeq_analy_eq1} over $r$, we obtain the order parameter given in Eq. (12) in the main text.

Similarly, we can analytically evaluate the condensation energy for the constant effective DOS as
\begin{align}
    E_{\text{cond}} &= \frac{\Delta_{0}^{2}}{U} - \frac{1}{\pi \epsilon_{R}^{2}}\int\dd{\text{Re}\epsilon}\dd{\text{Im}}\epsilon (\sqrt{\epsilon^{2}+\Delta_{0}^{2}} - \sqrt{\epsilon^{2}}) \notag \\
    &= \frac{\Delta_{0}^{2}}{U} - \frac{1}{\pi \epsilon_{R}^{2}}\int_{0}^{\epsilon_{R}}\dd{r}2\pi r R_{2}(r),
\end{align}
where
\begin{align}
    R_{2}(r) &= \frac{1}{2\pi r}\int_{-\pi}^{\pi} \dd{\theta} (r\sqrt{r^{2}e^{2i\theta}+\Delta_{0}^{2}} - r^{2}\sqrt{e^{2i\theta}}).
\end{align}
We proceed with the calculation by using the contour integration, obtaining
\begin{align}
     R_{2}(r) &= \begin{cases}
       \Delta_{0} - \frac{2r}{\pi}, & \quad (\Delta_{0} >r ),   \\
       &\rule{0pt}{0.5ex} \\ 
       \Delta_{0} +\frac{2}{\pi}\left[\sqrt{r^{2}-\Delta_{0} ^{2}}- \Delta_{0} \arctan(\frac{\sqrt{r^{2} - \Delta_{0} ^{2}}}{\Delta_{0} })\right]- \frac{2r}{\pi}, & \quad (\Delta_{0}\le r ).
     \end{cases}
\end{align}
Then, perfoming the integral over $r$, we get 
Eq. (13) in the main text. 
We note that we have introduced the dimensionless parameters in the main text as $E_{\text{cond}}'\equiv E_{\text{cond}}\pi/\epsilon_{R}$, $U^\prime\equiv U/(\pi \epsilon_{R})$, and $\Delta^\prime\equiv\Delta_{0}/\epsilon_{R}$.
For $\Delta^\prime\ll 1$, the condensation energy behaves asymptotically as $E_{\text{cond}}'\sim -\Delta'^{4}/6$, which is different from the results in the NH system for the constant DOS, where the condensation energy behaves as $E_{\text{cond}}^\prime\sim -\Delta^{\prime 2}$ for weak dissipation \cite{yamamoto19}. 

\section{Evaluation of the order parameter at $U\to0$ on the square lattice for $\gamma=1$}
We analytically evaluate the order parameter, which is numerically obtained in Fig. 6 in the main text. Though it is difficult to get the full analytic form of the order parameter, we can estimate the exponential decay at $U\to0$ limit for the square lattice by using 
\begin{equation}
    D(r)=\begin{cases}
        1/(4\pi^{2}r), & \; (r\le 2\pi), \\
        0, & \; (r>2\pi),
    \end{cases} 
\end{equation}
which reflects the singularity of the effective DOS at $|\epsilon|=0$. By using the fact that Eq.~\eqref{sdBCS_App_gapeq_analy_R1_eq} is applicable to this case and performing the integration over $r$, the NH gap equation is rewritten as
\begin{equation}
    \frac{1}{U} \sim \begin{cases}
      \frac{1}{2\Delta_{0}}, &\; (\Delta_{0}> 2\pi), \\
      &\rule{0pt}{0.5ex} \\ 
      \frac{1}{2\Delta_{0}} - \frac{1}{2\pi^{2}\Delta_{0}}\left[ 2\pi\arctan(\frac{\sqrt{4\pi^{2} -\Delta_{0}^{2}}}{\Delta_{0}}) - \Delta_{0}\log(\frac{2\pi+\sqrt{4\pi^{2} -\Delta_{0}^{2}}}{\Delta_{0}}) \right], &\; (\Delta_{0}\le 2\pi).
    \end{cases}
\end{equation}
For $\Delta_{0}\ll 1$, we get
\begin{equation}
    \frac{1}{U} \sim -\frac{1}{2\pi^{2}}\log \Delta_{0}.
\end{equation}
Then, we obtain $\Delta_{0}= \exp(-2\pi^{2}/U)$, which correctly captures the exponential decay at $U\to0$. This indicates that the order parameter becomes finite for arbitrarily small $U$.

\end{document}